\documentclass[11pt]{article}
\usepackage{fullpage,amsmath, amsfonts, amsthm,xr}
\usepackage{graphicx, algorithmic, algorithm, enumerate,bm}
\usepackage{natbib}
\usepackage[bbgreekl]{mathbbol}
\usepackage{multirow}
\usepackage[toc,page]{appendix}
\usepackage{float}

\usepackage[dvipsnames]{xcolor}

\newcommand{\new}{\textcolor{black}}

\begin{document}

\title{Generalized Single Index Models and Jensen Effects on Reproduction and Survival}
\author{Zi Ye, Giles Hooker and Stephen P. Ellner}
\date{}

\maketitle

\begin{abstract}
Environmental variability often has substantial impacts on natural populations and communities through its effects on the performance of individuals. Because organisms' responses to environmental conditions are often nonlinear (e.g., decreasing performance on both sides of an optimal temperature), the mean response is often
different from the response in the mean environment. \citet*{ye2019jensen} proposed testing for the presence of such variance effects on individual or population growth rates by estimating the ``Jensen Effect'', the difference in average growth rates under varying versus fixed environments, in functional single index models for environmental effects on growth. In this paper, we extend this analysis to effect of environmental variance on reproduction and survival, which have count and binary outcomes. In the standard generalized linear models used to analyze such data the direction of the Jensen Effect is tacitly assumed \emph{a priori} by the model's link function. Here we extend the methods of \cite{ye2019jensen} using a generalized single index model to test whether this assumed direction is contradicted by the data. We show that our test has reasonable power under mild alternatives, but requires sample sizes that are larger than are often available. We demonstrate our methods on a long-term time series of plant ground cover on the Idaho steppe.
\end{abstract}

\section{Introduction}\label{intro}

Effects of environmental variability on different organisms have many important impacts on ecological systems \citep{vasseur-mccann-2007}. Nonlinear responses to a varying environmental factor can either increase population growth \citep{drake2005population,koons2009life} or decrease it \citep{lewontin1969population} depending on the shape of the response to the varying environmental factor, and thus can determine how changes in environmental conditions under climate change will affect species distributions \citep{vasseur-etal-2014}. These differing effects can also play a role in maintaining coexistence among competing species \citep[see for example][]{hutchinson1961paradox, chesson1981environmental, ellner1987alternate, chesson1994multispecies, chesson2000general, chesson2000mechanisms}. Consequently, a key challenge is to understand whether environmental variability is beneficial or detrimental to individual species' growth rates. Classically, this analysis is obtained by examining the curvature of the reaction norm and deducing its effect by an application of Jensen's inequality. Specifically, for an environmental variable $Z$, a convex reaction norm function $g\left(Z\right)$ describing growth in different environmental conditions results in $Eg\left(Z\right) > g\left(EZ\right)$ and a positive effect of environmental variability with the reverse holding under concave responses. The difference between $Eg\left(Z\right)$ and $g\left(EZ\right)$ is the \emph{Jensen Effect} of environmental variability on the response.

In many parametric statistical models the assumed form of $g\left(\cdot\right)$ predetermines its curvature -- for example, positive curvature in a logistic regression when the fitted probability is low, negative curvature when the fitted probability is high. Recent interest has therefore focused on nonparametric methods to assess $g\left(\cdot\right)$. \cite{ye2018local} considered situations where the environment $Z$ is a single index representation of climate summaries: $Z = X\beta$ or as a functional single index $Z = \int X(t) \beta(t) \mathrm{d}t$, where high frequency climate history $X(t)$ is available. These were then used within a nonparametric single index model \citep{hardle1993optimal} $Y = g\left(X\beta\right) + \epsilon$. The curvature of $g$ determines the sign of the Jensen Effect. Because nonparametric estimates of curvature have high variability, \cite{ye2019jensen} proposed to instead estimate the effect of Jensen's inequality directly, the Jensen Effect   $E\hat{g}_{\lambda} \left(X \hat{\beta}\right) - \hat{g}_{\lambda}\left(E X \hat{\beta}\right)$. Here $\hat{g}_{\lambda}$ is obtained from a smoothing spline with smoothing parameter $\lambda$. \cite{ye2019jensen} proposed to test the sign of the effect by an extension of the SiZer method of \cite{chaudhuri1999sizer}, representing the estimated effect as a Gaussian process over $\lambda$ and examining its maximum deviation from zero.

In this paper, we extend the analysis of \cite{ye2019jensen} to generalized single index models in which a natural link function implies the sign of the Jensen Effect by default. This behavior results from the use of nonlinear link functions to impose natural boundaries in the response: positivity for counts, or constraining probabilities to $\left[0,1\right]$ for binary outcomes. Within this framework, we write $EY = h(g(X\beta))$ where $h$ is a fixed, user-chosen function. Extending results in \cite{ye2019jensen} requires accounting for the curvature of $h$ both within estimates of variability and within the bias of the resulting estimator. In the case of the identity link employed in \cite{ye2019jensen}, a second derivative penalty tends to shrink the Jensen Effect towards zero, making the analysis conservative. In our context, a strong second derivative penalty will make $\hat{g}_{\lambda}(\cdot)$ nearly linear yielding a Jensen Effect corresponding to the default behavior of $h(\cdot)$. We thus cannot readily test whether the Jensen Effect is different from zero, but we can ask whether it has the opposite sign to what would be implied by $h(\cdot)$, and we develop such one-sided tests below.

Our analysis can also apply to transformations used for statistical validity. \cite{ye2018local} examined the year-to-year growth of the sagebrush {\em Artemesia tripartita} using a functional single index model for the log growth (where growth is the ratio of final and initial plant sizes); the log transformation was introduced as a variance stabilizing transformation. The resulting model was largely concave. Our interest, however, lies in expected growth, rather than expected log growth and we thus examine the Jensen Effect of $\exp\left( \hat{g}_{\lambda}\left( X \hat{\beta} \right) \right)$. Noting that the exponential is convex, we ask whether the composite model retains the negative Jensen Effect.

 While exponential transforms appear naturally in transformed data and Poisson regression models, models for the binary response of survival present further challenges. The commonly-used logistic transform may be convex or concave depending on the predicted probability, making a default sign for the Jensen effect less evident. Instead, we test whether the estimated effect differs from the value obtained by linear logistic regression. Further, our models may include covariates other than those that describe the environment; in our case study example, plant area is a significant determinant of reproduction and survival. In a partially linear model, the inclusion of further covariates $A$ as additive terms, $Y = A \gamma + g(Z) + \epsilon$, does not interact alter the Jensen effect; similarly when using an exponential transform, examining $E\exp(g(Z)) -  \exp(g(EZ))$ provides an estimate of the Jensen effect at each value $A$ up to a multiplicative scale. In a logistic model, however, the size, and even the sign, of the effect can depend on additional covariates. \new{For such models, we therefore define the Jensen Effect to be the average effect of replacing $Z$ with its average for each data point; when only environment covariates are present in the model, this definition reduces to the Jensen effect as previously defined.}

In section \ref{exp_model} below we first review the procedures developed in \cite{ye2019jensen}, and then develops extensions to the case where we are interested in transformations of the response variable. Section \ref{gsim} then develops these methods for generalized single index models more generally and applies these to Poisson responses. 
Section \ref{log_model} then develops tests specifically for logistic regression. Simulation results and example data analyses accompany each of the sections.

\section{The Jensen Effect and Transformed Responses} \label{exp_model}

In this section we review the Jensen Effect as developed in \cite{ye2019jensen} and extend its application to transformed responses. For example, \cite{ye2018local} examined the effect of environmental history (daily temperature and precipitation over the preceding $36$ months) on the relative growth rate (RGR) of the sagebrush {\em Artemesia tripartita}. RGR is the change in plant size relative to its initial size, equivalent to the ratio between final size and initial size, and is a widely-used measure of plant performance  \citep[e.g.,][]{grime-1975,garnier-1992,cornelissen-1996,diaz-2004,agren-2012}
In their analyses \cite{ye2018local} used log-transformation as a variance-stabilizing mechanism, fitting a model
\begin{align}
\log Y = g\left(X\beta\right) + \epsilon.
\label{FSIM_revisit}
\end{align}
where $Y$ is the ratio between a plant's size at two successive annual censuses.
However, it is the effect of variable environments on the original scale
($Y$ rather than $\log Y$) which is of interest,
\begin{align}
Y = \exp\left[ g\left(X\beta\right) + \epsilon\right],
\label{eSIM}
\end{align}
and which we pursue here.

Below, we first develop estimates of $g$ and $\beta$, corresponding exactly to those employed in \cite{ye2019jensen}, and then extend inference methods to test the sign of $\delta = E \exp( g(X \beta)) - \exp (g(EX \beta))$.

\subsection{Model Formulation and Estimation}\label{exp_model_intro}
To ensure the identifiability of the model, we require that $\left\|\beta\right\| = 1$ and $\beta_1 > 0$. We will estimate $\beta$ and $g$ through the machinery of smoothing splines. We use a $K_1$-dimensional B-spline basis for the link function $g$. Then for any $s$ in the domain of $\mathcal{I} \doteq \left\{X_i\hat{\beta},i = 1,\ldots,n\right\}$, the link function $g$ has the form
\begin{align}
g\left(s\right) = \phi\left(s\right)^{\top}\bm{d},\label{g_basis}
\end{align}
where $\phi\left(\cdot\right)$ and $\bm{d}$ are $K_1$-dimensional column vectors, $\phi$ consisting of all basis functions evaluated at $s$ Here we note that the range of $\mathcal{I}$ depend on $\beta$, but that the $\|\beta\|=1$ constraint allows us to specify that $\left|s\right| < \max_{ij} \left|X_{ij}\right|$.

The coefficients $\beta$ and $\bm{d}$ are estimated by minimizing a penalized sum of squares
\begin{align}
\text{PLS} \doteq & \sum\limits_{i=1}^n \left[Y_i^{\ast} - g\left( X_i\beta\right)\right]^2 + \lambda \int_{\mathcal{I}} \left(g^{\left(2\right)}\left(s\right)\right)^2\mathrm{d}s  \label{exp_PLS}\\
=& \sum\limits_{i=1}^n \left[Y_i^{\ast} - \phi\left(X_i\beta\right)^{\top}\bm{d}\right]^2 + \lambda \bm{d}^{\top}\mathbb{P}_g\bm{d}. \nonumber
\end{align}
This is a nonlinear optimization problem, which we solved numerically using the BFGS algorithm as implemented in the {\tt R} function {\tt optim}. We implement the norm constraint on $\beta$ by standardizing within the function evaluation. We initialize the optimization at $g\left(s\right) = s$ with the linear model for $\beta$ estimated by unpenalized least squares. We ran {\tt optim} twice, initializing the second run from the result of the first run to ensure convergence; see \cite{ye2019jensen} for further details. The resulting estimates are $\hat{\beta}_{\lambda}$ and $\hat{g}_{\lambda}(s) = \phi(s)^{\top} \hat{\bm{d}}_{\lambda}$.

\subsection{Jensen Effect and its Variance} \label{exp_model_jensen}

To investigate the Jensen Effect, we need to estimate $\delta$, and conduct inference about its sign. We define $\delta$ in the exponential model as
\begin{equation}
\delta \doteq
 \frac{1}{n}\sum\limits_{i=1}^n \exp\left[g\left( X_i \beta\right)\right] - \exp\left[g\left( \bar{X}\beta\right)\right],
 \label{deltaExp}
\end{equation}
where $\bar{X} = \frac{1}{n}\sum\limits_{i=1}^n X_i$.

Inference is based on using our estimates from above in eqn. \eqref{deltaExp}, for a series of
different $\lambda$ values. Define a $\left(n+1\right)$-dimensional column vector $\bm{a} \doteq
\left(\frac{1}{n},\cdots,\frac{1}{n},-1\right)^{\top}$ and the augmented set of evaluation
points $\mathbf{e} \doteq \left(E_1,\ldots,E_n,\bar{E}\right)^{\top}$ with corresponding evaluation matrix $\Phi^+$, where $E_i = X_i\hat{\beta}$ (at each environment value) and $\bar{E} = \frac{1}{n}\sum\limits_{i=1}^n X_i\hat{\beta}$ (averaged across all environment values), we express
\[
\hat{\delta}_{\lambda} \doteq \bm{a}^{\top} \exp\left(\Phi^+_{\lambda} \hat{\bm{d}}_{\lambda}\right).
\]
We use these $\hat{\delta}$ values to perform inference on $\delta$, as follows:
\begin{enumerate}
    \item We develop a covariance function estimate $\mbox{cov}(\hat{\delta}_{\lambda_1}, \hat{\delta}_{\lambda_2})$ for $\hat{\delta}$ thought of as a function of $\lambda$.

    \item We use the covariance function to evaluate critical values for $\min_{\lambda}  \hat{\delta}_{\lambda}/\sqrt{\mbox{var}(\hat{\delta}_{\lambda})}$ which constitutes our test statistic.
\end{enumerate}
Specifically, for any $\lambda_1$ and $\lambda_2$, we use a Delta method approximation (\cite{casella2002statistical}):
\begin{align}
\Sigma_{\delta}\left(\lambda_1,\lambda_2\right) \doteq & \text{cov}\left(\hat{\delta}_{\lambda_1},\hat{\delta}_{\lambda_2}\right) \label{exp_cov_delta}\\
=& \left[\bm{a} \cdot \exp\left(\Phi^+_{\lambda_1} \hat{\bm{d}}_{\lambda_1} \right)\right]^{\top}  \Phi^+_{\lambda_1} \text{cov}\left(\hat{\bm{d}}_{\lambda_1}, \hat{\bm{d}}_{\lambda_2}\right) \left(\Phi^+_{\lambda_2}\right)^{\top} \left[\bm{a} \cdot \exp\left(\Phi^+_{\lambda_2} \hat{\bm{d}}_{\lambda_2} \right)\right], \nonumber
\end{align}
where $\text{cov}\left(\hat{\bm{d}}_{\lambda_1}, \hat{\bm{d}}_{\lambda_2}\right)$ is calculated in \cite{ye2019jensen} and we review this here. Recall \eqref{exp_PLS}, the penalized least squares criterion to be minimized, and define the matrix of evaluations of the link function bases and derivatives $\Phi^{(k)}_{ij} \doteq \phi_j^{(k)}\left( X_i\hat{\beta}\right)$ and covariates matrix $\mathbb{X}$. We derive the gradient of \eqref{exp_PLS} as
\begin{align}
\left(
\begin{array}{c c}
\bigtriangledown_{\bm{d}}  \\
\bigtriangledown_{\beta}
\end{array}\right)
= \left(\begin{array}{c}
\Phi^{\top}\left(\bm{Y}^{\ast} - \Phi^{\top}\bm{d}\right) + \lambda \mathbb{P}_g \bm{d} \\
  \mathbb{X}^{\top}\text{diag}\left[\left(\Phi^{(1)}\right)^{\top}\bm{d}\right]\left(\bm{Y}^{\ast} - \Phi^{\top}\bm{d}\right)
\end{array}\right) \doteq \left( \begin{array}{c} \mathbb{Z}_g \\ \mathbb{Z}_\beta \end{array} \right) (\bm{Y}^{\ast} - \Phi^{\top}\bm{d}) + \left( \begin{array}{c} \lambda \mathbb{P}_g \bm{d} \\ \bm{0}_p \end{array} \right),\label{deriv_PLS_exp}
\end{align}
and expected Hessian as
\begin{align}
 \mathbb{H} \doteq \left( \begin{array}{cc}  \mathbb{Z}_g^{\top} \mathbb{Z}_g + \lambda \mathbb{P}_g & \mathbb{Z}_g^{\top} \mathbb{Z}_\beta \\  \mathbb{Z}_\beta^{\top} \mathbb{Z}_g  &
 \mathbb{Z}_\beta^{\top}  \mathbb{Z}_\beta \end{array} \right).\label{exp_Hessian}
\end{align}
We can now obtain the sandwich covariance estimate
\begin{align}
\text{cov}\left(
\begin{array}{c c}
\bm{d} \\
\beta
\end{array}\right) = \hat{\sigma}^2\mathbb{H} ^{-1}\left(
\begin{array}{c c}
\mathbb{Z}_g \\
\mathbb{Z}_\beta
\end{array}\right)^{\top}\left(
\begin{array}{c c}
\mathbb{Z}_g \\
\mathbb{Z}_\beta
\end{array}\right)
\mathbb{H} ^{-1}\label{exp_cov_dc}
\end{align}
in which the variance of the random error $\sigma^2$ is estimated using the model selected by GCV and
\begin{align}
\hat{\sigma}^2 = \frac{1}{\text{df}_{\text{res}}}\sum\limits_{i=1}^n\left[Y^{\ast}_i - \hat{g}_{\lambda_{\mbox{gcv}}}\left( X_i\hat{\beta}\right)\right]^2,\label{exp_sigma}
\end{align}
where the residual degrees of freedom is defined as
\begin{align}
\text{df}_{\text{res}} = n - 2\text{tr}\left(\mathbb{S}\right) + \text{tr}\left(\mathbb{S}\mathbb{S}^{\top}\right) - p,\label{df_exp}
\end{align}
as in \cite{ruppert2003semiparametric} with $\mathbb{S} \doteq \mathbb{S}\left(\lambda\right)$ being the approximate smoother matrix defined by the model.

This derivation follows exactly the calculations in \cite{ye2019jensen} with only the addition of a Delta method approximation to account for the exponential transform in \eqref{exp_cov_delta}. \new{In practice, we have found} that including the single index parameters $\nabla \beta$ within the covariance calculation has negligible effect on the covariance for $\hat{\delta}_\lambda$ and we have dropped these terms from our calculations in Section \ref{gsim} in order to ease the notational burden.

We can now form the normalized $\hat{\delta}_\lambda$ process, $t_\lambda = \hat{\delta}_{\lambda}/\sqrt{\mbox{var}{\hat{\delta}_{\lambda}}}$. This process has covariance approximated by
\begin{equation}
\text{cov}\left(t_{\lambda_1},t_{\lambda_2}\right) = \frac{\text{cov}\left(\hat{\delta}_{\lambda_1},\hat{\delta}_{\lambda_2}\right)}{\sqrt{\text{var}\left(\hat{\delta}_{\lambda_1}\right)}\sqrt{\text{var}\left(\hat{\delta}_{\lambda_2}\right)}},\label{cov_t_exp1}.
\end{equation}
These combine to form the variance-covariance matrix for  $t_\lambda$ over
a grid of $\lambda$ values:
\begin{equation}
 \Sigma_t = \left[\text{diag}\left(\Sigma_{\delta}\right)\right]^{-\frac{1}{2}}\Sigma_{\delta} \left[\text{diag}\left(\Sigma_{\delta}\right)\right]^{-\frac{1}{2}}. \label{cov_t_exp2}
\end{equation}

Even when it is true that $\delta = 0$, we will get $E t_{\lambda} > 0$ for large values of $\lambda$, because
the penalty forces $\hat{g}_{\lambda}(\cdot)$ to be close to linear, and the convexity of the exponential transform produces a positive Jensen Effect. We thus consider a one-sided alternative, and test for $\delta < 0$. To carry this out, we use $\min_{\lambda} t_\lambda$ as our test statistic. In practice, the minimum is evaluated at a grid of $m \approx 20$ values of $\lambda$, and we obtain critical values of the minimum by simulating from the multivariate normal distribution $\bm{t}_{\text{null}} \sim \mathrm{N}\left(\bm{0}_m,\Sigma_t\right)$ and taking the $\alpha$ quantile of $\min \bm{t}_{\text{null}}$.

\subsection{Simulation Results}\label{exp_model_sim}

We evaluated the hypothesis test introduced in the previous section in a series of simulation studies.
Testing for a nonzero Jensen Effect requires estimating the quantity $\delta$ with a range of smoothing parameters $\lambda$. We expect that our estimate $\delta_{\lambda}$ will exhibit high variance at small $\lambda$ because little smoothing is done on the link function estimate, and large positive bias at high $\lambda$ when the estimated link function is strongly shrunk towards linear, but that if the underlying relationship is generally concave, we will be able to detect a negative Jensen Effect.

Our simulation study starts with $p = 5$ covariates generated uniformly on $\left[0, 0.5\right]$. Since we will eventually need to take a log-transformation, these covariates must be restricted to the positive real line. The true coefficients are $\beta = \frac{1}{\sqrt{p}} \bm{1}_p$ so that $\left\| \beta \right\| = 1$.

We considered $4$ different link functions, (1) $g^{\ast}\left(s\right) = \exp\left(s\right)$, (2) $g^{\ast}\left(s\right) = \sqrt{s}$, (3) $g^{\ast}\left(s\right) = \sin\left(s\right)$, (4) $g^{\ast}\left(s\right) = s$, where the domain is $s \in \left[0,0.5\right]$. For the estimate of the link function $g$ in the single index model we used a $25$-dimensional quintic B-spline basis. For each link function, we simulated $1000$ data sets of size $1000$, for error standard deviation $0.01$. We obtained critical values for our test by simulating $5000$ samples from the multivariate Gaussian null distribution. The method's performance at such a low level of noise can be viewed as a ``sanity check'' on our method -- if it failed in this setting, we need to look for a different approach. Fortunately, it did not.

\begin{figure}
\centering
\begin{tabular}{cc}
\includegraphics[width =0.45\textwidth,height = 0.3\textheight]{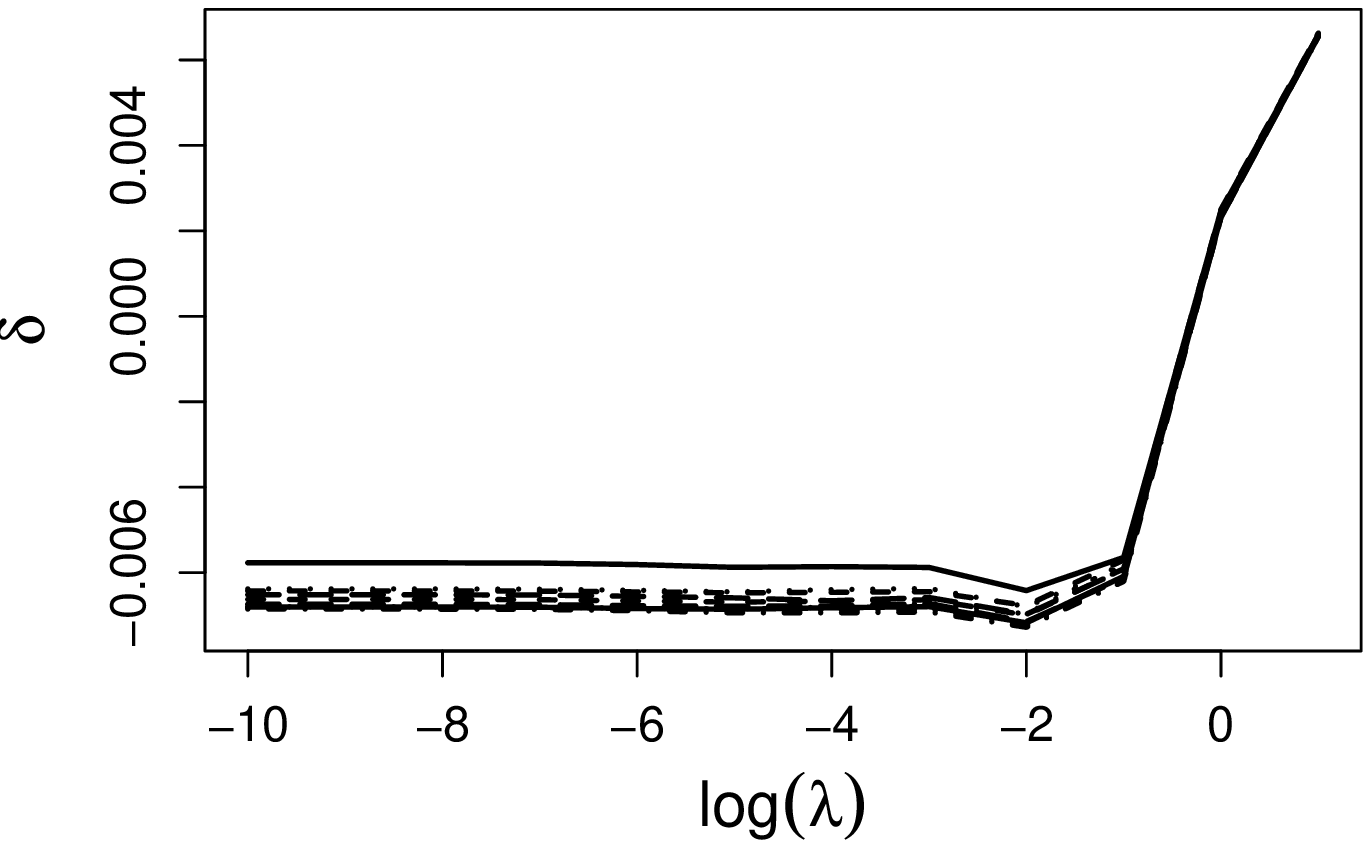} 
\includegraphics[width =0.45\textwidth,height = 0.3\textheight]{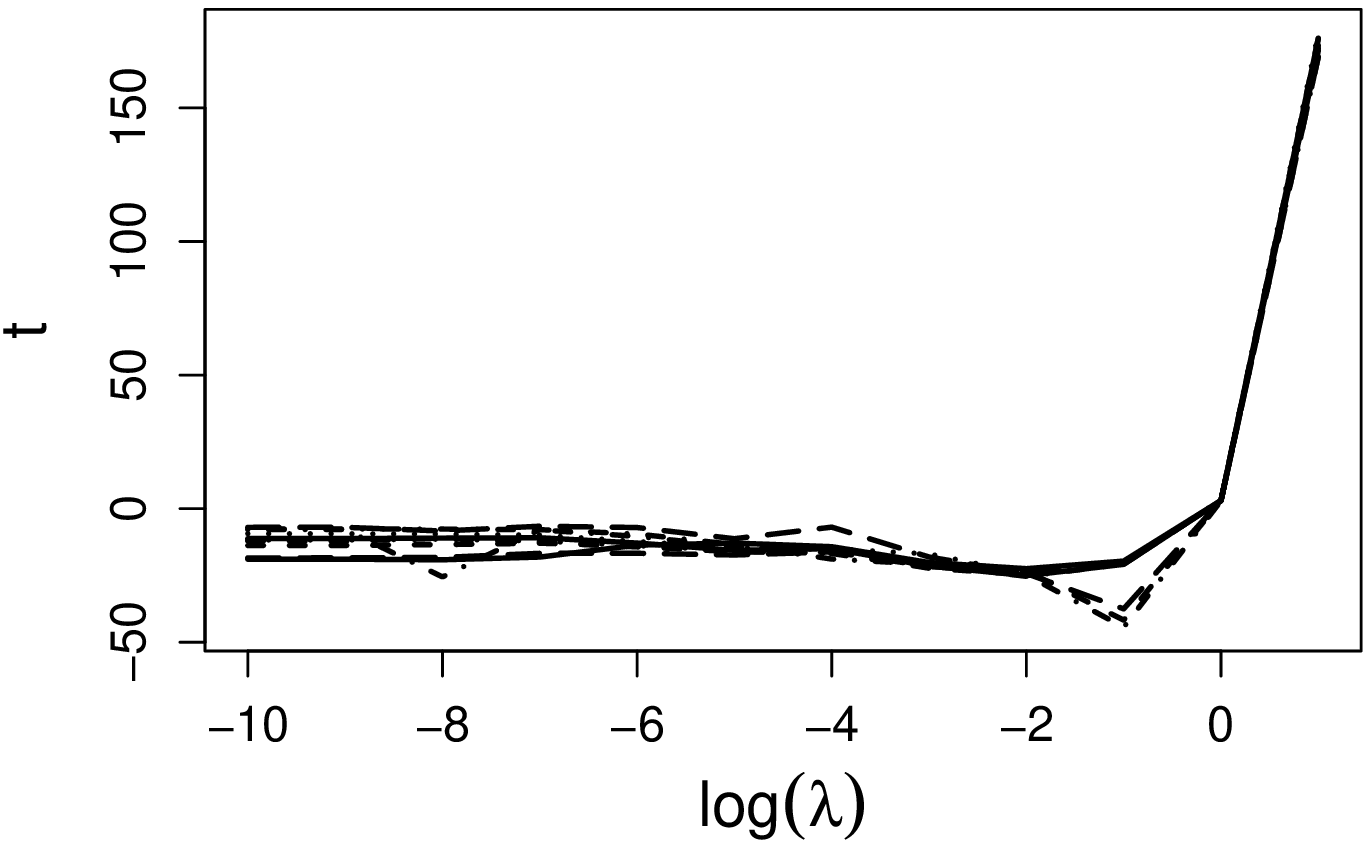}
\end{tabular}
\caption{Illustration of the Jensen Effect for a concave function in the exponential single index model, where the link function is $g^{\ast}\left(s\right) = \sqrt{s}$. Left: a sample of the Jensen Effect values $\delta_{\lambda}$ as a function of smoothing parameters $\lambda$. Right: the corresponding t-statistics $t_{\lambda}$ functions.}
\label{figure:SIM_poly}
\end{figure}

Figure \ref{figure:SIM_poly} presents a sample of $\delta_{\lambda}$ and $t_{\lambda}$ functions versus $\log(\lambda)$ for link function $g^{\ast}\left(s\right) = \sqrt{s}$; plots for the other link functions are in Appendix \ref{appendix:sim}. The rejection rates for the four link functions were $0\%$, $100\%$, $98.3\%$, $0.1\%$ respectively. Note that the $0\%$ rejection rate for the first case is expected as $\delta > 0$ in this case, while we tested for the opposite sign. In the case of concave composite link functions, we correctly reject with high power, because the noise level in this case was very low. In the final case with a linear link function, the result is conservative relative to the nominal $5\%$ rejection rate when the null hypothesis holds.

\subsubsection{Power Analysis}\label{exp_sim:power}
To investigate the statistical power of our hypothesis test, we considered a series of values of $\sigma$, the standard deviation of the random errors, along with four sample sizes. We considered $\sigma$ values
of $\left\{0.03,0.05,0.07,0.09,0.1\right\}$,
and sample sizes of $100$, $200$, $500$ and $1000$. For each combination we
generated $400$ data sets from which to calculate the rejection frequency.

\begin{figure}
\centering
\begin{tabular}{cc}
\includegraphics[width =0.45\textwidth,height = 0.3\textheight]{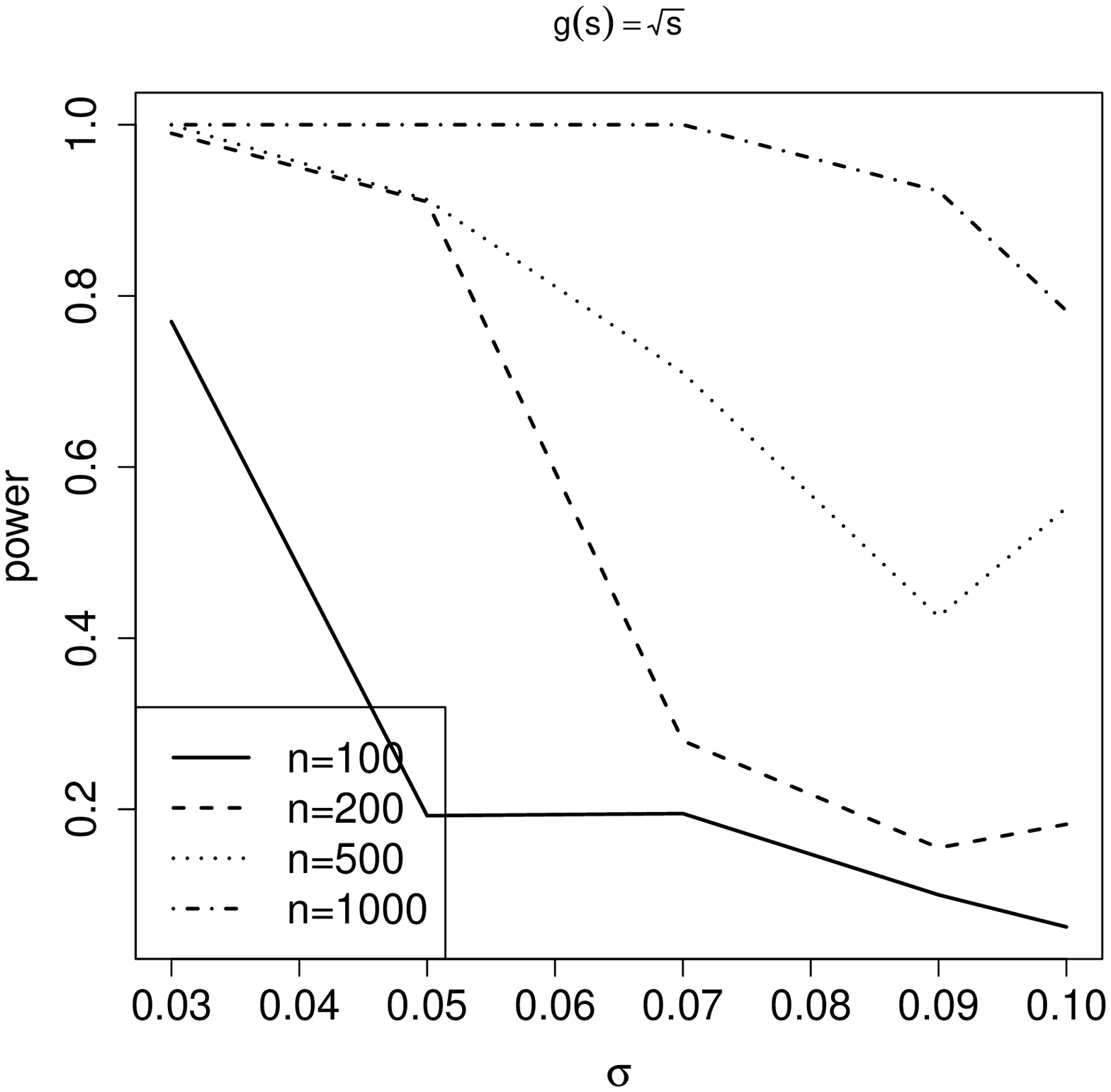}
\includegraphics[width =0.45\textwidth,height = 0.3\textheight]{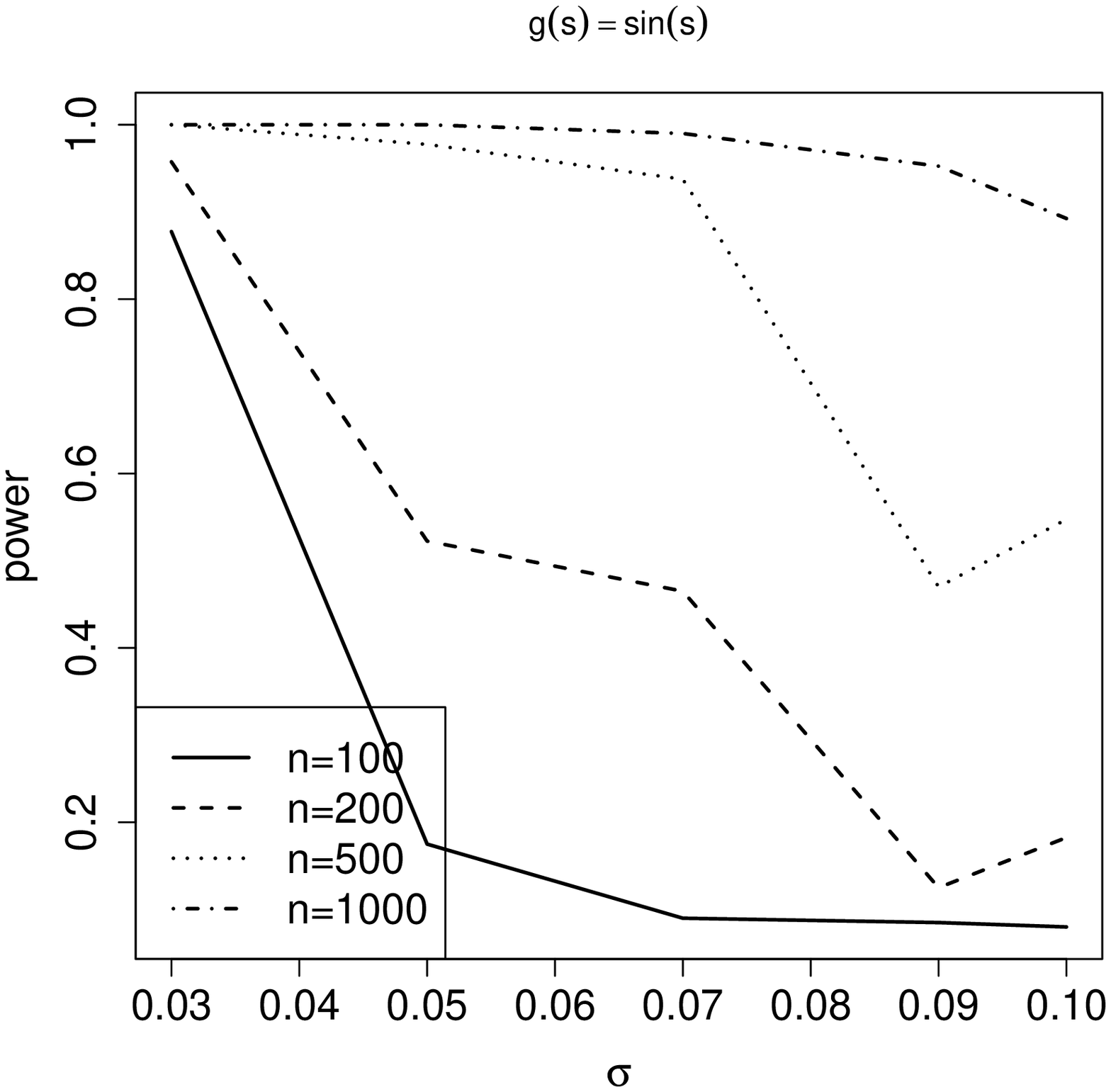}
\end{tabular}
\caption{Statistical power of the hypothesis test for the Jensen Effect being negative in the exponential single index model. The rejection rates of the Jensen Effect test are plotted against $\sigma$ values as the standard deviation of the random error. Left: concave function $g^{\ast}\left(s\right) = \sqrt{s}$. Right: concave function $g^{\ast}\left(s\right) = \sin\left(s\right)$.}
\label{fig:exp_power}
\end{figure}

Figure \ref{fig:exp_power} presents the rejection rates plotted against $\sigma$ value for different
sample sizes. We observe a sharp decrease as $\sigma$ increases unless the sample size is large. As we expected, decreasing noise and increasing sample size increase the rejection rate.

\subsubsection{Empirical Application}\label{exp_real}
Returning to our motivating ecological study, we again analyzed the growth of the sagebrush \emph{Artemisia tripartita} (ARTR) observed at the U.S. Sheep Experiment Station (USSES) between $1926$ and $1957$.
USSES is located $9.6$ km north of Dubois, Idaho ($44.2^{\circ}$N, $112.1^{\circ}$W), $1500$m above sea level. Ecologists at the USSES established $26$ $1$ m$^2$ quadrats between $1926$ and $1932$. Among them, $18$ quadrats were distributed among $4$ ungrazed exclosures, and $8$ were distributed in $2$ pastures grazed at medium intensity spring through fall. All quadrats were located on similar topography and soils.

During the period of data collection, mean annual precipitation was $270$ mm and mean temperatures ranged from $-8^{\circ}$C (January) to $21^{\circ}$C (July). The vegetation is dominated by the
shrub, \emph{Artemisia tripartita}, and the C$3$ perennial bunchgrasses \emph{Pseudoroegneria spicata},
\emph{Hesperostipa comata}, and \emph{Poa secunda}. These $4$ species, the focus of our analyses here, comprised over $70\%$ of basal cover (grasses) and $60\%$ of canopy cover (shrubs and forbs).

\begin{figure}
\centering
\begin{tabular}{cc}
\includegraphics[width =0.45\textwidth,height = 0.3\textheight]{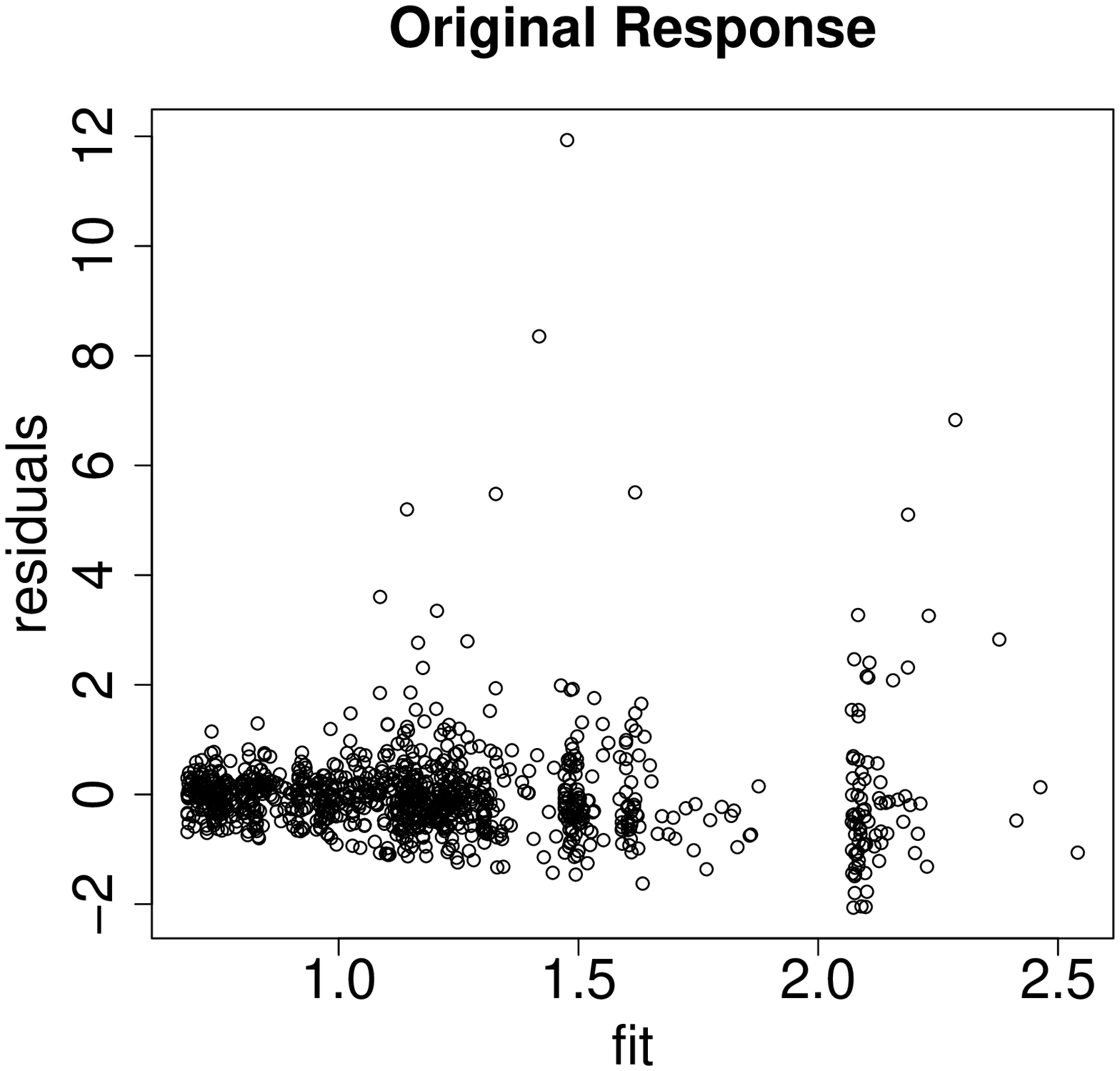}
\includegraphics[width =0.45\textwidth,height = 0.3\textheight]{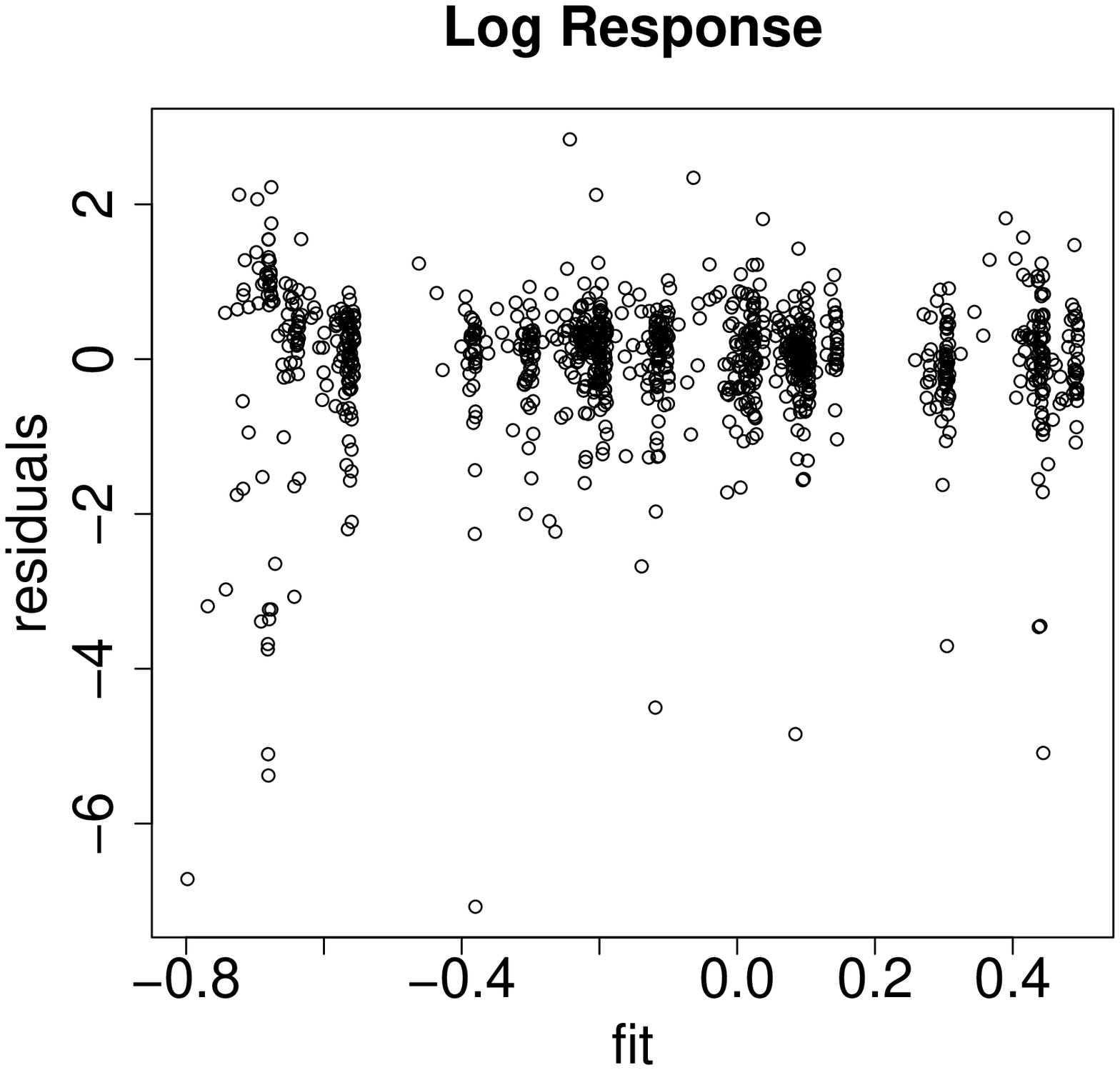}
\end{tabular}
\caption{Fitted values versus residuals for a principal components regression for the shrub {\em Artemisia tripartita}. Left: untransformed size data. Right: log-transformed size data. We can see that log-transformation greatly reduces the variance heteroscedasticity. Note that clumping in the fitted values is due to multiple observations from the same year having the same climate history.}
\label{fig:exp_qqplot}
 \end{figure}

Each individual \emph{A. tripartita} plant was mapped in each of $26$ $1 \mathrm{m}^2$ ungrazed quadrats. Mapping was note done every year, however plant growth over one year (the increase in ground area covered by the plant canopy) could measured for $22$ year-to-year transitions. The response of interest is the relative change in area between two successive annual censuses. Figure \ref{fig:exp_qqplot} plots residuals versus fitted values for a principal components regression using either the original measurements or logged values where we see that there is some heteroscedasticity in the response variable that is corrected after taking the log. 

We have a total of $1003$ year-to-year transitions across all individual plants (genets), quadrats, and years. Because not all plants survive each year, and new plants establish naturally each year, not all plants are observed for all $22$ years. For simplicity, the model we use here does not include individual plant effects, which tend to be small relative to year-to-year variability. For each observed annual transition, we record the following variables:
\begin{enumerate}
    \item $\Delta$ \texttt{logarea} -- the change in the logarithm of total area from one year to the next.

    \item \texttt{W} --- a measure of competition given by a weighted sum of the area occupied by conspecific individuals in a series of concentric annuli extending from the genet's center.

    \item \texttt{t(s)} -- daily temperature history (average of daily minimum and maximum temperatures) over the
    $36$ months prior to the second census ($s$ giving historical time). A small number of missing values in the temperature record were filled in by spline interpolation.

    \item \texttt{p(s)} -- daily precipitation over the previous $36$ months.
\end{enumerate}
Note that all records from a given year have exactly the same temperature and precipitation history; see \citet{teller2016linking} for further details. We also include the competition variable $W$ as a further scalar covariate with its own coefficient. Note that no intercept is included as this is confounded with the estimate of $g$. The resulting model is
\begin{align}
\log\left(A_{t_1}\right) - \log\left(A_{t_0}\right) = g\left(\alpha W + \int p\beta_1 + \int t\beta_2\right),\label{log:real}\\
\frac{A_{t_1}}{A_{t_0}} = \exp\left[g\left(\alpha W + \int p\beta_1 + \int t\beta_2\right)\right],\label{exp:real}
\end{align}
where $\log\left(A_{t_1}\right) - \log\left(A_{t_0}\right)$ is the difference of plant's logarithm of area between time $t_0$ and $t_1$. \eqref{exp:real} is an exponential functional single index model, and the intercept $\alpha$, the functions $g$, $\beta_1$ and $\beta_2$ need to be estimated. Note that the left-hand side of  \eqref{exp:real} differs by $1$ from the plant's relative growth rate $\left(A_{t_1} - A_{t_0}\right)/A_{t_0}$. So to investigate the Jensen Effect on relative growth rate, we need to get an estimate of the composite link function $g^{\ast} \doteq \exp\left(g\right)$.

We use a $15$-dimensional Fourier basis for the two coefficient functions $\beta_1$ and $\beta_2$. We generated initial values to solve the nonlinear optimization problem by assuming that the composite link function is linear, such that $g^{\ast}\left(s\right) = s$. Our analysis rejects the null hypothesis $\delta > 0$, thus concluding that environmental variability reduces relative growth rate in {\em Artemesia}. This confirms the findings in \cite{ye2018local}, even after accounting for the exponential transform.

\begin{figure}
\centering
\begin{tabular}{cc}
\includegraphics[width =0.45\textwidth,height = 0.3\textheight]{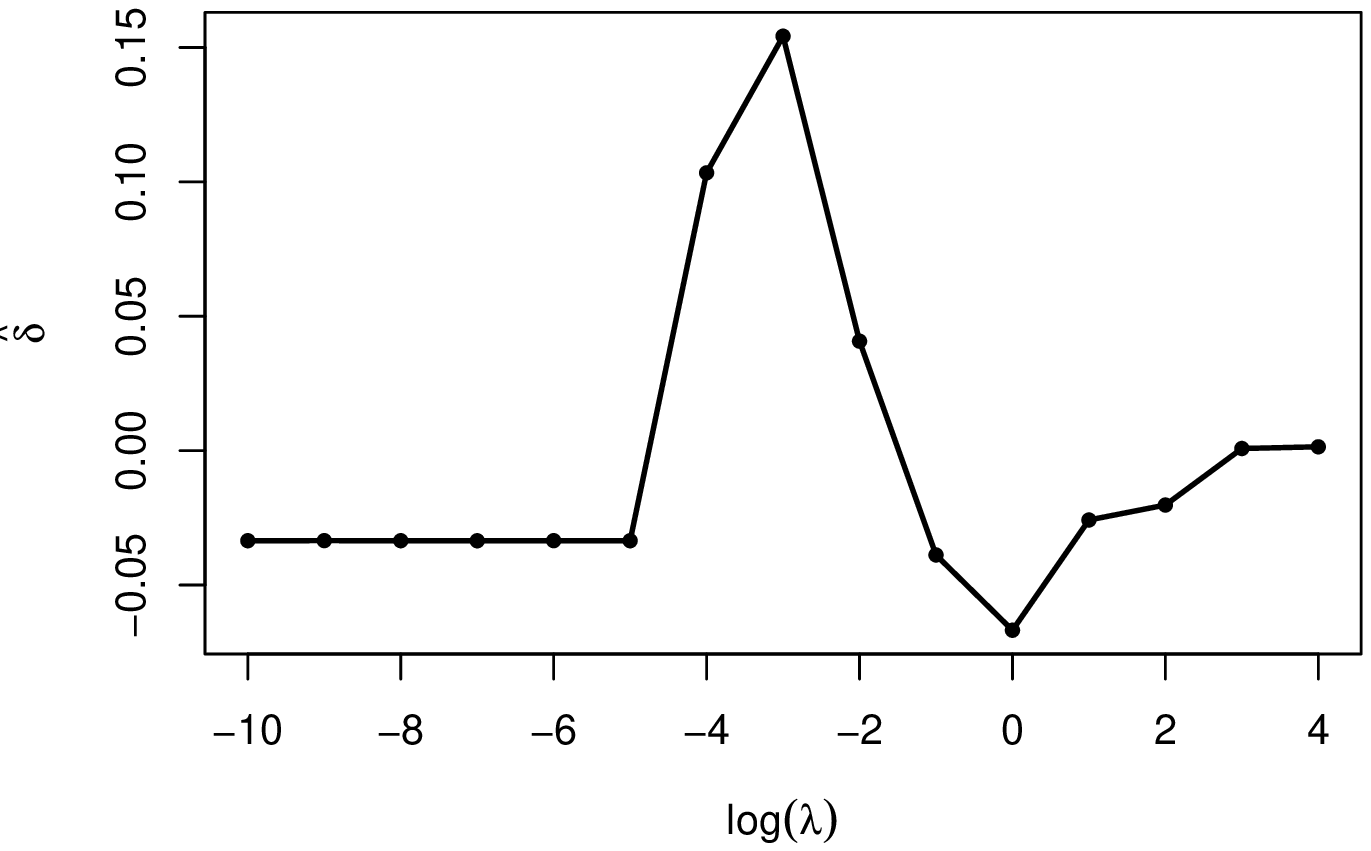} 
\includegraphics[width =0.45\textwidth,height = 0.3\textheight]{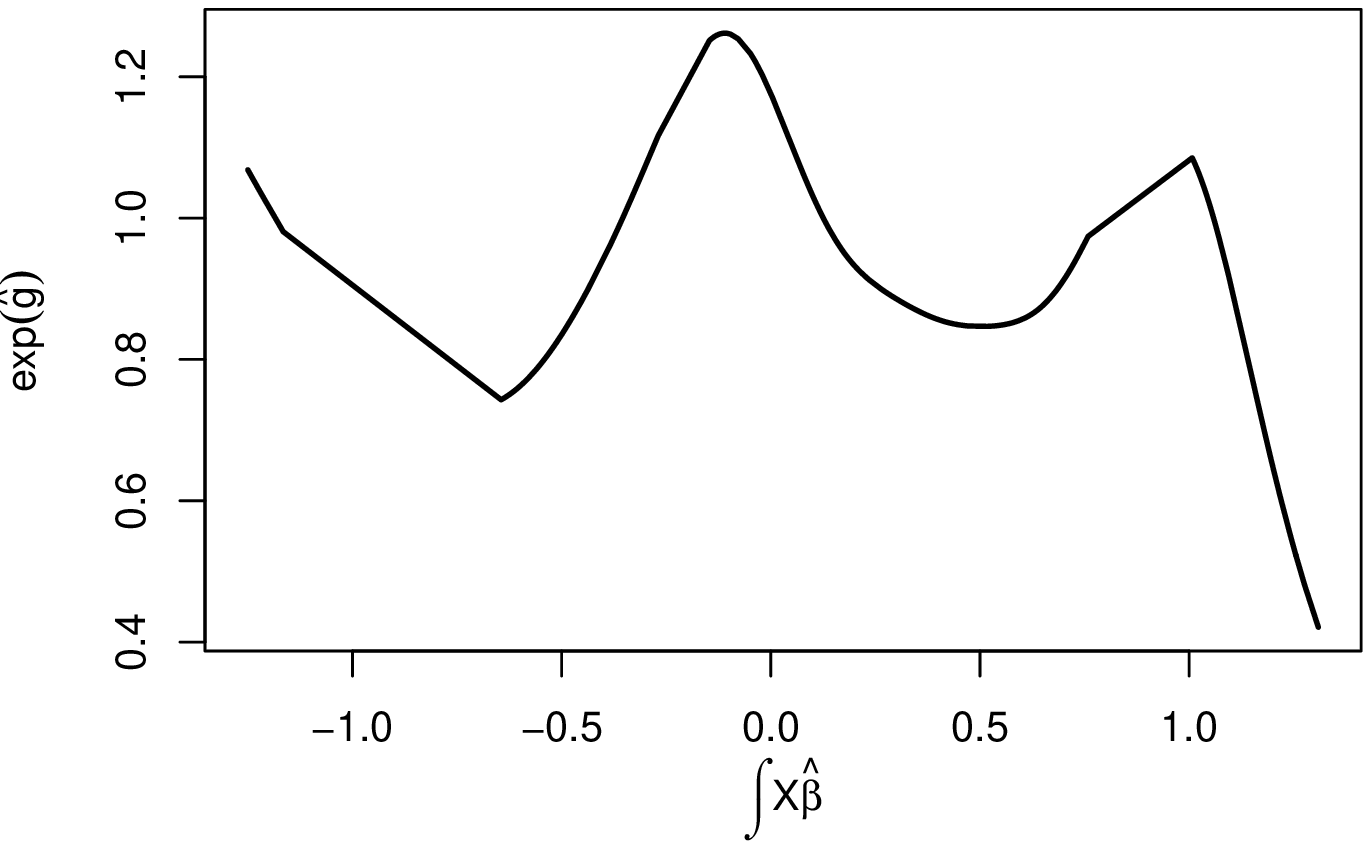}\\
\includegraphics[width =0.45\textwidth,height = 0.3\textheight]{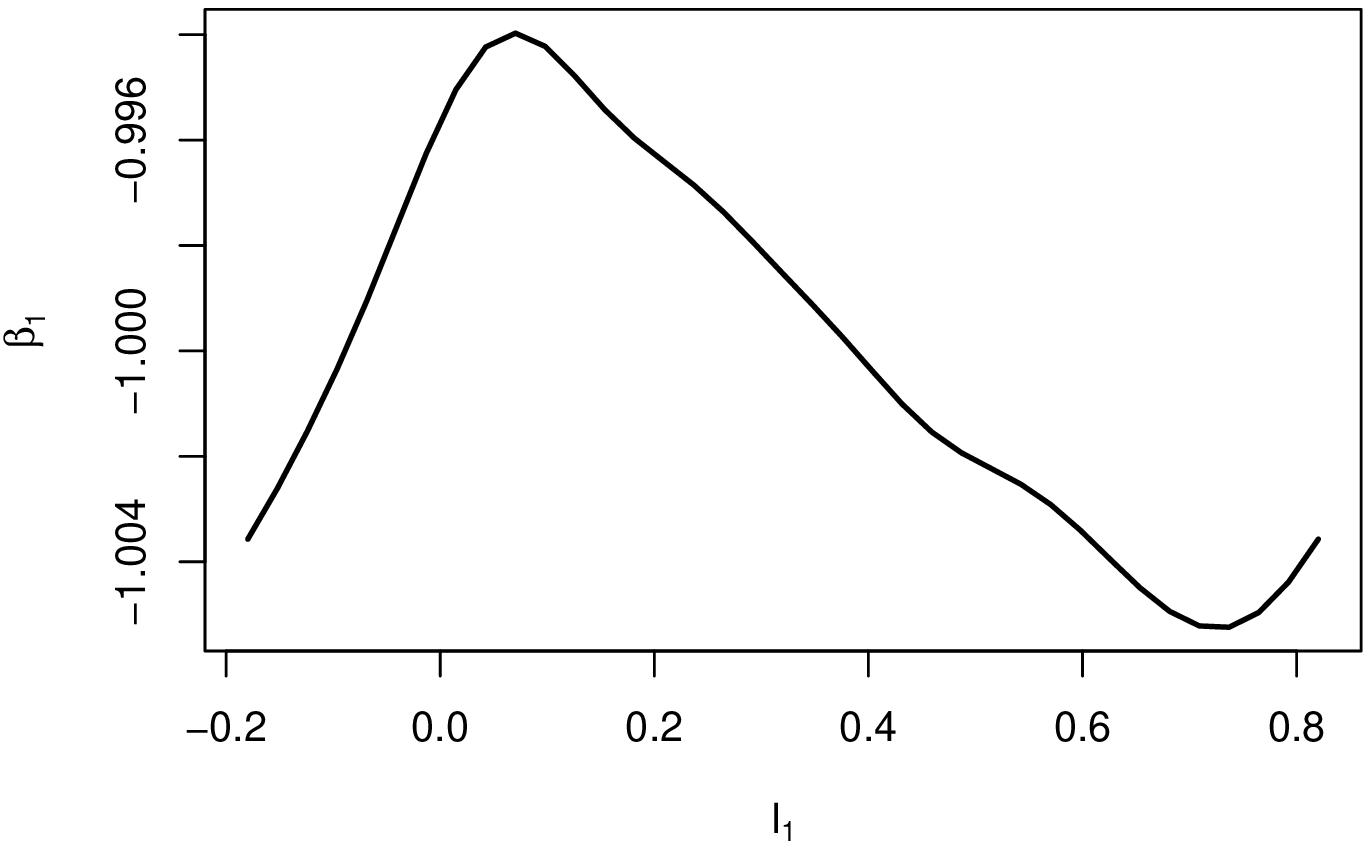} 
\includegraphics[width =0.45\textwidth,height = 0.3\textheight]{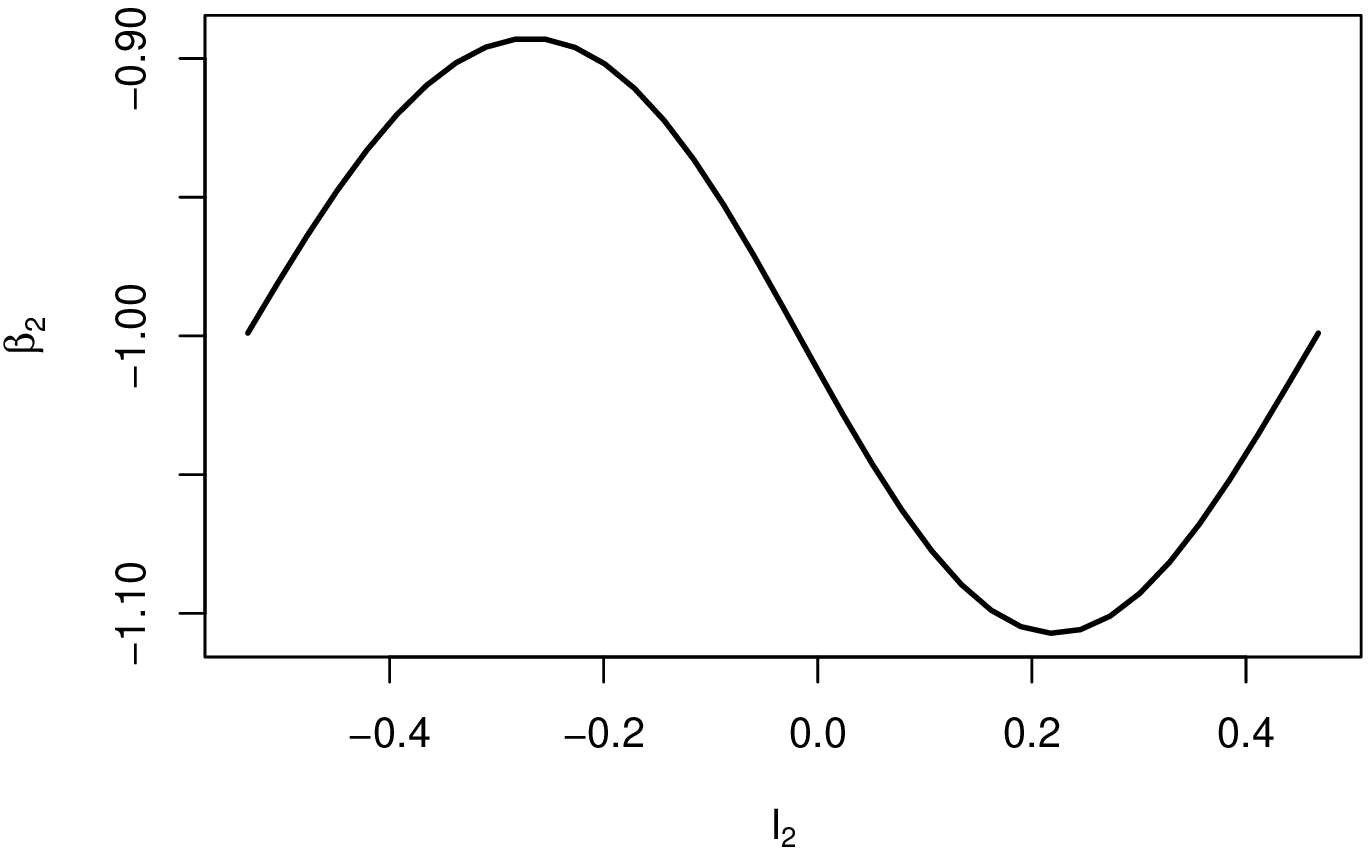}\\
\end{tabular}
\caption{Jensen Effect for relative growth rate of the shrub \emph{Artemisia tripartita}, using the exponential single index model formulation. Top-left: plot of the Jensen Effect estimates $\delta$ against the logarithm of smoothing parameter, $\log\left(\lambda\right)$. Top-right: plot of the estimate of the link function $g^{\ast}$ corresponding to the $\lambda$ selected by GCV. Bottom: plots of estimates of the coefficient functions $\beta_1$ and $\beta_2$ selected by GCV.}
\label{fig:exp_real}
 \end{figure}

We observe that most of the Jensen Effect $\delta$ values are negative in Figure \ref{fig:exp_real}, when they are estimated by different $\lambda$ values. This lends credence to the conclusion that $g^{\ast}$ is concave in its effects over the range of observed environments, and environmental variability is harmful for growth. The remaining plots in Figure \ref{fig:exp_real} are estimates of functions $g^{\ast}$, $\beta_1$ and $\beta_2$, where the smoothing parameter $\lambda$ is selected by generalized cross-validation.

\section{Generalized Single Index Model}\label{gsim}
We now extend the methods of the previous section to generalized linear single index models, to allow for Poisson distributed responses (such as annual number of offspring), or binary outcomes such as survival.

We define a Generalized Single Index model as
\begin{eqnarray}
\eta &\doteq & g\left(X\beta\right), \label{glm_link} \\
\mu &=& h\left(\eta\right) = h\left[g\left(X\beta\right)\right], \label{glm} \\
Y &\sim & \text{Distribution}\left(\mu\right). \label{glm_Y}
\end{eqnarray}
Here, $\eta$ is the response from a single index model, $h$ is the GLM link function, and we are interested in a possible Jensen Effect on the composite link function defined as $h\circ g$.

In particular, we examine the two most common generalized linear model applications. In Section \ref{pois_model}, we examine a Poisson regression framework to model the number of offspring produced by {\em Artemesia} in a given year. Since the Poisson model shares the exponential link function $h$ with the exponential model, we can borrow some ideas from the exponential model to assess the Jensen Effect: whether environmental variability will increase or decrease the number of offspring of a species.

\subsection{Poisson Model}\label{pois_model}

In a Poisson single index model, we observe a count response $Y$ drawn from a Poisson$\left(\mu\right)$ distribution with $\mu = \exp(\eta)$  where $\eta$ is generated by a single index model:
\begin{eqnarray}
\eta &:= & g\left(X\beta\right), \label{def:eta} \\
\mu &=& \exp\left(\eta\right) = \exp\left[g\left(X\beta\right)\right]. \label{pois_link}
\end{eqnarray}
Recall that $\mu$ is the mean and variance of a Poisson distribution and the canonical GLM link function $h$ is exponential.

We assume $n$ observations $\left(X_i,Y_i\right)$ with $Y_i$ generated from the Poisson distribution above. The Jensen Effect $\delta$ is defined as
\begin{align}
\delta = \frac{1}{n}\sum\limits_{i=1}^n \exp\left[g\left(X_i\beta\right)\right] - \exp\left[g\left(\bar{X}\beta\right)\right], \label{true_delta}
\end{align}
where $\bar{X} = \frac{1}{n}\sum\limits_{i=1}^n X_i$.

In a generalized linear model, we replace squared error by the log-likelihood, given in the Poisson model by
\begin{align}
\ell = \sum\limits_{i=1}^n \left[Y_i\eta_i - \exp\left(\eta_i\right)\right]
= \sum\limits_{i=1}^n \left\{Y_i g\left(X_i\beta\right) - \exp\left[g\left(X_i\beta\right)\right]\right\}. \label{loglik}
\end{align}
As above, we use B-spline basis to represent the link function $g$, such that $g\left(s\right) = \phi^{\top}\left(s\right)\bm{d}$ and $\mathbb{P}_g$ as the second derivative penalty matrix for $g$. The penalized log-likelihood is
\begin{align}
\text{PLL} = & -\ell + \lambda \bm{d}^{\top}\mathbb{P}_g\bm{d}  \label{PLL} \\
=& -\sum\limits_{i=1}^n \left\{Y_i \phi^{\top}\left(X_i\beta\right)\bm{d} - \exp\left[\phi^{\top}\left(X_i\beta\right)\bm{d}\right]\right\} + \lambda \bm{d}^{\top}\mathbb{P}_g\bm{d}.\nonumber
\end{align}

For a fixed smoothing parameter $\lambda$, we estimate the coefficients $\beta$ and $\bm{d}$ by minimizing the penalized log-likelihood \eqref{PLL}. Denote the resulting estimates by $\hat{\beta}_{\lambda}$ and $\hat{\bm{d}}_{\lambda}$, we have $\hat{\bm{g}}_{\lambda} = \phi^{\top}\left(X\hat{\beta}_{\lambda}\right)\hat{\bm{d}}_{\lambda} = \Phi_{\lambda}\hat{\bm{d}}_{\lambda}$ and $\hat{\bm{\mu}}_{\lambda} = \exp\left(\hat{\bm{g}}_{\lambda}\right)$, where $\hat{\bm{g}}_{\lambda}$ and $\hat{\bm{\mu}}_{\lambda}$ are $n$-dimensional column vectors evaluated at each observation using the estimated $\hat{\beta}$.

Using the notation from Section \ref{exp_model_jensen}, we can write
\begin{align}
\hat{\delta}_{\lambda} = \frac{1}{n} \sum\limits_{i=1}^n \exp\left[\hat{g}_{\lambda}\left(E_i\right)\right] - \exp\left[\hat{g}_{\lambda}\left(\bar{E}\right)\right] = \bm{a}^{\top} \exp\left(\Phi^+_{\lambda} \hat{\bm{d}}_{\lambda} \right). \label{est_delta}
\end{align}
By the Delta method \citep{casella2002statistical}, for smoothing parameters $\lambda_1$ and $\lambda_2$, the covariance between $\delta_{\lambda_1}$ and $\delta_{\lambda_2}$ is
\begin{align}
\Sigma_{\delta}\left(\lambda_1,\lambda_2\right) = &\text{cov}\left(\hat{\delta}_{\lambda_1},\hat{\delta}_{\lambda_2}\right) \label{cov_delta}\\
=& \left[\bm{a} \cdot \exp\left(\Phi^+_{\lambda_1} \hat{\bm{d}}_{\lambda_1} \right)\right]^{\top}  \Phi^+_{\lambda_1} \text{cov}\left(\hat{\bm{d}}_{\lambda_1}, \hat{\bm{d}}_{\lambda_2}\right) \left(\Phi^+_{\lambda_2}\right)^{\top} \left[\bm{a} \cdot \exp\left(\Phi^+_{\lambda_2} \hat{\bm{d}}_{\lambda_2} \right)\right]. \nonumber
\end{align}
Applying standard arguments, the covariance between $\hat{\bm{d}}_{\lambda_1}$ and $\hat{\bm{d}}_{\lambda_2}$ is
\begin{align}
\text{cov}\left(\hat{\bm{d}}_{\lambda_1},\hat{\bm{d}}_{\lambda_2}\right) \approx& \left(\frac{\partial^2 \left(\text{PLL}\right)}{\partial \hat{\bm{d}}_{\lambda_1}^2}\right)^{-1} \text{cov}\left( \frac{\partial \left(\text{PLL}\right)}{\partial \hat{\bm{d}}_{\lambda_1}}, \frac{\partial \left(\text{PLL}\right)}{\partial \hat{\bm{d}}_{\lambda_2}} \right)\left(\frac{\partial^2 \left(\text{PLL}\right)}{\partial \hat{\bm{d}}_{\lambda_2}^2}\right)^{-1} \label{cov_d}\\
\approx & \left( \Phi^{\top}_{\lambda_1}\mathbb{W}_{\lambda_1}\Phi_{\lambda_1} + \lambda_1 \mathbb{P}_g\right)^{-1} \Phi^{\top}_{\lambda_1}\text{cov}\left(Y\right)\Phi_{\lambda_2}\left( \Phi^{\top}_{\lambda_2}\mathbb{W}_{\lambda_2}\Phi_{\lambda_2} + \lambda_2 \mathbb{P}_g\right)^{-1}. \nonumber
\end{align}
where $\mathbb{W}_{\lambda} = \text{diag}\left(\hat{\bm{\mu}}_{\lambda}\right)$ is an adjusted weight matrix.
Finally, $\text{cov}\left(Y\right) = \bm{\mu}$ and this does not change with $\lambda$.

To estimate \ref{cov_d}, we use $\mathbb{W}_{\lambda}$ with $\lambda$ selected by the GCV criterion adapted to a penalized generalized linear model as in \cite{wood2006generalized}:
\begin{align}
\text{GCV}\left(\lambda\right) = \frac{n\left\|\sqrt{\mathbb{W}_{\lambda}}\left(\bm{z}_{\lambda}-\Phi_{\lambda}\hat{\bm{d}}_{\lambda}\right)\right\|^2}{\left[n-\text{tr}\left(\mathbb{S}_{\lambda}\right)\right]^2}, \label{GCV2}
\end{align}
where the smoother matrix is
\begin{align}
\mathbb{S}_{\lambda} = \Phi^{\top}_{\lambda} \left( \Phi^{\top}_{\lambda}\mathbb{W}_{\lambda}\Phi _{\lambda}+ \lambda \mathbb{P}_g\right)^{-1} \Phi^{\top}_{\lambda}\mathbb{W}_{\lambda}. \label{smooth_mtx}
\end{align}
$\lambda_{\mbox{gcv}}$ is selected by minimizing GCV$\left(\lambda\right)$.

Therefore, the covariance matrix is estimated as
\begin{align}
\text{cov}\left(\hat{\bm{d}}_{\lambda_1},\hat{\bm{d}}_{\lambda_2}\right) \approx \left( \Phi^{\top}_{\lambda_1}\mathbb{W}_{\lambda_1}\Phi_{\lambda_1} + \lambda_1 \mathbb{P}_g\right)^{-1} \Phi^{\top}_{\lambda_1}\mathbb{W}_{\lambda_{\mbox{gcv}}}\Phi_{\lambda_2}\left( \Phi^{\top}_{\lambda_2}\mathbb{W}_{\lambda_2}\Phi_{\lambda_2} + \lambda_2 \mathbb{P}_g\right)^{-1}. \label{cov_d_final}
\end{align}
Putting \eqref{cov_delta} and \eqref{cov_d_final} together, we have an estimate of the covariance matrix of $\hat{\delta}$. The covariance estimate for the t-statistic vector $\bm{t} = \text{diag}\Sigma_{\delta}^{-\frac{1}{2}}\hat{\bm{\delta}}$ is
\begin{align}
&\text{cov}\left(t_{\lambda_1},t_{\lambda_2}\right) = \frac{\text{cov}\left(\hat{\delta}_{\lambda_1},\hat{\delta}_{\lambda_2}\right)}{\sqrt{\text{var}\left(\hat{\delta}_{\lambda_1}\right)}\sqrt{\text{var}\left(\hat{\delta}_{\lambda_2}\right)}},\\
& \Sigma_t = \left[\mbox{diag}\left(\Sigma_{\delta}\right)\right]^{-\frac{1}{2}}\Sigma_{\delta} \left[\mbox{diag}\left(\Sigma_{\delta}\right)\right]^{-\frac{1}{2}}. \nonumber
\end{align}
The corresponding null distribution is simulated as a multivariate normal distribution $\bm{t}_{\text{null}} \sim \mathrm{N}\left(\bm{0},\Sigma_t\right)$, with critical value obtained from the quantiles of $\min_{\lambda} \bm{t}$.

\subsubsection{Simulation Results}\label{pois_sim:sim}
As with the exponential model, we test for the Jensen Effect by calculating the quantity $\delta$ over a range of smoothing parameters $\lambda$. In the Poisson single index model, if the composite link function $g^{\ast} = \exp\left(g\right)$ is convex, the $\delta$ estimate will be positive for most $\lambda$ values; in other cases we expect a positive bias as $\lambda$ increases. For each simulation, we simulate the null distribution from a Gaussian process, and conduct a one-sided hypothesis test to see if $\delta < 0$.

\begin{figure}
\centering
\begin{tabular}{cc}
\includegraphics[width =0.45\textwidth,height = 0.3\textheight]{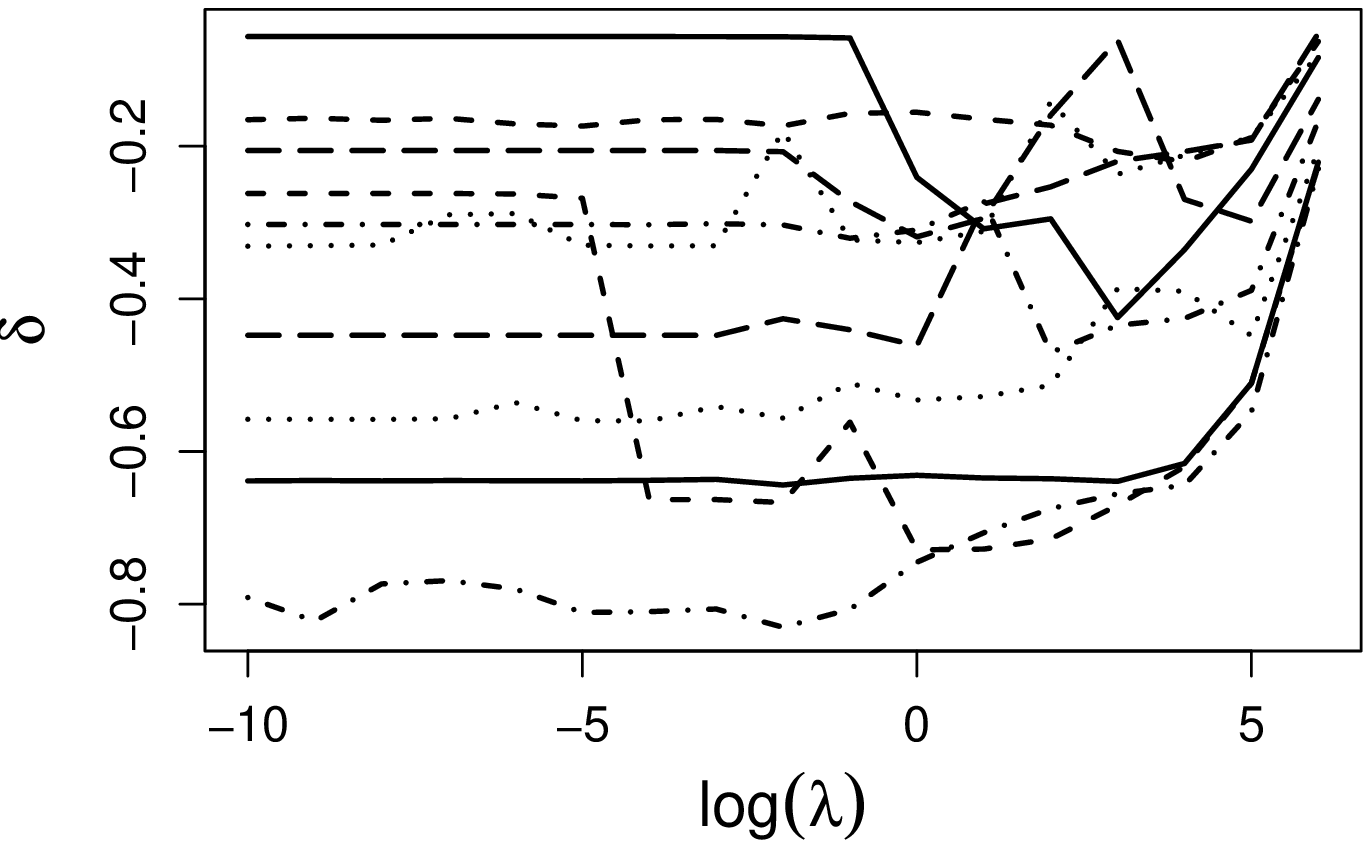} 
\includegraphics[width =0.45\textwidth,height = 0.3\textheight]{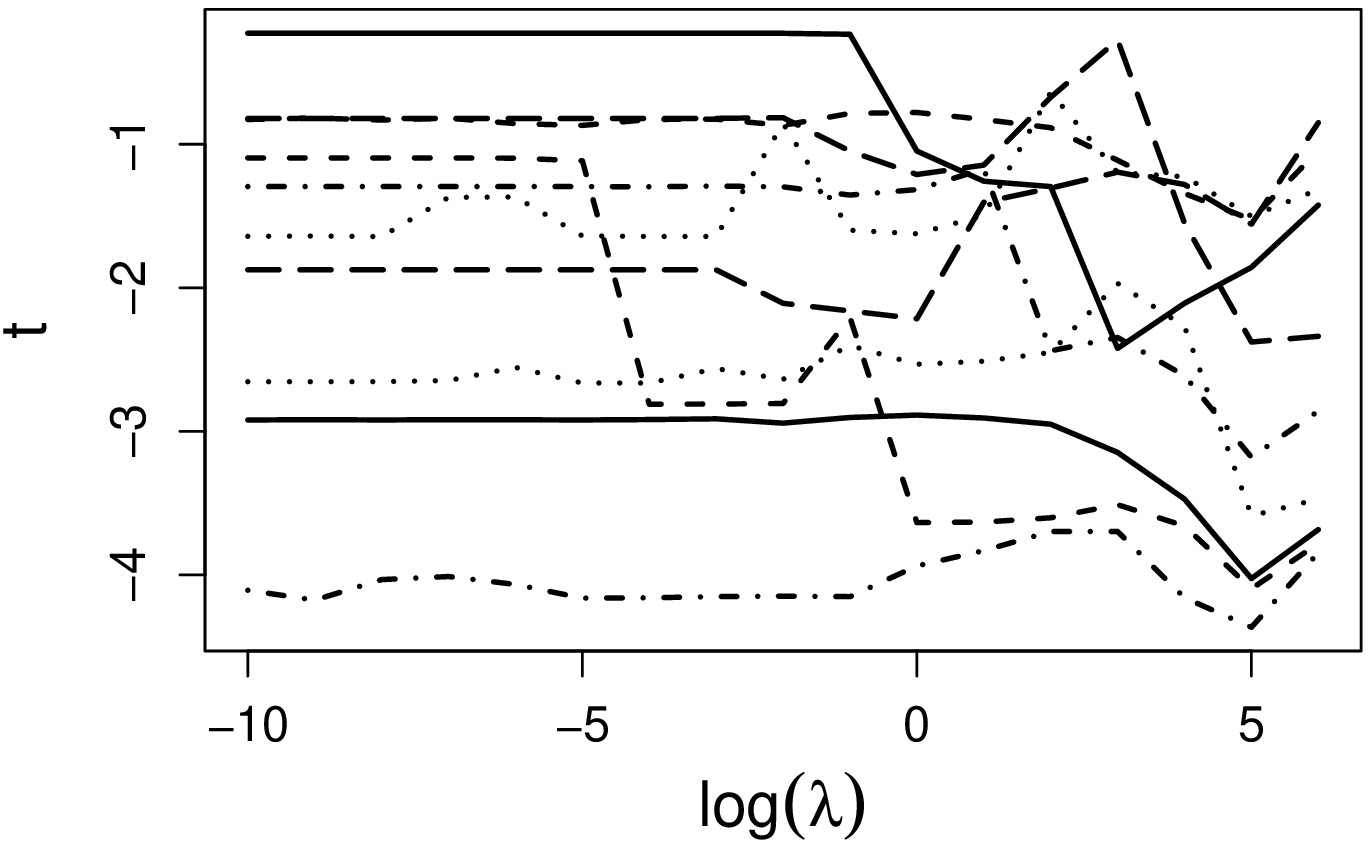}
\end{tabular}
\caption{Illustration of the Jensen Effect for a  concave composite link function in the Poisson single index model, where the link function is $g^{\ast}\left(s\right) = \frac{30}{1+\exp\left(-\frac{s}{8}\right)}$. Left: a sample of the Jensen Effect values $\delta_{\lambda}$ as a function of smoothing parameters $\lambda$. Right: the corresponding t-statistics $t_{\lambda}$ functions.}
\label{figure:concave_pois}
\end{figure}

Our simulation study starts with $p = 5$ covariates generated uniformly on $\left[0, 20\right]$. The coefficients $\beta = \frac{1}{\sqrt{p}} \bm{1}_p$ so that $\left\|\beta\right\| = 1$.
To illustrate the Jensen Effect, we choose $3$ different link functions, (1) $g^{\ast}\left(s\right) = \exp\left(\frac{s}{8}\right)$, (2) $g^{\ast}\left(s\right) = \frac{30}{1+\exp\left(-\frac{s}{8}\right)-15}$, (3) $g^{\ast}\left(s\right) = s$. In each case we represent $g$ by a $25$-dimensional quintic B-spline basis. For each link function, we simulated $200$ data sets of size $1000$. We obtained critical values for our test by simulating $5000$ normal samples from the null distribution. Figure \ref{figure:concave_pois} presents a sample of $\delta_{\lambda}$ and $t_{\lambda}$ functions versus $\log(\lambda)$ for $g^{\ast}\left(s\right) = \frac{30}{1+\exp\left(-\frac{s}{8}\right)}-15$; plots for the other link functions are in Appendix \ref{appendix:sim:pois}. The rejection rates for these functions are: $0\%$, $82.5\%$, $1.5\%$ respectively. The convex function has positive Jensen Effect value $\delta$ so that we observe a zero rejection rate as expected.

\subsubsection{Power Analysis}\label{pois:power}
\begin{figure}
\centering
\begin{tabular}{cc}
\includegraphics[width =0.45\textwidth,height = 0.3\textheight]{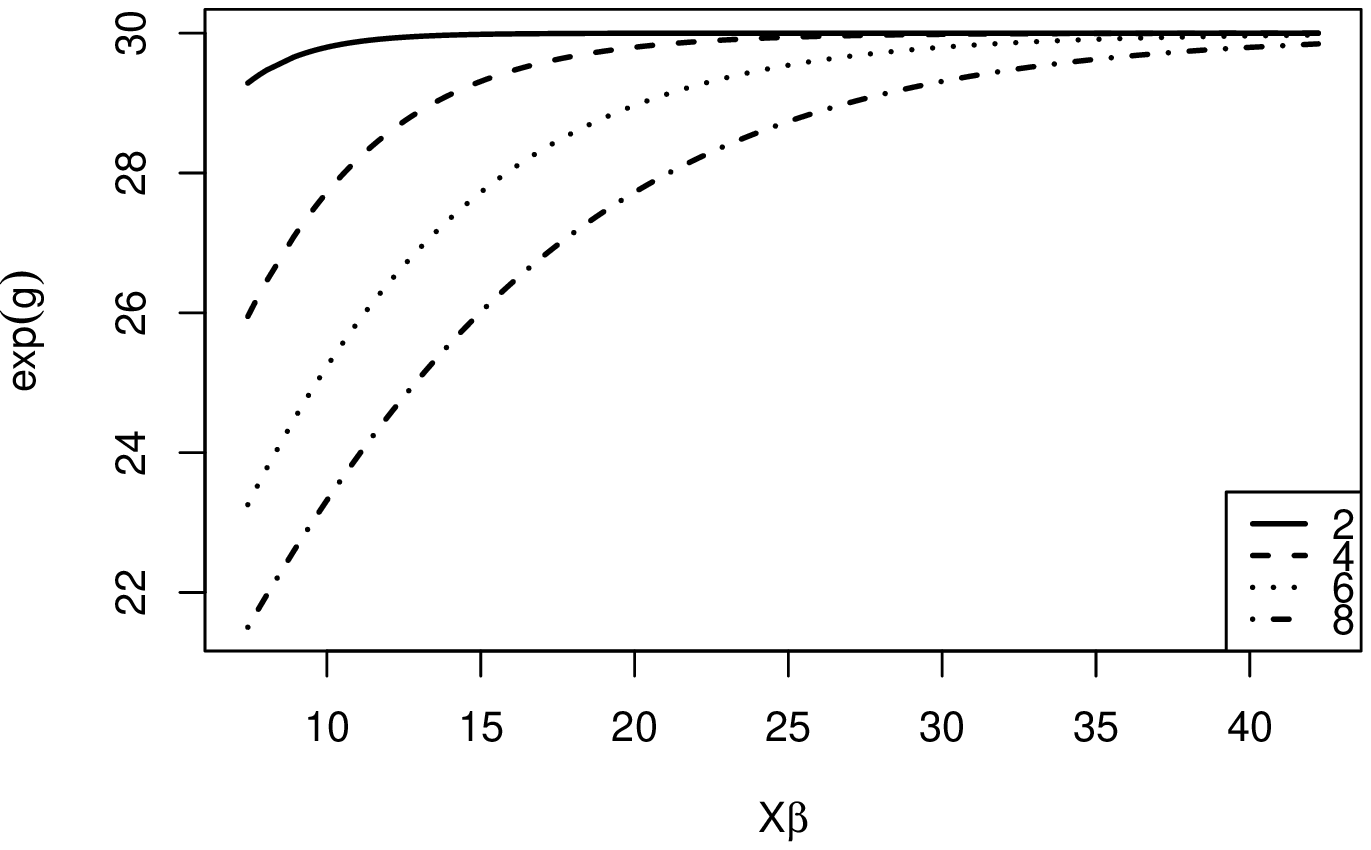} 
\includegraphics[width =0.45\textwidth,height = 0.3\textheight]{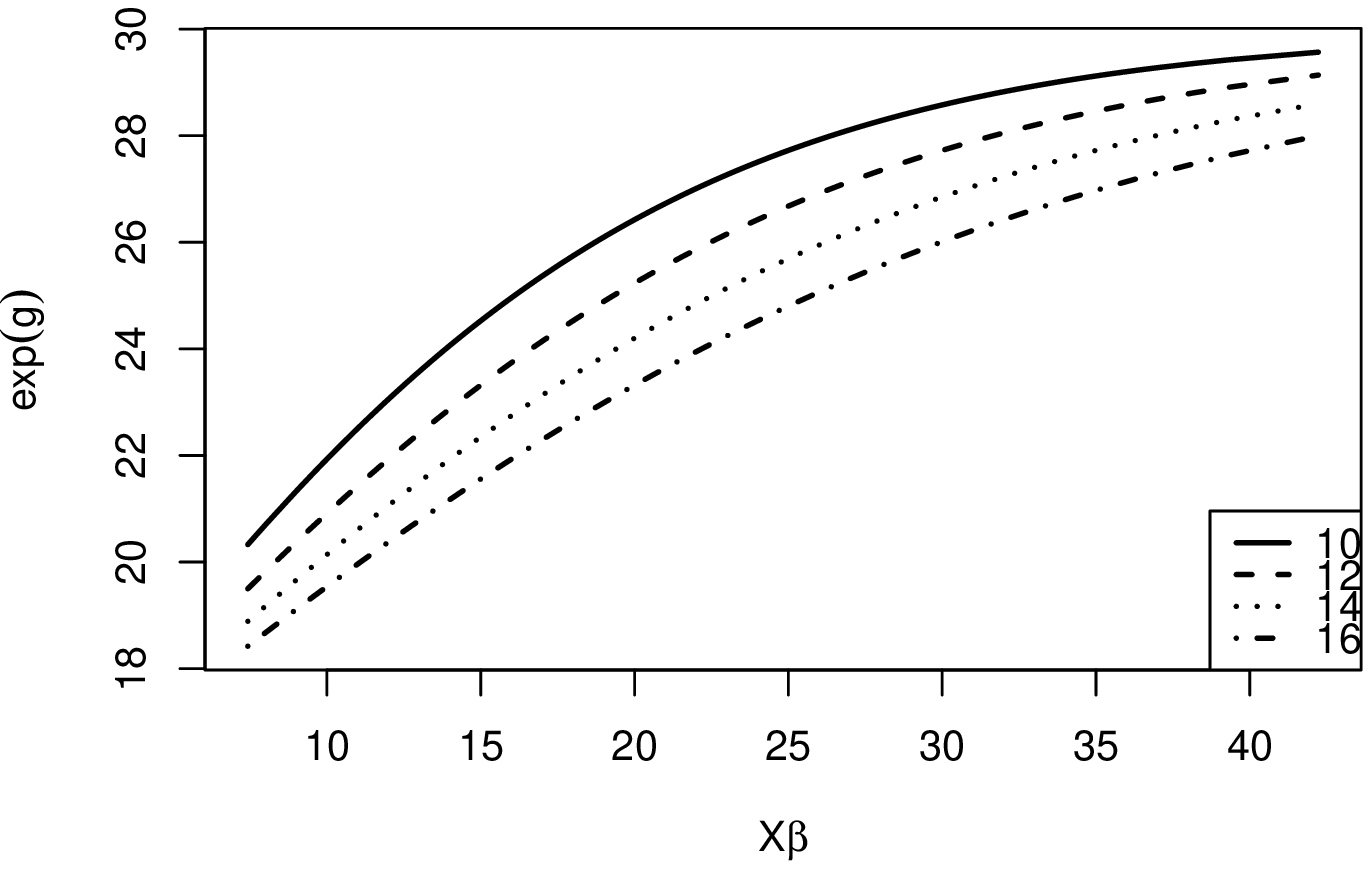}
\end{tabular}
\caption{Concave link functions $g^{\ast}$ in a Poisson single index model, with $\left(h\circ g\right)\left(s\right) = \exp\left(g\right)\left(s\right) = \frac{30}{1+\exp\left(-\frac{s}{a}\right)}-15$. Left: the values of $a$ are $\left\{2,4,6,8\right\}$. The link function becomes more concave as the value of $a$ increases. Right: the values of $a$ are $\left\{10,12,14,16\right\}$. The link function becomes less concave as the value of $a$ increases.}
\label{figure:pois_expg}
\end{figure}

\begin{figure}
\centering
\begin{tabular}{cc}
\includegraphics[width =0.45\textwidth,height = 0.3\textheight]{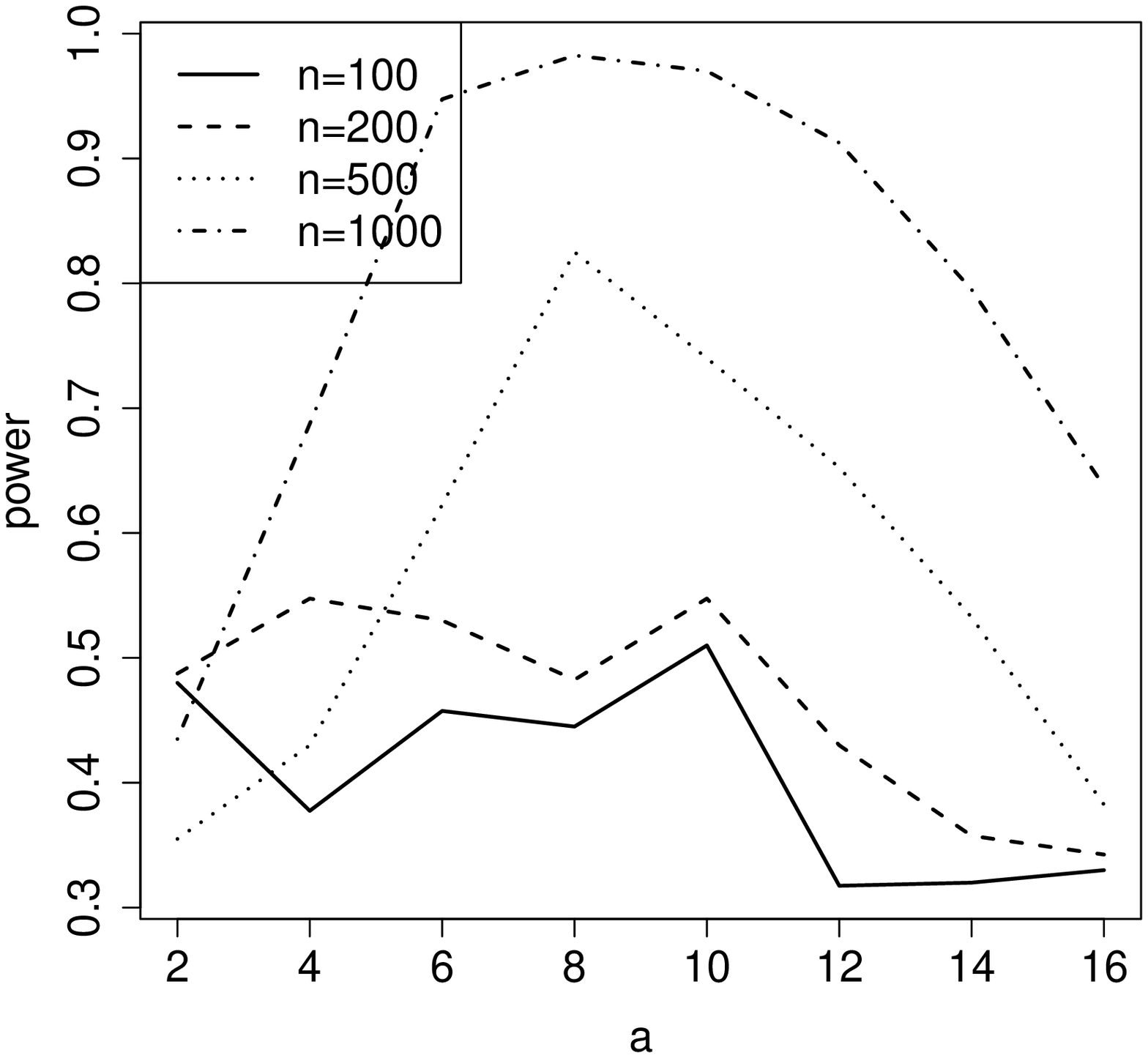}
\includegraphics[width =0.45\textwidth,height = 0.3\textheight]{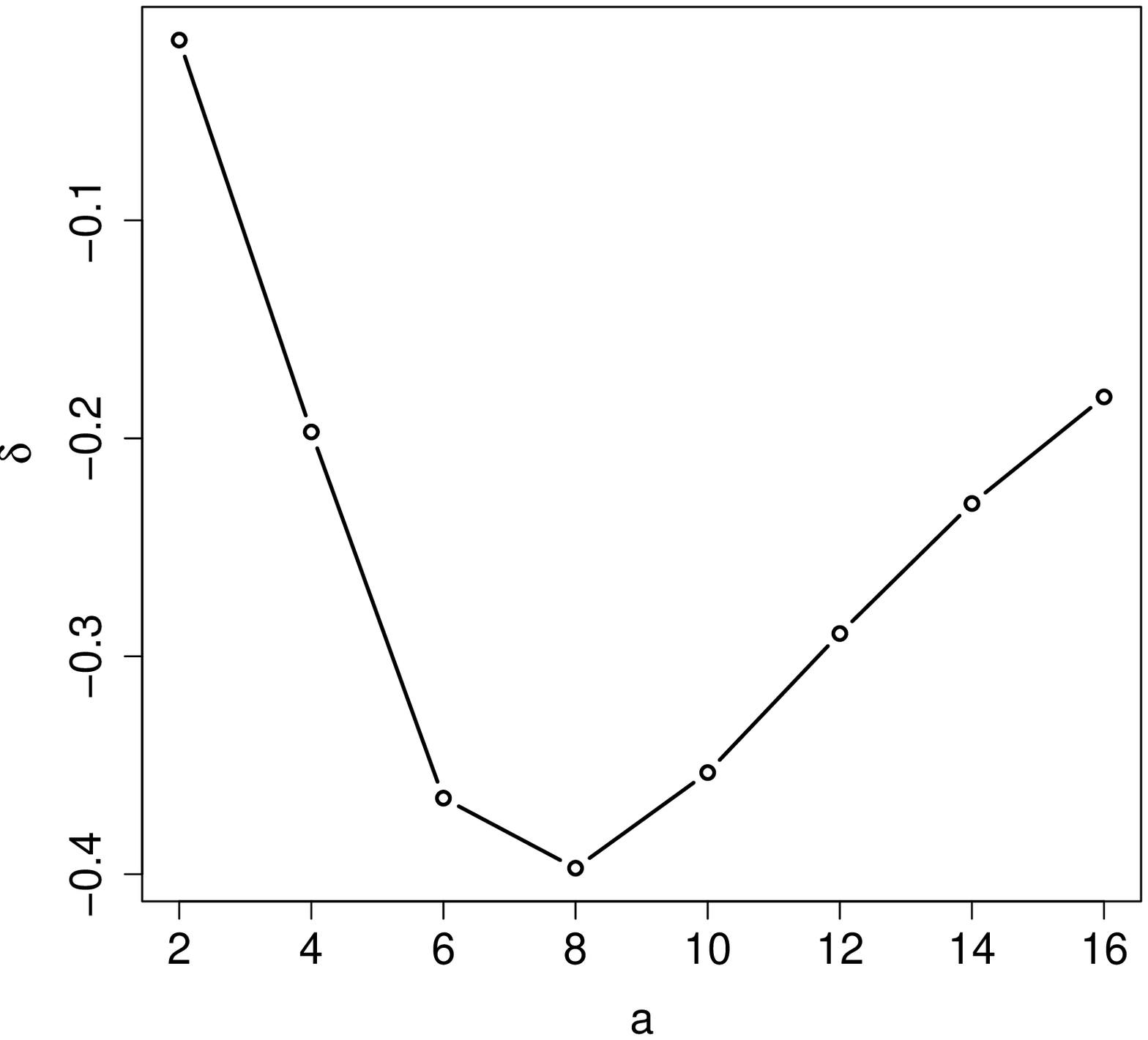}
\end{tabular}
\caption{Power analysis based on 400 simulations of a test for concavity in a Poisson single index model with link function $ \exp\left(g\right)\left(s\right) = \frac{30}{1+\exp\left(-\frac{s}{a}\right)}-15$, where the values of $a$ determine concavity. Left: the rejection rates plotted against the values of $a$. Right: the true Jensen Effect $\delta$ values plotted against the values of $a$.}
\label{figure:pois_reject_delta}
\end{figure}

To investigate the power of the hypothesis test in a Poisson single index model, we use the concave link function
\begin{align}
\left(h\circ g\right)\left(s\right) = \exp\left(g\right)\left(s\right) =  \frac{30}{1+\exp\left(-\frac{s}{a}\right)}-15. \label{pois_power_link}
\end{align}
and employed values of $a$ in $\left\{2,4,6,8,10,12,14,16\right\}$ along with data sets of size $100$, $200$, $500$ and $1000$. From the plots of $g^{\ast}$ in Figure \ref{figure:pois_expg}, we can observe that $g^{\ast}$ becomes more concave for the first half of the $a$ values, and becomes less concave for the second half. The rejection rates and the true Jensen Effect $\delta$ values are in Figure \ref{figure:pois_reject_delta}. We see that as the link function becomes more concave, the rejection rate increases to $1$, and when it becomes less concave, the rejection rate decreases.


\subsubsection{Empirical application} \label{sec:Pois_real}
We return to the USSES data to examine different endpoints. The vegetation in the exclosures is dominated by the
shrub, \emph{Artemisia tripartita} (ARTR), and the C$3$ perennial bunchgrasses \emph{Pseudoroegneria spicata},
\emph{Hesperostipa comata}, and \emph{Poa secunda} (PSSP, HECO, and POSE, respectively). These four species, the focus of our analyses here, comprised over $70\%$ of basal cover (grasses) and $60\%$ of canopy cover (shrubs and forbs).

For each species, we use the number of seedlings in each quadrat as the response variable, and employ five covariates:  $\tt{totParea}$ (total area currently covered by the species), $\tt{ppt1}$, $\tt{ppt2}$ (total precipitation in each of the past two years), $\tt{TmeanSpr1}$, $\tt{TmeanSpr2}$ (mean Spring temperature in each of the past two years). Unlike our growth analysis, we have restricted to a relatively simple model without functional effects. This is because recruitment data are only available at the quadrat level (i.e., it is impossible to know which parent plant produced a particular new recruit), which significantly reduces the number of data points for each year in the study.

Because the response variable is the count of new seedlings, we use a Poisson single index model for the data. Define $\eta$ as
\begin{eqnarray}
\eta = \gamma A + g\left(X\beta\right), \label{glm_link2}
\end{eqnarray}
where $A$ denotes the total area, and $X$ is the covariate vector for temperature and precipitation. \new{Here we continue to measure
\[
\delta_{\lambda} = \frac{1}{n} \sum \left(\exp\left( g_{\lambda} X \beta\right) \right) - \exp\left( g_{\lambda}\left(\bar{X} \beta \right)\right)
\]
noting that the prediction model is $\exp(\gamma A) \exp \left( g \left( X \beta \right) \right) $ and hence $\delta_{\lambda}$ provides a scaled measure of the Jensen effect at each $A$ and its sign is unaffected.}

\begin{figure}[H]
\centering
\begin{tabular}{cc}
\includegraphics[width =0.45\textwidth,height = 0.3\textheight]{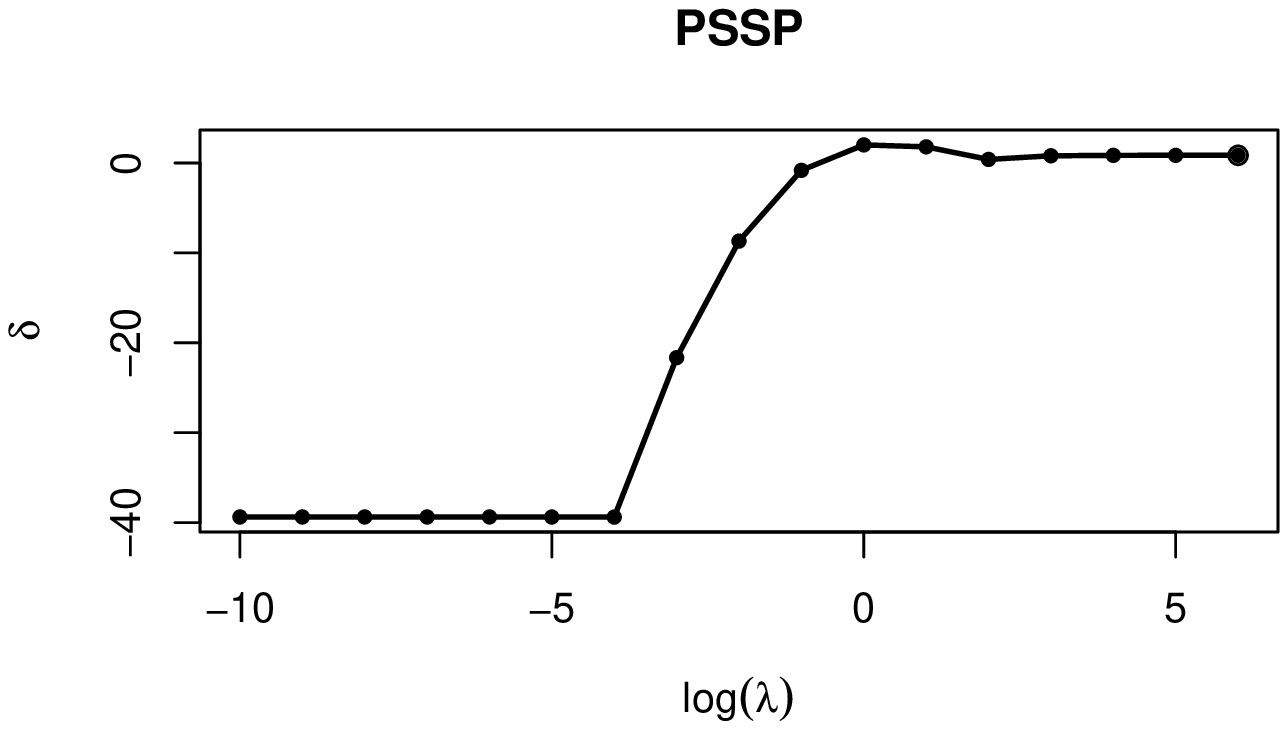} 
\includegraphics[width =0.45\textwidth,height = 0.3\textheight]{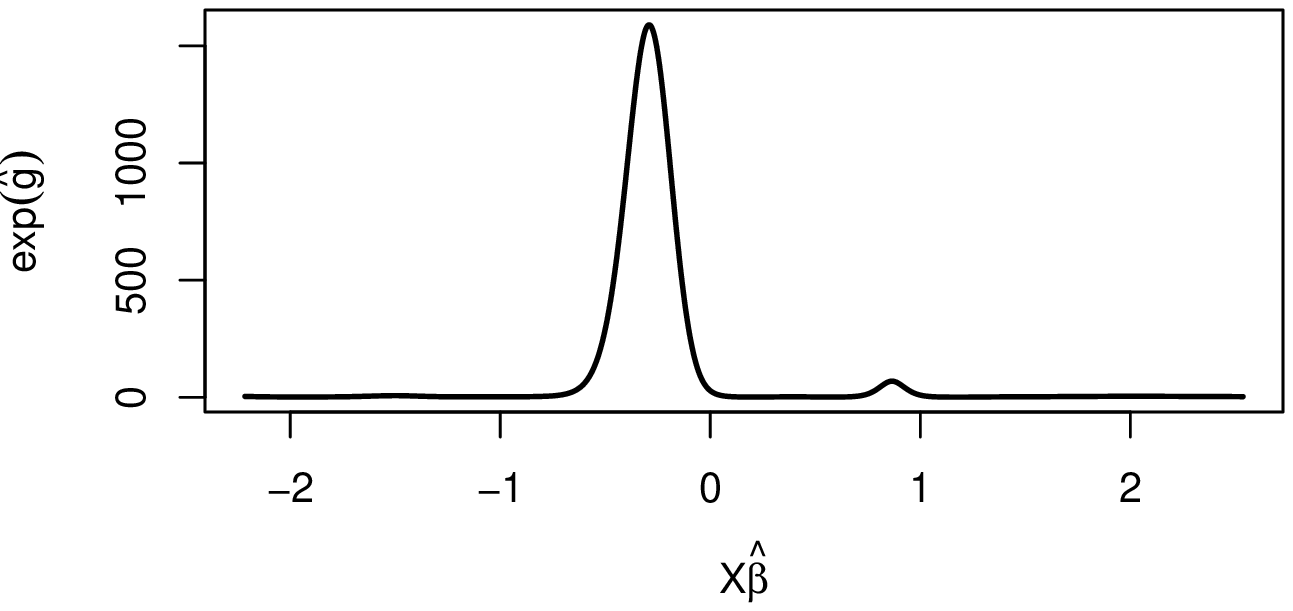}\\
\end{tabular}
\caption{Jensen Effect for reproduction by the grass \emph{Pseudoroegneria spicata} (PSSP) in USSES data set, where we model the seedling count data using Poisson single index model. Left: the Jensen Effect values $\delta$ plotted against the values of $\lambda$. Right: the curve of the exponential link function $\exp\left(\hat{g}\right)$, where $\lambda$ is corresponding to the minimum of $\delta$.}
\label{fig:PSSP}
 \end{figure}

Our hypothesis test rejects the null hypothesis for \emph{Hesperostipa comata} (HECO) and \emph{Pseudoroegneria spicata} (PSSP): environmental variability has a detrimental effect on the average number of HECO and PSSP seedlings. In Figures \ref{fig:PSSP} and \ref{fig:HECO}, we can observe that nearly all $\delta_{\lambda}$ values are negative, confirming that a less variable environment would benefit their reproduction. The right-hand panels in these figures illustrate an example of $\exp\left(g\right)$ estimate, corresponding to the maximum $\lambda$ for which $\hat{\delta}_{\lambda}$ was significantly different from zero. For PSSP the estimated response $\exp\left(\hat{g}\right)$ says that seedling production is maximized at intermediate values of the single index -- a ``Goldilocks'' zone of intermediate
conditions favorable for reproduction -- and environmental variability results in many
years being outside that zone. For HECO there are several peaks in
the estimated response, which are hard to explain biologically, but they may be a result
of the GCV-selected estimate being somewhat under-smoothed. Nonetheless, the overall
pattern is for reproduction to be highest at intermediate values of the environment index.

\begin{figure}[H]
\centering
\begin{tabular}{cc}
\includegraphics[width =0.45\textwidth,height = 0.3\textheight]{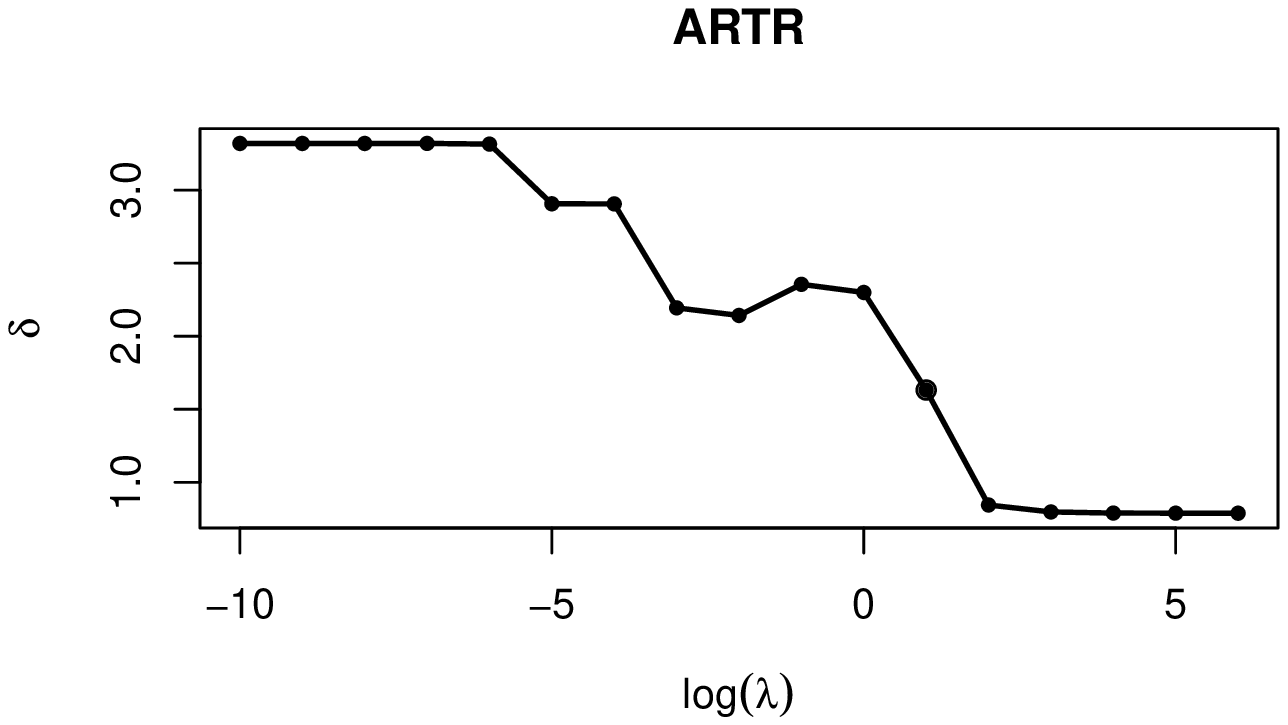} 
\includegraphics[width =0.45\textwidth,height = 0.3\textheight]{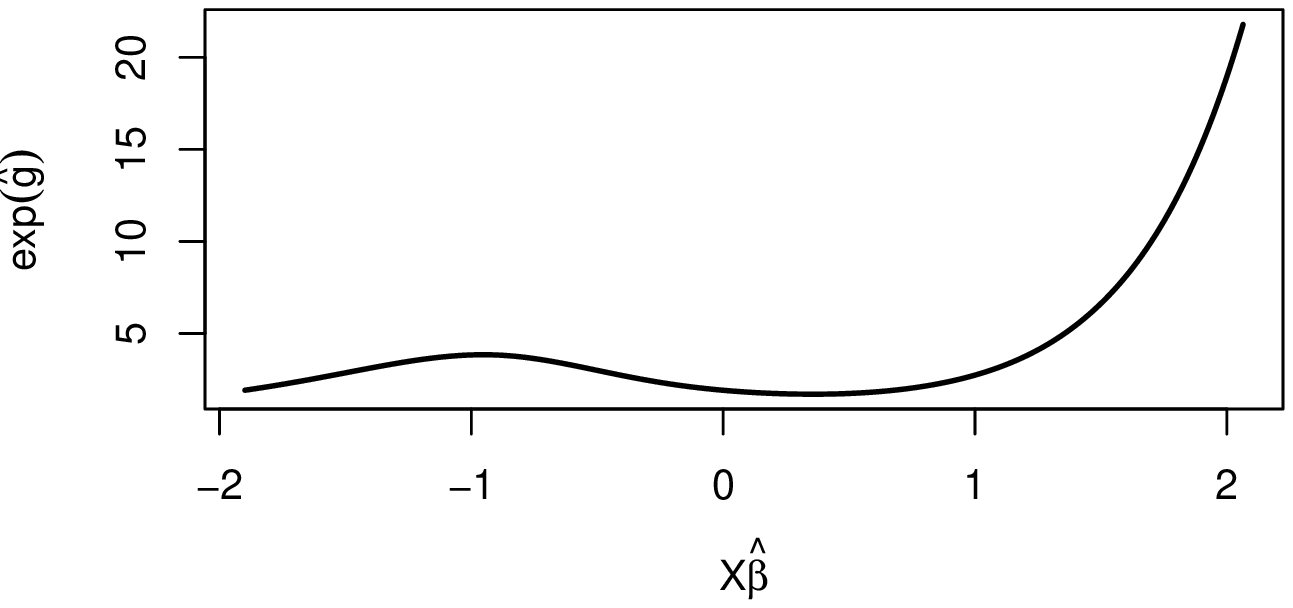}\\
\end{tabular}
\caption{Jensen Effect for reproduction by the shrub \emph{Artemisia tripartita} (ARTR) in USSES data set, where we model the data using a Poisson single index model. Left: the Jensen Effect values $\delta$ plotted against the values of $\lambda$. Right: the curve of the exponential link function $\exp\left(\hat{g}\right)$, where $\lambda$ is selected by GCV.}
\label{fig:ARTR}
 \end{figure}

For the two other species, the null hypothesis is not rejected, and most of the $\delta_{\lambda}$ estimates appear to be positive: environmental variability will increase the average number of seedlings. Figures \ref{fig:ARTR} and \ref{fig:POSE} in Appendix \ref{appendix:real:pois} show the estimated response functions $\exp\left(g\right)$ with $\lambda$ selected by generalized cross-validation. In both cases, the estimated
response functions say that reproduction is highest at one extreme end of the range of
the environment index $X\beta$, and much lower under any other conditions. Any substantial reduction in environmental variability would eliminate the rare years that are favorable for reproduction. Thus, temporal niche partitioning between HECO and PSSP (reproducing best under
intermediate, near-average conditions) versus ARTR and POSE (reproducing best under extreme,
atypical conditions) may be a factor contribute to coexistence of these species in the community.

\section{Logistic Model}\label{log_model}


A final common endpoint in demographic studies is survival: did this individual survive from one census to the next? This is usually modeled by a logistic regression in which the curvature of the response is positive when $\mathrm{P}\left(Y=1\right)$ lies below $0.5$ and negative above it. The default sign of the Jensen Effect, that we wish to challenge through our hypothesis test, therefore depends on the distribution of the environmental index. This makes it difficult to obtain large power in a test against the default: if the environmental index varies enough for accurate estimation of $g$, it is likely to include regions with different curvature in the logistic.
We therefore develop a test that the Jensen Effect is different from the expected value assuming a linear-logistic response, as well as a test of difference in sign when that can be determined.

Unlike the case of exponential models, in this model we cannot separate the effect of variability in environmental covariates from variability in other covariates.  However, as noted above, for this case we define the Jensen effect as the average effect of replacing the environmental covariates with their average value, while keeping the remaining features at their observed values. That is, given observed values $(A_1,X_1),\ldots,(A_n,X_n)$, with $A_i$ being covariates that on included within environmental variation, we define the Jensen Effect to be
\begin{equation} \label{true_delta_logit}
\delta = \frac{1}{n} \sum_{i=1}^n h\left(g(A_i\gamma+X_i\beta)\right) - h\left(g(A_i \gamma + \bar{X} \beta) \right)
\end{equation}
where $s(\cdot)$ is the logistic function. This change is straightforward to implement within our framework and we note that when there are no $A_i$ present, this more general definition reduces to the definition of the Jensen Effect in the models above.

\subsection{Model Formulation and Estimation}\label{log_intro}

The survival status (binary response) of a species, $Y \in \left\{1,0\right\}$, is generated by a Bernoulli$\left(\pi\right)$ distribution, where $\pi$ is the probability of survival, given by
\begin{eqnarray}
\eta &\doteq& g\left(A\gamma + X\beta\right), \label{def:eta:logit} \\
\pi &=& h\left(\eta\right) = \frac{1}{1+\exp\left[-g\left(X\beta\right)\right]}. \label{logit_link}
\end{eqnarray}
with Bernoulli responses $Y_1,\ldots,Y_n$ corresponding to the covariates above.

As above, we represent $g$ by a basis expansion, $g\left(s\right) = \phi^{\top}\left(s\right)\bm{d}$ and estimate parameters $(\gamma,\beta)$ and $\bm{d}$ by maximizing the penalized logistic log-likelihood
\begin{align}
\text{PLL} \doteq & -\ell + \lambda \bm{d}^{\top}\mathbb{P}_g\bm{d}  \label{PLL_logit} \\
=& -\sum\limits_{i=1}^n \left\{Y_i \phi^{\top}\left(A_ \gamma + X_i\beta\right)\bm{d} - \log\left[1+\exp\left(\phi^{\top}\left(A_i \gamma + X_i\beta\right)\bm{d}\right)\right]\right\} + \lambda \bm{d}^{\top}\mathbb{P}_g\bm{d}.\nonumber
\end{align}
At fixed $\lambda$, our estimate of the Jensen Effect is
\begin{align}
\hat{\delta}_{\lambda} = \frac{1}{n} \sum\limits_{i=1}^n h\left[\hat{g}_{\lambda}\left(A_i \gamma + E_i\right)\right] - h\left[\hat{g}_{\lambda}\left(A_i \gamma + \bar{E}\right)\right] = \bm{a}^{\top} h\left(\Phi^+_{\lambda} \hat{\bm{d}}_{\lambda} \right). \label{est_delta_logit}
\end{align}
where we redefine $\Phi^+_\lambda$ to be the evaluation matrix for $A_1 \gamma_\lambda + X_1 \beta_\lambda, A_1 \gamma_\lambda + \bar{X} \beta_{\lambda}, \ldots, A_n \gamma_\lambda + X_n \beta_\lambda,  A_n \gamma_\lambda + \bar{X} \beta_{\lambda}$ and $\bm{a} = \left(1,-1,\ldots,1,-1\right)$.

We can now obtain an estimate of $\mbox{cov}\left(\hat{\delta}_{\lambda_1}, \hat{\delta}_{\lambda_2}\right)$ as in \eqref{cov_delta}, \eqref{cov_d} with the substitute that for $\hat{\bm{\pi}}_{\lambda} = \text{logistic}\left(\hat{\bm{g}}_{\lambda}\right)$ we define the weight matrix  $\mathbb{W}_{\lambda} \doteq \text{diag}\left[\hat{\bm{\pi}}_{\lambda}\left(\bm{1}_n-\hat{\bm{\pi}}_{\lambda}\right)\right]$ which also gives the variance of $Y$.

In the logistic model, we have found that $\Sigma_{\delta}$ can have non-positive eigenvalues, likely associated with numerical error. When this is the case, we form the SVD
\begin{align}
\Sigma_{\delta} = \mathbb{U}_{\delta}\mathbb{V}_{\delta} \mathbb{U}_{\delta}^{\top}. \label{svd}
\end{align}
and truncate negative entries of $\mathbb{V}_{\delta}$ at $0$, yielding $\mathbb{V}_{\delta}^+$ and reformulate the covariance of $\delta$ to be
\begin{align}
\Sigma_{\delta} = \mathbb{U}_{\delta}\mathbb{V}_{\delta}^+ \mathbb{U}_{\delta}^{\top}. \label{svd+}
\end{align}
The t-statistics are again obtained from $\bm{t} \doteq \text{diag}\Sigma_{\delta}^{-\frac{1}{2}}\hat{\bm{\delta}}$. The null distribution is simulated as a multivariate normal distribution $\bm{t}_{\text{null}} \sim \mathrm{N}\left(\bm{0},\Sigma_t\right)$ with a critical value obtained from the quantiles of $\max_{\lambda} \bm{t}$.

\subsection{Testing an Alternative Null}

Because the curvature of the logistic model changes sign, there will not necessarily be a natural sign to test against. Here we develop a test for the quantitative size of the Jensen effect being different from that obtained from standard
logistic regression on the environment index (i.e., different from the value that
would obtain if $g$ is linear). We can frame this as a test of the null hypothesis that
\[
\delta^{+}_\lambda = \delta_{\lambda} - \delta_{\infty} = 0
\]
because, for the second derivative smoothing penalty that we use, $\delta_{\infty}$ corresponds to the Jensen effect when $g$ is linear. In fact, we implement
the calculation of $\delta_{\infty}$ by carrying out standard logistic regression
with linear $g$, and calculating
\[
\hat{\delta}^{+}_\lambda = \bm{a}^\top \left[\mbox{logistic}(\Phi^{+}_\lambda \hat{\bm{d}}_\lambda) - \mbox{logistic}(X \hat{\beta}_{\infty})\right]
\].

The procedures for the test now carry through as before, except that we require an estimate of $\mbox{cov}\left(\hat{\delta}^{+}_{\lambda_1}, \hat{\delta}^{+}_{\lambda_2}\right)$. For this, we employ a Taylor expansion for the difference in fitted values between the single index and logistic regression models as
\[
\hat{Y}_\lambda - \hat{Y}_\infty \approx \left[ \mathbb{W}_\lambda (\Phi^\top_\lambda \mathbb{W}_\lambda \Phi^\top_\lambda + \lambda \mathbb{P}_g)^{-1}\Phi^\top_\lambda - \mathbb{W}_\infty( X^\top \mathbb{W}_\infty X)^{-1} X^\top \right] Y = H^{+}_{\lambda}Y
\]
from which
\[
\mbox{cov}\left(\hat{\delta}^{+}_{\lambda_1}, \hat{\delta}^{+}_{\lambda_2}\right) \approx \bm{a}^\top H^{+}_{\lambda_1} \mathbb{W}_{\lambda_{\mbox{gcv}}} H^{+\top}_{\lambda_2} \bm{a}
\]
and our test proceeds as above.

Note that this same framework can be developed to test for a departure from the Jensen effect that would result under particular parametric models in both the exponential and Poisson cases considered above. We did not pursue such extensions, because each of those models includes a natural sign for the Jensen effect, and so a difference in sign is most likely to be the quantity of interest.

\subsection{Power Analysis}\label{log_model_sim}

As in the Poisson model, we test for the Jensen Effect by calculating the quantity $\delta$ over a range of smoothing parameters $\lambda$. \new{For comparison with previous models, our simulation assumes all covariates are associated with environmental conditions, resulting in the same Jensen effect definition.} We generate $p = 5$ covariates uniformly on $\left[0, 0.5\right]$. The coefficient $\beta = \frac{1}{\sqrt{p}} \bm{1}_p$ so that $\left\| \beta \right\| = 1$ and note that $\bm{x} \beta > 0$ in all cases so that the logistic model is concave across the range of the single index and we test whether $\delta > 0$.  To estimate $g$ and $\beta$, we represent $g$ by a $25$-dimensional quintic B-spline basis.

\begin{figure}
\centering
\begin{tabular}{cc}
\includegraphics[width =0.45\textwidth,height = 0.3\textheight]{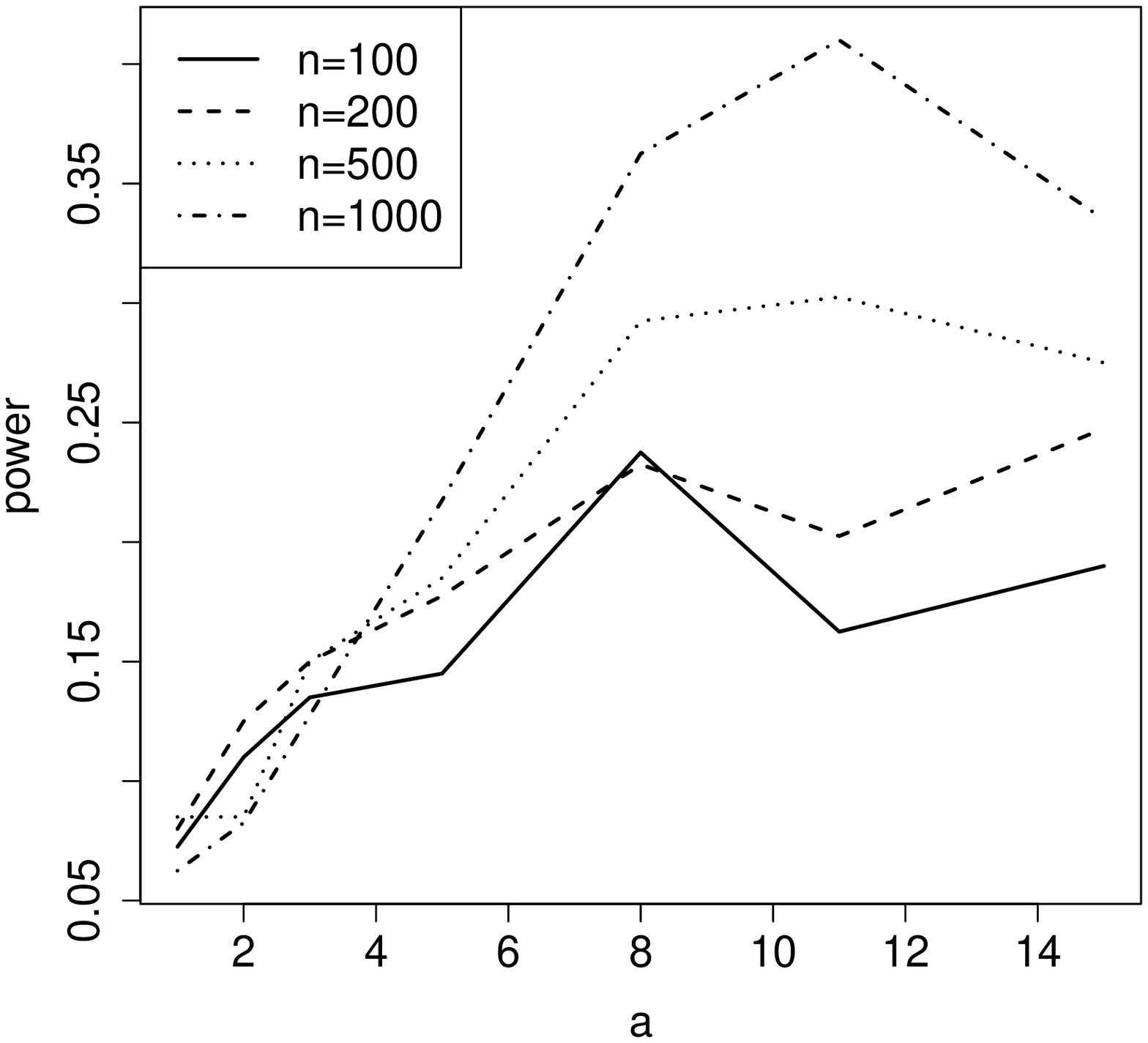}
\includegraphics[width =0.45\textwidth,height = 0.3\textheight]{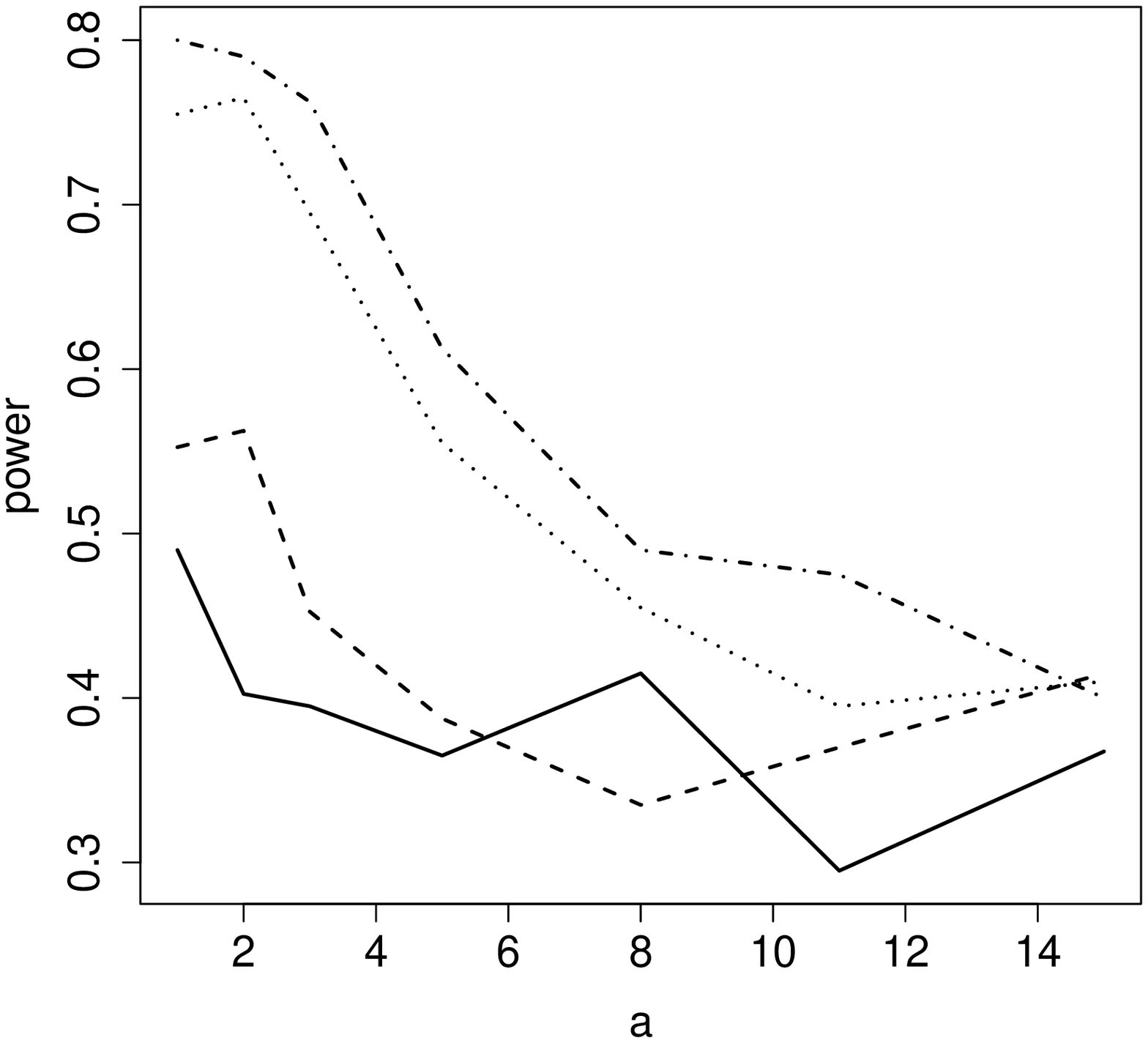}
\end{tabular}
\caption{Statistical power of the hypothesis test for the Jensen Effect in the logistic single index model under generating response \eqref{log_link}. Left: power of the model to detect a difference in sign from the default logistic model. Right: the rejection rate of a test that the quantitative value of the Jensen effect differs from the default logistic model.
}
\label{figure:log_power}
\end{figure}

To illustrate the statistical power of the hypothesis test, we choose the single index to produce the convex composite link
\begin{align}
\left(h\circ g\right)\left(s\right) \doteq \text{logistic}\left(g\right)\left(s\right) = \frac{\exp\left(-as\right) - \exp(-a)}{2(1-\exp(-a))} + \frac{1}{2}, \label{log_link}
\end{align}
which, for $s > 0$ always lies in $[0.5,\ 1]$ on which the logistic function is naturally concave. Our first test is therefore of the positivity of the Jensen Effect. We employed values of $a \in \{1,  2,  3,  5 , 8, 11, 15\}$ and sample sizes $100$, $200$, $500$ and $100$, conducting $400$ simulations at each value combination. There is relatively little power for this test (fig. \ref{figure:log_power}, left) due to the restricted range of the probabilities and the corresponding limited range of $\delta$ values that can be achieved.

Using the same simulated data sets, we investigated our second hypothesis, that the numeric value of the Jensen Effect differs which would obtain under a standard logistic
regression model (linear $g$). Here we find the power of the test unexpectedly decreasing with $a$ (fig. \ref{figure:log_power}, right). This may be due to the increasing steepness of the logistic model that best approximates the true relationship, moving the observed $X \beta$ further from the origin, where the curvature is smaller, as is the corresponding Jensen effect.


\subsection{Empirical Application}

\new{We again employ the USSES data for models of plant survival. We employ the same species and climate covariates as in Section \ref{sec:Pois_real} and also record the area covered by each plant, and a measure of competition given by a weighted sum of the areas covered by conspecifics in concentric annuli around the plant \citep{teller2016linking}. We then measure survival from one year to the next as our outcome. We elected to use the four-variable climate summary employed in Section \ref{sec:Pois_real} rather than as functional covariates due to the smoothness of the effects for the growth model and the lowered power associated with logistic regression. After removing two observations with extreme measures of local competition, we had $5410$ observations of {\em Pseudoregneria spciata}, $2394$ of {\em Hesperostipa comata}, $3648$ for {\em Poa secunda} and $1154$ of {\em Artemesia tripartita}. Within our models, we regarded both climate and competition as environmental covariates, but kept the logarithm of plant area as an additional covariate. This corresponds to our choices in Sections \ref{exp_real} and \ref{sec:Pois_real}; from the perspective of any one plant the location and size of its competitors is an exogenous factor like rainfall, but we recognize that one can argue for or against including local competition as an ``environmental'' variable. }

\new{In all species, a linear logistic model resulted in most survival probabilities being estimated above 50\% and a correspondingly negative Jensen Effect. The Jensen effect from the single index model was also estimated to be negative at every level of smoothness for all species, which also failed to be statistically distinguishable from the value found from a linear logistic model. Plots the estimated adjustment to the linear logistic model and corresponding Jensen Effect over smoothing parameters have been reserved for Appendix \ref{appendix:real:logit}. }

\section{Discussion}

The analysis of demographic data has tended to follow classical statistical paradigms, employing generalized linear models, and using data transformations to ensure reasonable correspondence with statistical assumptions. However, when those models are used to explore the effects of environmental variability, the conclusion is often
predetermined by the link function used in the fitted model. Effects of environmental
variability are central to many classic and current questions in ecology and
evolution \citep{teller2016linking}, so it is important to assess those effects using models where any
outcome is possible, depending on the data.

In this paper, we have extended our prior analysis of nonparametric estimation of Jensen Effects, making it possible to assess the impact of environmental variability on reproduction and survival, as well as accounting for common data transforms in the analysis of growth. Specifically, the methods presented here allow us to test whether the curvature (and thus the sign of Jensen Effects) tacitly assumed by a standard GLM model is contradicted by the data. Because our methods are based on searching over a wide range of smoothing parameters, we are also able to detect relationships that might not be picked up at the one smoothing parameter value chosen by a model selection criterion.

In our empirical applications, we found that environmental variability
is harmful for growth of the shrub \emph{Artemisia tripartita} in western US sagebrush
steppe, but essential for reproduction which only occurs under rare, extreme
values of the environmental index. Variability is also essential for reproduction
by the grass \emph{Poa seconda}, but harmful for reproduction by the grasses \emph{Pseudoroegneria spicata} and \emph{Hesperostipa comata}. However, variability is harmful for the survival of individuals of all species.


Due to the bias towards linearity in smoothed function estimates, we can only conduct a one-sided test to determine if the Jensen Effect is opposite to that implied by the natural link function. While that is often the relevant question, it may also be useful to test whether the Jensen Effect is different from zero. Smoothing bias prevents us from conducting that test in our framework. The alternative approach developed here is to test whether the Jensen Effect is different from that implied by the link function. In the USSES data it appears that survival in two species is, in fact, harmed by environmental variability more than a logistic model would suggest, although this is not statistically significant.

\new{We note that while our analysis has focused on single index models as the most parsimonious nonparametric representation of the reaction to multiple environmental factors affecting individual performance, a very similar analysis can be applied to more general smoothing models for effects of environmental covariates. Employing a smoothing penalty for the composite link $h \circ g$ may also be possible, but its statistical properties are not well understood.}

\new{In our own analysis of the USSES data, we have included plant area either as an additional linear term outside of the single index, in situations where structure of the link function meant that we could separate environmental effects from other covarites, and otherwise as an element of the single index. Applying our generalized definition of the Jensen Effect in Section \ref{log_model}, we could have made either choice for any model. We could also have employed separate smoothing models for temperature and precipitation if we felt that the data would support that level of model complexity. Finally, we could partition the Jensen Effect into components due to temperature, precipitation and competition. These can all be accomplished within the same framework as our current tests, albeit requiring the implementation of new models and the corresponding estimates of $\delta$.}

\bibliographystyle{chicago}
\bibliography{poisson.bib}

\begin{thebibliography}{}

\bibitem[\protect\citeauthoryear{Agren, Wetterstedt, and Billberger}{Agren
  et~al.}{2012}]{agren-2012}
Agren, G.~I., J.~A.~M. Wetterstedt, and M.~F.~K. Billberger ({2012}, {JUN}).
\newblock Nutrient limitation on terrestrial plant growth - modeling the
  interaction between nitrogen and phosphorus.
\newblock {\em New Phytologist\/}~{\em {194}\/}({4}), {953--960}.

\bibitem[\protect\citeauthoryear{Casella and Berger}{Casella and
  Berger}{2002}]{casella2002statistical}
Casella, G. and R.~L. Berger (2002).
\newblock {\em Statistical inference}, Volume~2.
\newblock Duxbury Pacific Grove, CA.

\bibitem[\protect\citeauthoryear{Chaudhuri and Marron}{Chaudhuri and
  Marron}{1999}]{chaudhuri1999sizer}
Chaudhuri, P. and J.~S. Marron (1999).
\newblock Si{Z}er for exploration of structures in curves.
\newblock {\em Journal of the American Statistical Association\/}~{\em
  94\/}(447), 807--823.

\bibitem[\protect\citeauthoryear{Chesson}{Chesson}{1994}]{chesson1994multispecies}
Chesson, P. (1994).
\newblock Multispecies competition in variable environments.
\newblock {\em Theoretical population biology\/}~{\em 45\/}(3), 227--276.

\bibitem[\protect\citeauthoryear{Chesson}{Chesson}{2000a}]{chesson2000general}
Chesson, P. (2000a).
\newblock General theory of competitive coexistence in spatially-varying
  environments.
\newblock {\em Theoretical population biology\/}~{\em 58\/}(3), 211--237.

\bibitem[\protect\citeauthoryear{Chesson}{Chesson}{2000b}]{chesson2000mechanisms}
Chesson, P. (2000b).
\newblock Mechanisms of maintenance of species diversity.
\newblock {\em Annual review of Ecology and Systematics\/}~{\em 31\/}(1),
  343--366.

\bibitem[\protect\citeauthoryear{Chesson and Warner}{Chesson and
  Warner}{1981}]{chesson1981environmental}
Chesson, P.~L. and R.~R. Warner (1981).
\newblock Environmental variability promotes coexistence in lottery competitive
  systems.
\newblock {\em The American Naturalist\/}~{\em 117\/}(6), 923--943.

\bibitem[\protect\citeauthoryear{Cornelissen, Diez, and Hunt}{Cornelissen
  et~al.}{1996}]{cornelissen-1996}
Cornelissen, J., P.~Diez, and R.~Hunt ({1996}, {OCT}).
\newblock Seedling growth, allocation and leaf attributes in a wide range of
  woody plant species and types.
\newblock {\em Journal of Ecology\/}~{\em {84}\/}({5}), {755--765}.

\bibitem[\protect\citeauthoryear{Diaz, Hodgson, Thompson, Cabido, Cornelissen,
  Jalili, Montserrat-Marti, Grime, Zarrinkamar, Asri, Band, Basconcelo,
  Castro-Diez, Funes, Hamzehee, Khoshnevi, Perez-Harguindeguy, Perez-Rontome,
  Shirvany, Vendramini, Yazdani, Abbas-Azimi, Bogaard, Boustani, Charles,
  Dehghan, de~Torres-Espuny, Falczuk, Guerrero-Campo, Hynd, Jones, Kowsary,
  Kazemi-Saeed, Maestro-Martinez, Romo-Diez, Shaw, Siavash, Villar-Salvador,
  and Zak}{Diaz et~al.}{2004}]{diaz-2004}
Diaz, S., J.~Hodgson, K.~Thompson, M.~Cabido, J.~Cornelissen, A.~Jalili,
  G.~Montserrat-Marti, J.~Grime, F.~Zarrinkamar, Y.~Asri, S.~Band,
  S.~Basconcelo, P.~Castro-Diez, G.~Funes, B.~Hamzehee, M.~Khoshnevi,
  N.~Perez-Harguindeguy, M.~Perez-Rontome, F.~Shirvany, F.~Vendramini,
  S.~Yazdani, R.~Abbas-Azimi, A.~Bogaard, S.~Boustani, M.~Charles, M.~Dehghan,
  L.~de~Torres-Espuny, V.~Falczuk, J.~Guerrero-Campo, A.~Hynd, G.~Jones,
  E.~Kowsary, F.~Kazemi-Saeed, M.~Maestro-Martinez, A.~Romo-Diez, S.~Shaw,
  B.~Siavash, P.~Villar-Salvador, and M.~Zak (2004, {JUN}).
\newblock The plant traits that drive ecosystems: Evidence from three
  continents.
\newblock {\em Journal of Vegetation Science\/}~{\em {15}\/}({3}), {295--304}.

\bibitem[\protect\citeauthoryear{Drake}{Drake}{2005}]{drake2005population}
Drake, J.~M. (2005).
\newblock Population effects of increased climate variation.
\newblock {\em Proceedings of the Royal Society of London B: Biological
  Sciences\/}~{\em 272\/}(1574), 1823--1827.

\bibitem[\protect\citeauthoryear{Ellner}{Ellner}{1987}]{ellner1987alternate}
Ellner, S. (1987).
\newblock Alternate plant life history strategies and coexistence in randomly
  varying environments.
\newblock In {\em Theory and models in vegetation science}, pp.\  199--208.
  Springer.

\bibitem[\protect\citeauthoryear{Garnier}{Garnier}{1992}]{garnier-1992}
Garnier, E. ({1992}).
\newblock Growth analysis of congeneric annual and perennial grass species.
\newblock {\em Journal of Ecology\/}~{\em {80}\/}({4}), {665--675}.

\bibitem[\protect\citeauthoryear{Grime and Hunt}{Grime and
  Hunt}{1975}]{grime-1975}
Grime, J. and R.~Hunt ({1975}).
\newblock Relative growth rate - its range and adaptive significance in a local
  flora.
\newblock {\em Journal of Ecology\/}~{\em {63}\/}({2}), {393--422}.

\bibitem[\protect\citeauthoryear{Hardle, Hall, Ichimura, et~al.}{Hardle
  et~al.}{1993}]{hardle1993optimal}
Hardle, W., P.~Hall, H.~Ichimura, et~al. (1993).
\newblock Optimal smoothing in {S}ingle-{I}ndex {M}odels.
\newblock {\em The annals of Statistics\/}~{\em 21\/}(1), 157--178.

\bibitem[\protect\citeauthoryear{Hutchinson}{Hutchinson}{1961}]{hutchinson1961paradox}
Hutchinson, G.~E. (1961).
\newblock The paradox of the plankton.
\newblock {\em The American Naturalist\/}~{\em 95\/}(882), 137--145.

\bibitem[\protect\citeauthoryear{Koons, Pavard, Baudisch, Metcalf,
  et~al.}{Koons et~al.}{2009}]{koons2009life}
Koons, D.~N., S.~Pavard, A.~Baudisch, J.~E. Metcalf, et~al. (2009).
\newblock Is life-history buffering or lability adaptive in stochastic
  environments?
\newblock {\em Oikos\/}~{\em 118\/}(7), 972--980.

\bibitem[\protect\citeauthoryear{Lewontin and Cohen}{Lewontin and
  Cohen}{1969}]{lewontin1969population}
Lewontin, R.~C. and D.~Cohen (1969).
\newblock On population growth in a randomly varying environment.
\newblock {\em Proceedings of the National Academy of Sciences\/}~{\em
  62\/}(4), 1056--1060.

\bibitem[\protect\citeauthoryear{Ruppert, Wand, and Carroll}{Ruppert
  et~al.}{2003}]{ruppert2003semiparametric}
Ruppert, D., M.~P. Wand, and R.~J. Carroll (2003).
\newblock {\em Semiparametric regression}.
\newblock Cambridge University Press.

\bibitem[\protect\citeauthoryear{Teller, Adler, Edwards, Hooker, and
  Ellner}{Teller et~al.}{2016}]{teller2016linking}
Teller, B.~J., P.~B. Adler, C.~B. Edwards, G.~Hooker, and S.~P. Ellner (2016).
\newblock Linking demography with drivers: climate and competition.
\newblock {\em Methods in Ecology and Evolution\/}~{\em 7\/}(2), 171--183.

\bibitem[\protect\citeauthoryear{Vasseur, DeLong, Gilbert, Greig, Harley,
  McCann, Savage, Tunney, and O’Connor}{Vasseur
  et~al.}{2014}]{vasseur-etal-2014}
Vasseur, D.~A., J.~P. DeLong, B.~Gilbert, H.~S. Greig, C.~D.~G. Harley, K.~S.
  McCann, V.~Savage, T.~D. Tunney, and M.~I. O’Connor (2014).
\newblock Increased temperature variation poses a greater risk to species than
  climate warming.
\newblock {\em Proceedings of the Royal Society B - Biological Sciences\/}~{\em
  281}, 20132612.

\bibitem[\protect\citeauthoryear{Vasseur and McCann}{Vasseur and
  McCann}{2007}]{vasseur-mccann-2007}
Vasseur, D.~A. and K.~S. McCann (2007).
\newblock {\em The Impact of Environmental Variability on Ecological Systems}.
\newblock Springer Netherlands.

\bibitem[\protect\citeauthoryear{Wood}{Wood}{2006}]{wood2006generalized}
Wood, S.~N. (2006).
\newblock {\em {G}eneralized {A}dditive {M}odels: An Introduction with {R}}.
\newblock Chapman and Hall/CRC.

\bibitem[\protect\citeauthoryear{Ye and Hooker}{Ye and
  Hooker}{2020}]{ye2018local}
Ye, Z. and G.~Hooker (2020).
\newblock Local quadratic estimation of the curvature in a {F}unctional
  {S}ingle {I}ndex {M}odel.
\newblock {\em Scanadanavian Journal of Statistics\/}~{\em 47\/}(4),
  1307--1338.

\bibitem[\protect\citeauthoryear{Ye, Hooker, and Ellner}{Ye
  et~al.}{2020}]{ye2019jensen}
Ye, Z., G.~Hooker, and S.~Ellner (2020).
\newblock The {J}ensen {E}ffect and {F}unctional {S}ingle {I}ndex {M}odels:
  Estimating the ecological implications of nonlinear reaction norms.
\newblock {\em Annals of Applied Statistics\/}~{\em 14\/}(3), 1326--12341.

\end{thebibliography}

\appendix

\section{Diagnostic Plots for the Jensen Effect: Exponential Model} \label{appendix:sim}

\begin{figure}[H]
\centering
\begin{tabular}{cc}
\includegraphics[width =0.45\textwidth,height = 0.3\textheight]{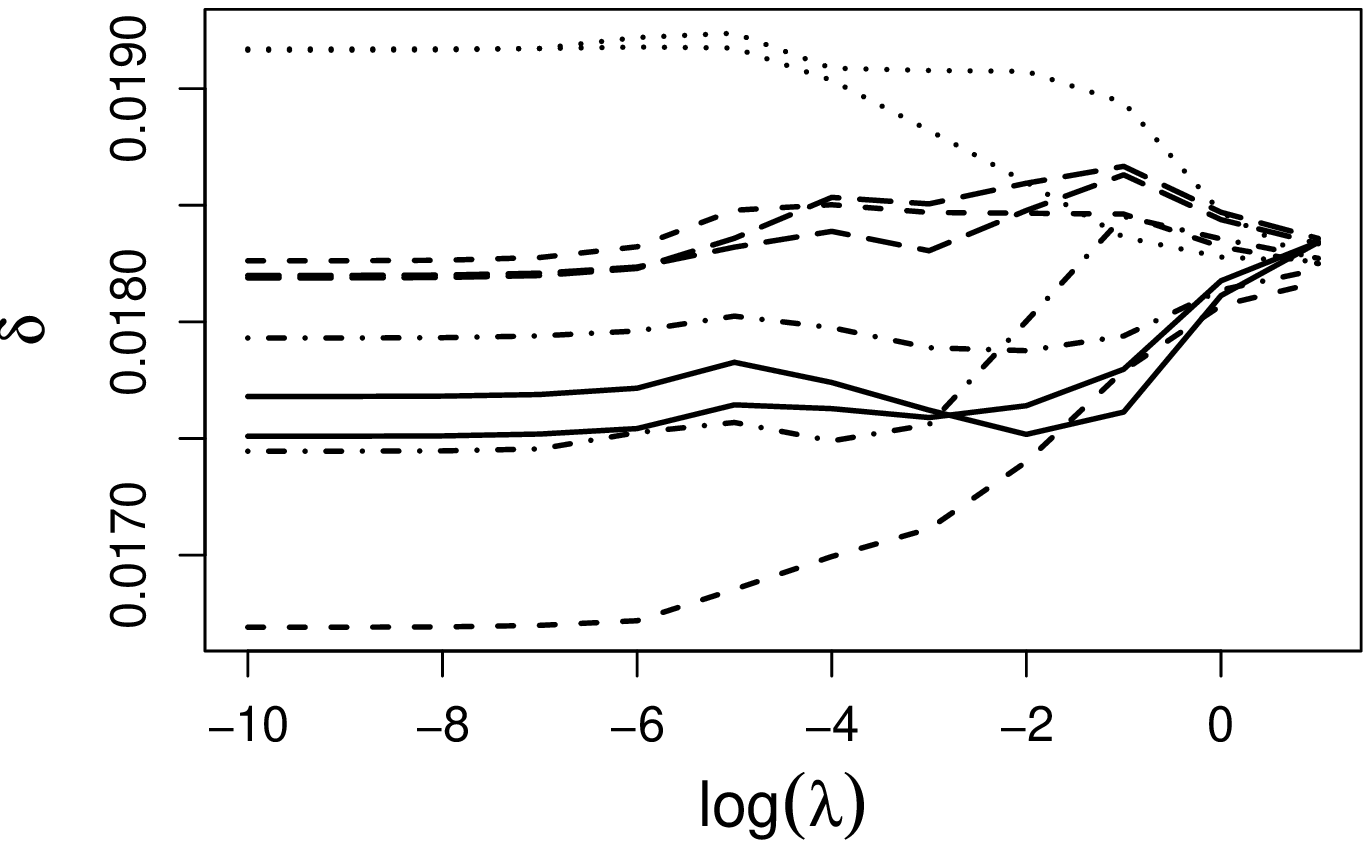} 
\includegraphics[width =0.45\textwidth,height = 0.3\textheight]{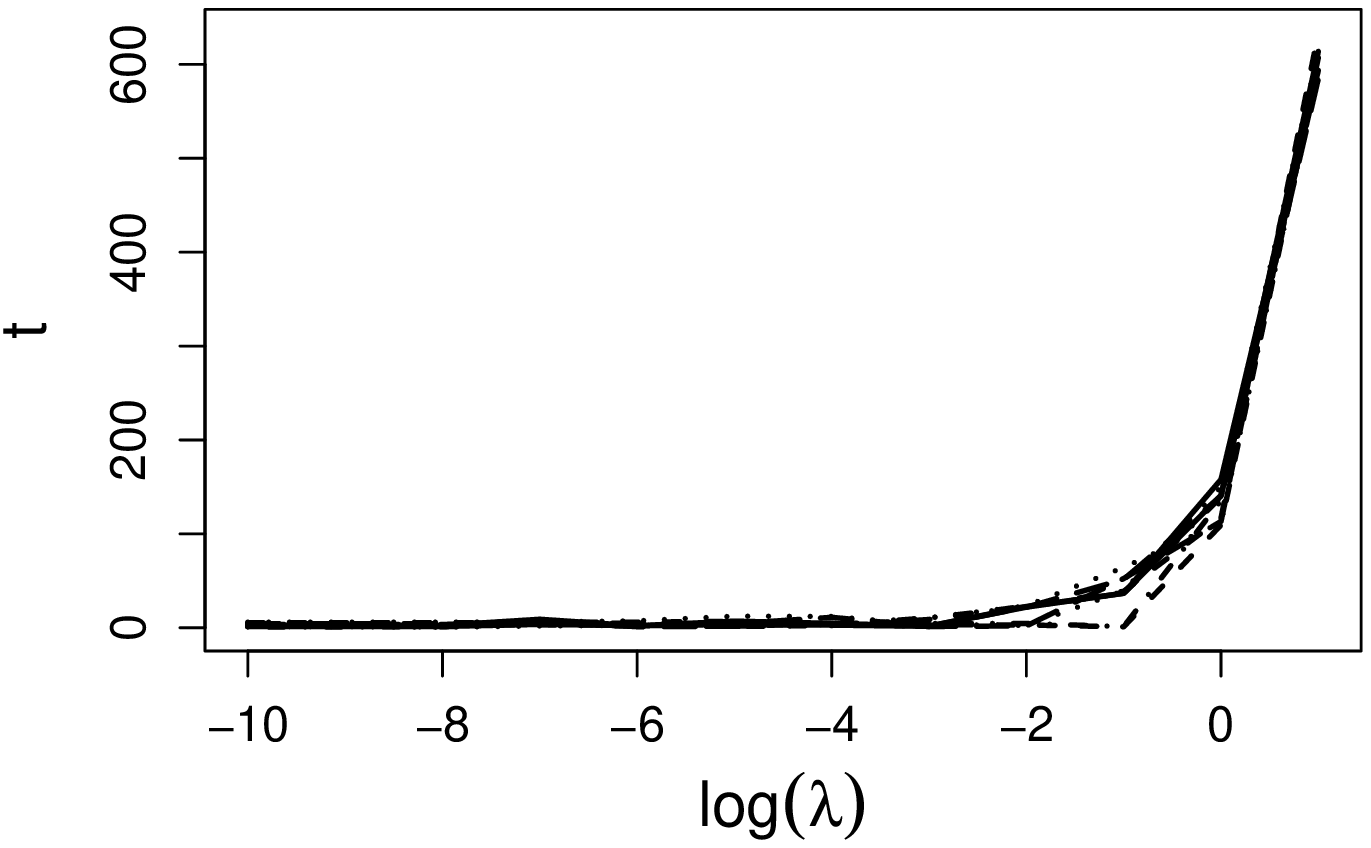}
\end{tabular}
\caption{Jensen Effect in a exponential single index model with composite link function $g^{\ast}\left(s\right) = \exp\left(s\right)$. Left: a sample of the Jensen Effect values $\delta_{\lambda}$ as a function of smoothing parameters $\lambda$. Right: the corresponding t-statistics $t_{\lambda}$ functions.}
\label{figure:SIM_convex}
\end{figure}

\begin{figure}[H]
\centering
\begin{tabular}{cc}
\includegraphics[width =0.45\textwidth,height = 0.3\textheight]{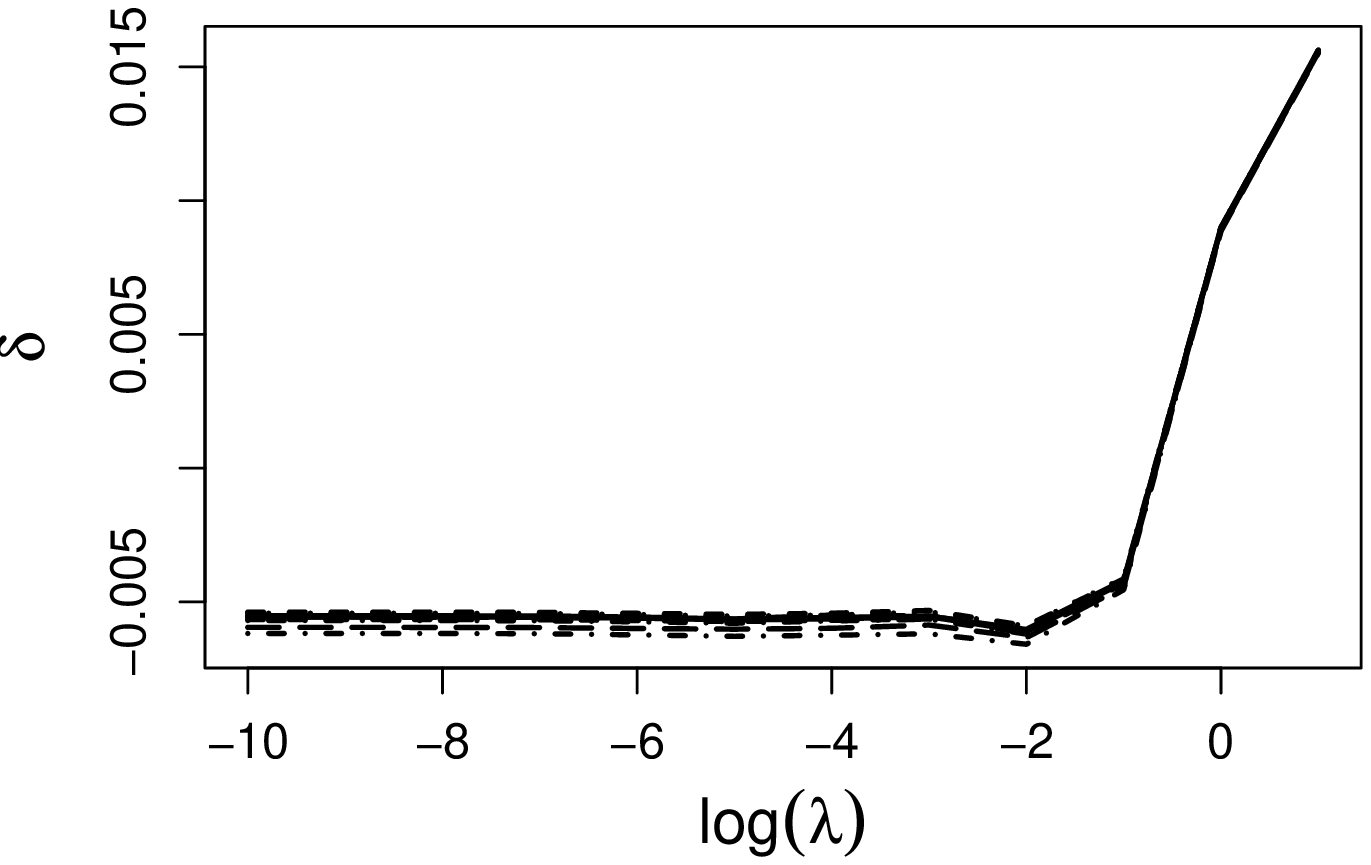} 
\includegraphics[width =0.45\textwidth,height = 0.3\textheight]{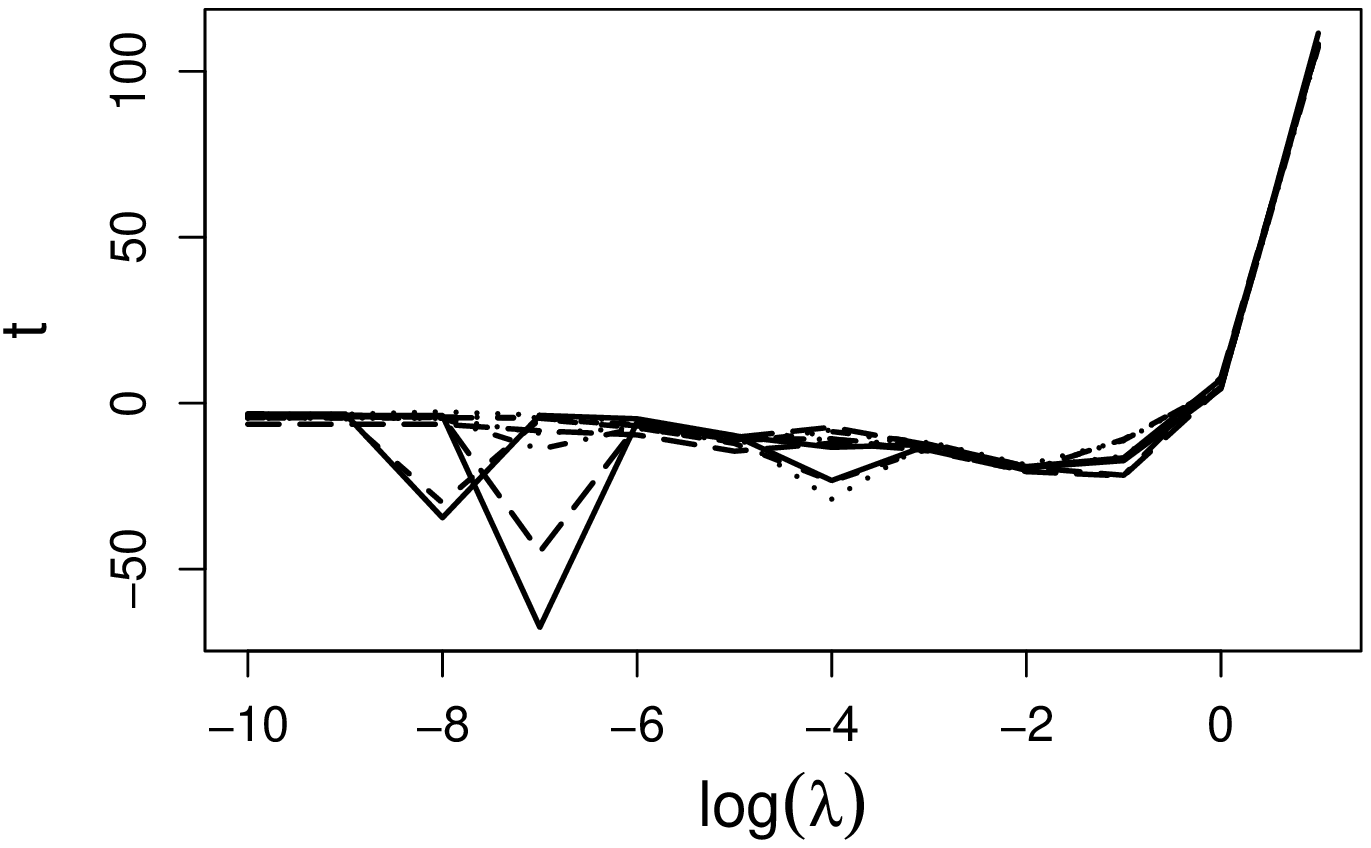}
\end{tabular}
\caption{Illustration of the Jensen Effect for a concave composite link function in the exponential single index model, where the link function is $g^{\ast}\left(s\right) = \sin\left(s\right)$. Left: a sample of the Jensen Effect values $\delta_{\lambda}$ as a function of smoothing parameters $\lambda$. Right: the corresponding t-statistics $t_{\lambda}$ functions.}
\label{figure:SIM_sin}
\end{figure}

\begin{figure}[H]
\centering
\begin{tabular}{cc}
\includegraphics[width =0.45\textwidth,height = 0.3\textheight]{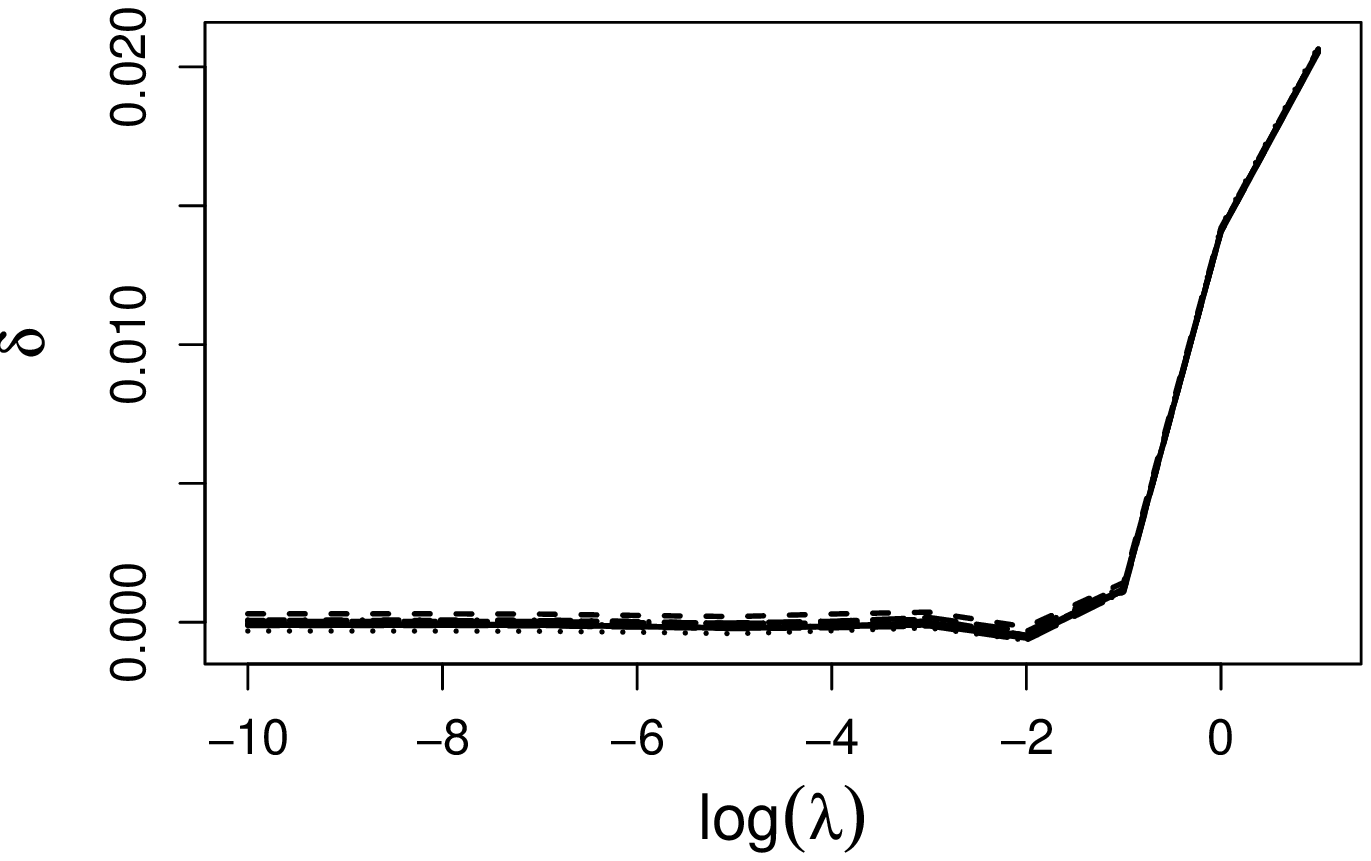}
\includegraphics[width =0.45\textwidth,height = 0.3\textheight]{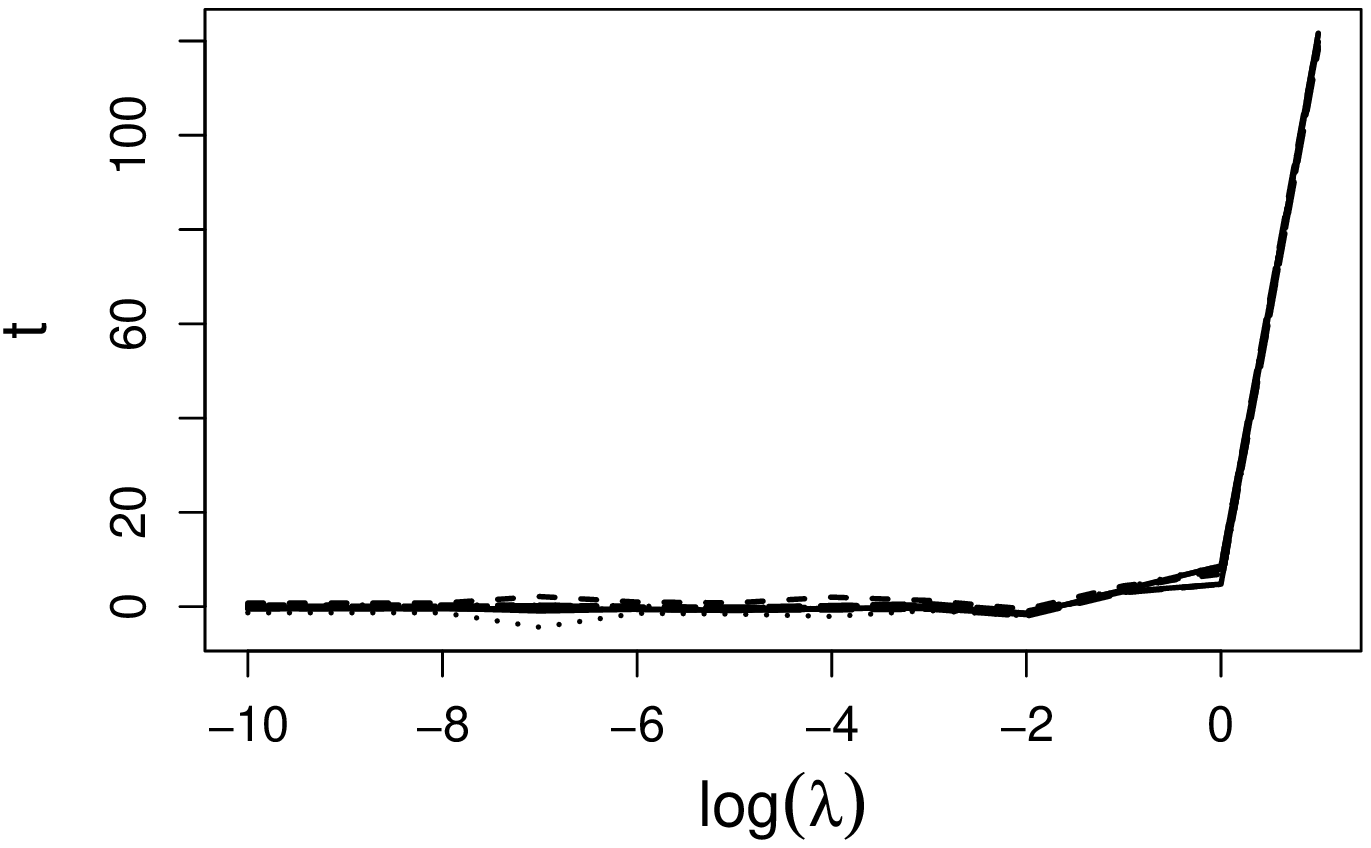}
\end{tabular}
\caption{Illustration of the Jensen Effect for a linear composite link function in the exponential single index model, where the link function is $g^{\ast}\left(s\right) = s$. Left: a sample of the Jensen Effect values $\delta_{\lambda}$ as a function of smoothing parameters $\lambda$. Right: the corresponding t-statistics $t_{\lambda}$ functions.}
\label{figure:SIM_linear}
\end{figure}

\section{Diagnostic Plots for the Jensen Effect: Poisson Model} \label{appendix:sim:pois}

\begin{figure}[H]
\centering
\begin{tabular}{cc}
\includegraphics[width =0.45\textwidth,height = 0.3\textheight]{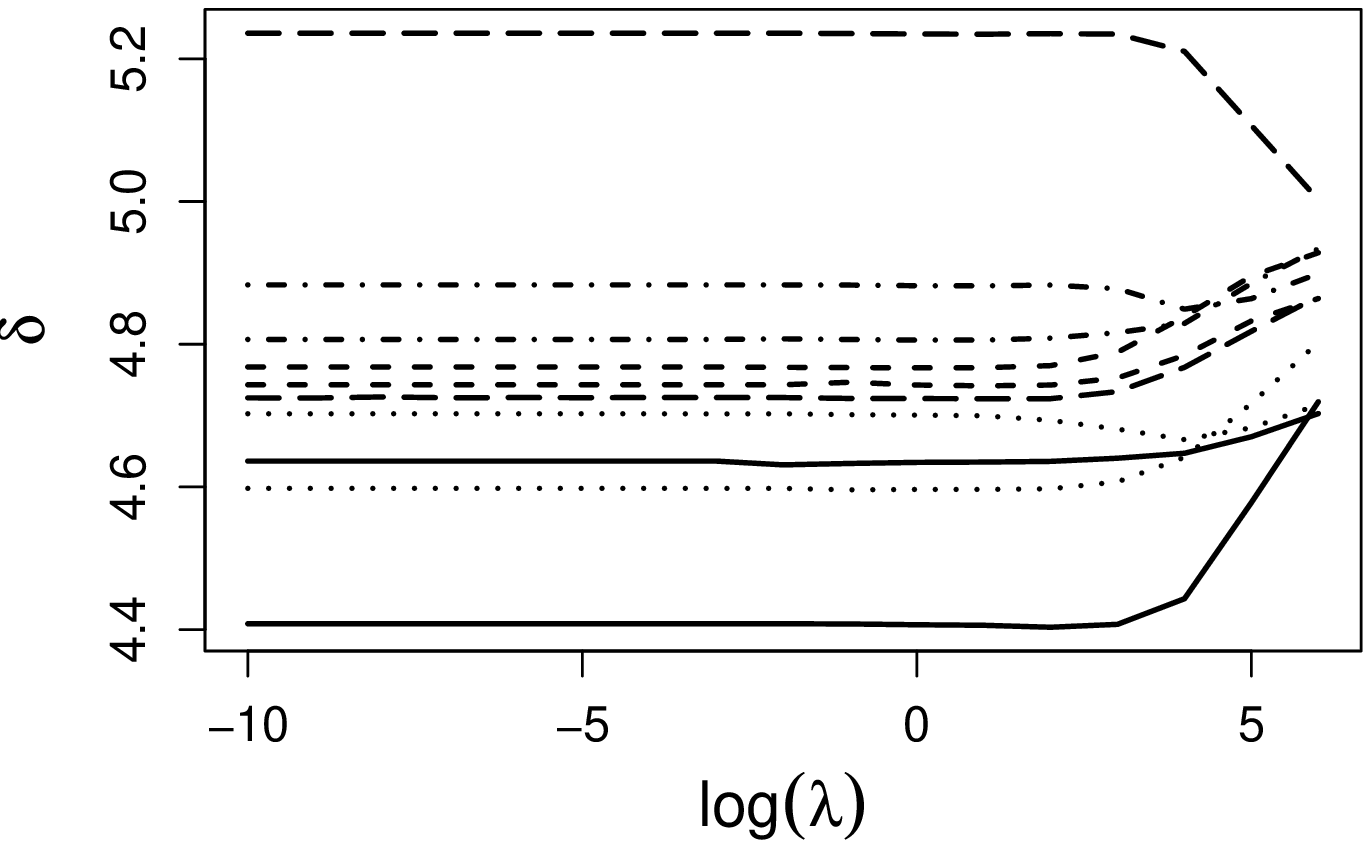} 
\includegraphics[width =0.45\textwidth,height = 0.3\textheight]{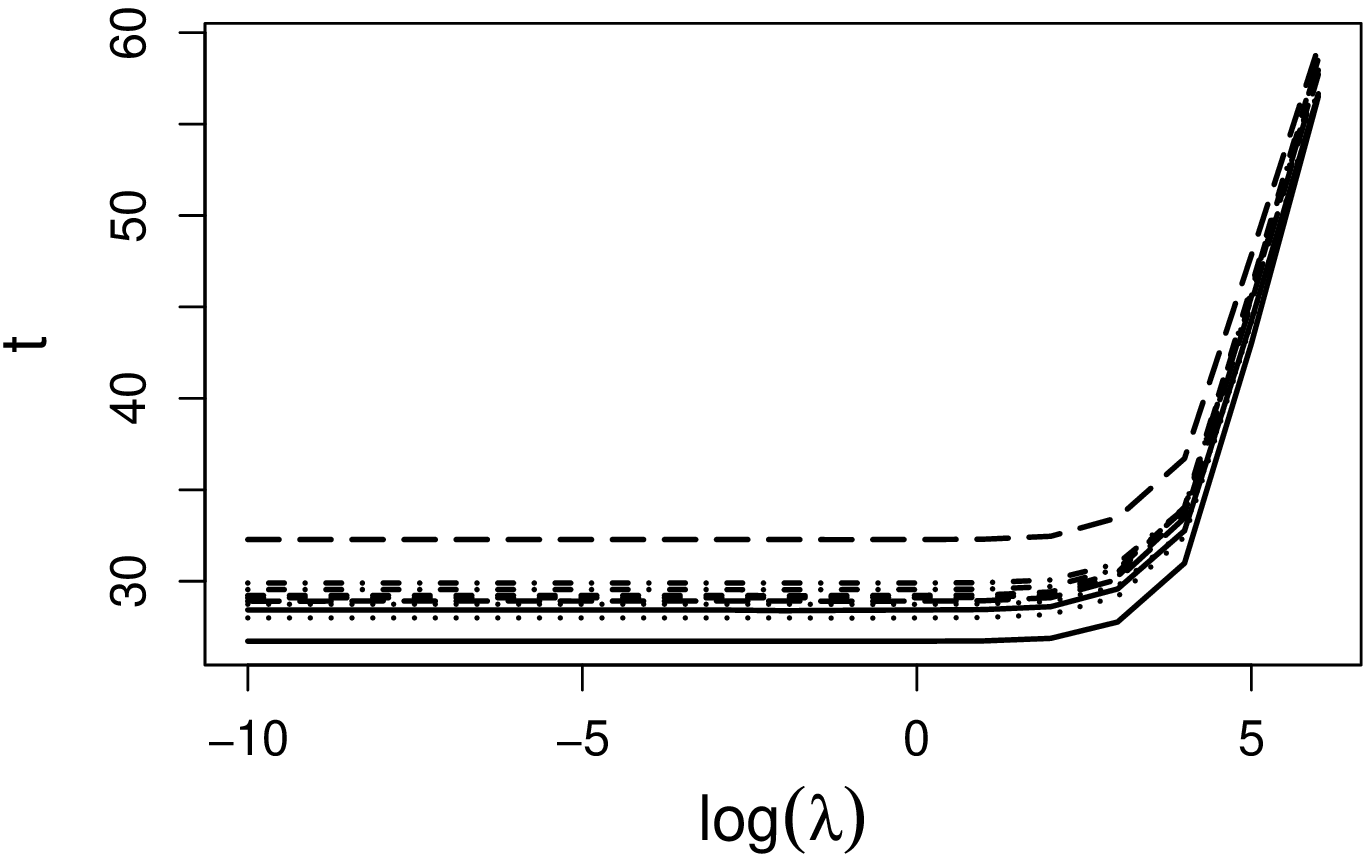}
\end{tabular}
\caption{Jensen Effect in Poisson regression with composite link $g^{\ast}\left(s\right) = \exp\left(\frac{s}{8}\right)$. Left: a sample of the Jensen Effect values $\delta_{\lambda}$ as a function of smoothing parameters $\lambda$. Right: the corresponding t-statistics $t_{\lambda}$ functions.}
\label{figure:SIM_convex_pois}
\end{figure}

\begin{figure}[H]
\centering
\begin{tabular}{cc}
\includegraphics[width =0.45\textwidth,height = 0.3\textheight]{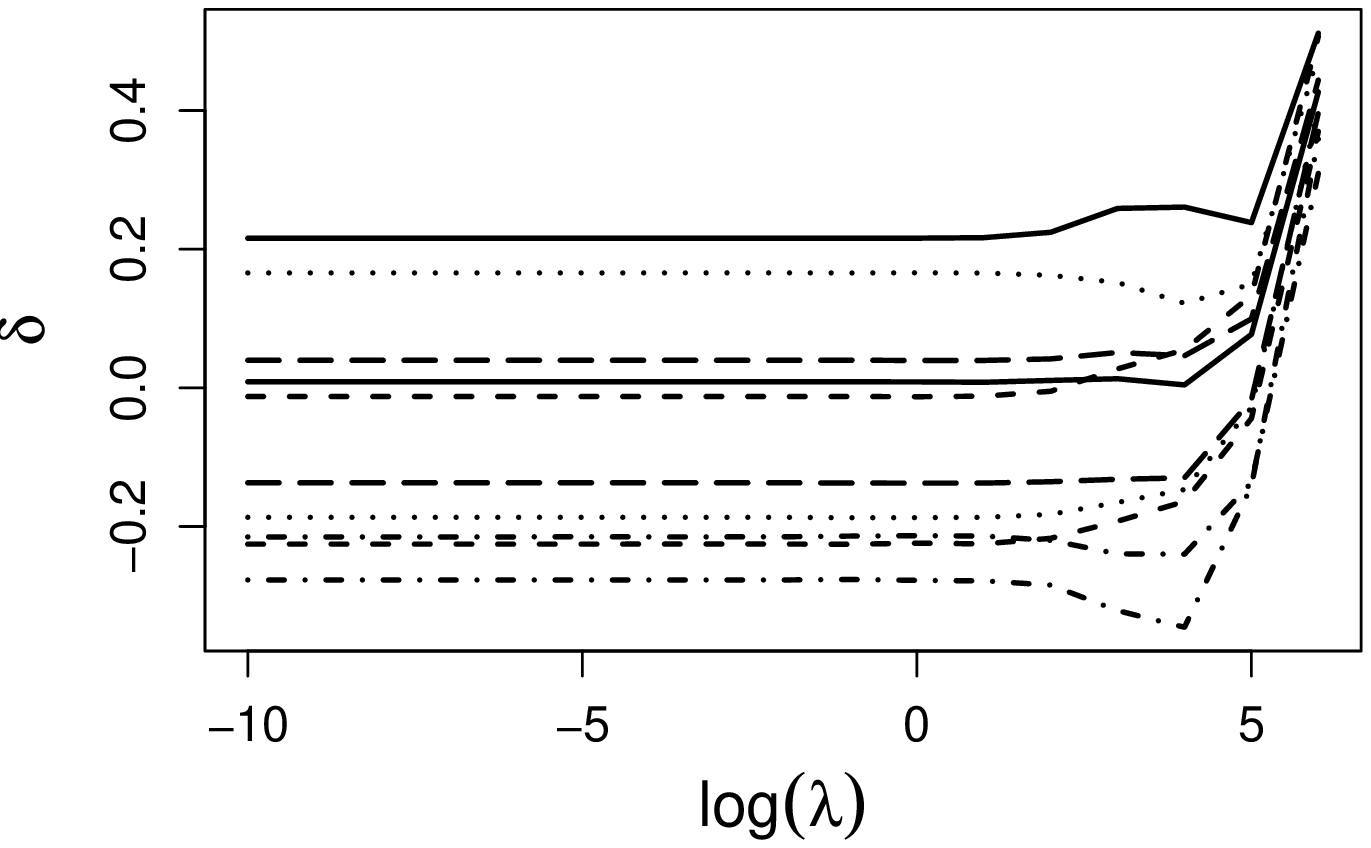} 
\includegraphics[width =0.45\textwidth,height = 0.3\textheight]{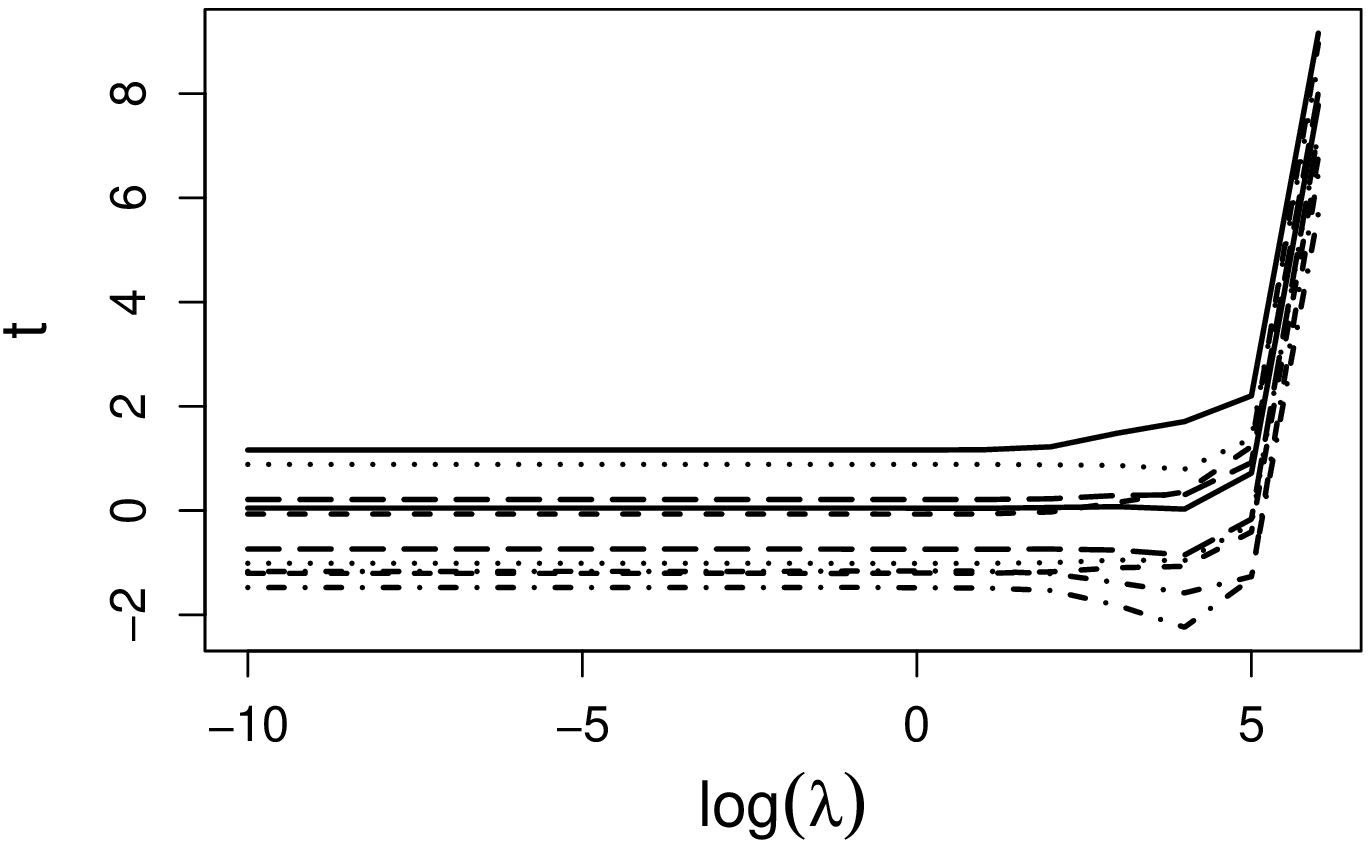}
\end{tabular}
\caption{Jensen Effect in Poisson regression with composite link $g^{\ast}\left(s\right) = s$. Left: a sample of the Jensen Effect values $\delta_{\lambda}$ as a function of smoothing parameters $\lambda$. Right: the corresponding t-statistics $t_{\lambda}$ functions.}
\label{figure:concave_linear}
\end{figure}

\section{Plots for USSES Plants data in Poisson Single Index Model} \label{appendix:real:pois}

\begin{figure}[H]
\centering
\begin{tabular}{cc}
\includegraphics[width =0.45\textwidth,height = 0.3\textheight]{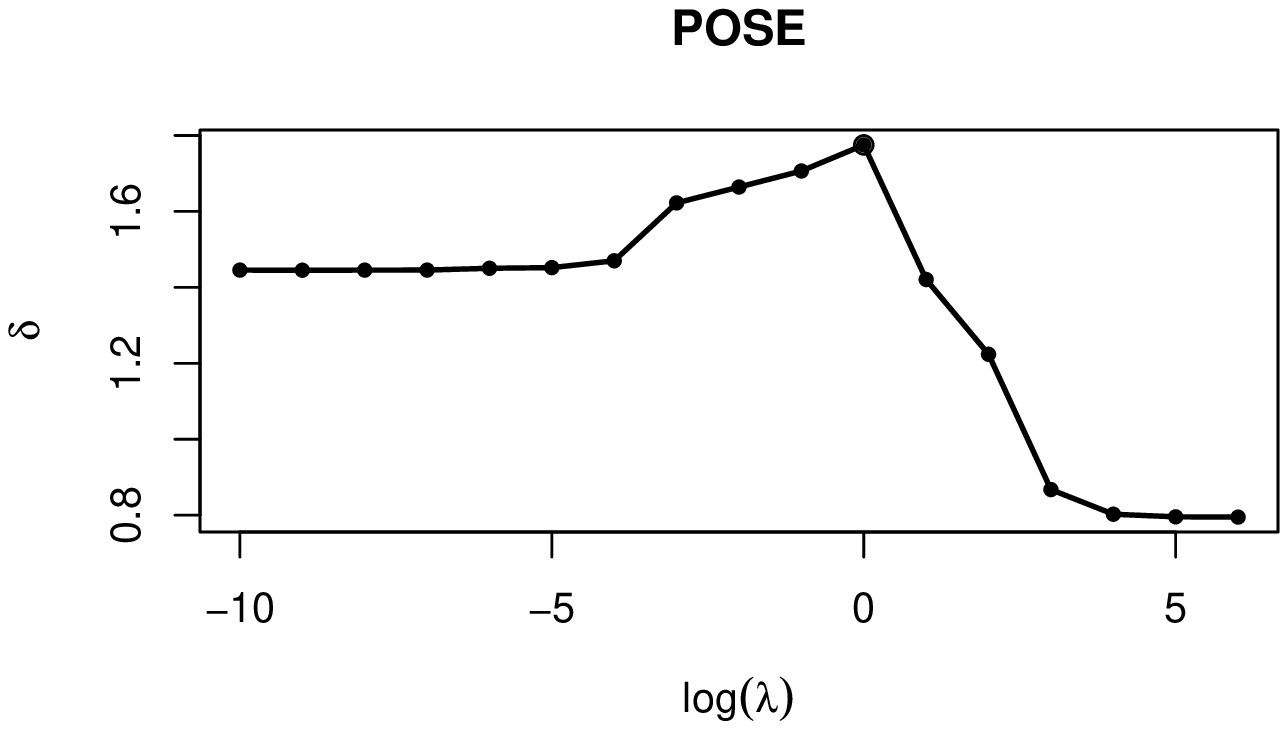} 
\includegraphics[width =0.45\textwidth,height = 0.3\textheight]{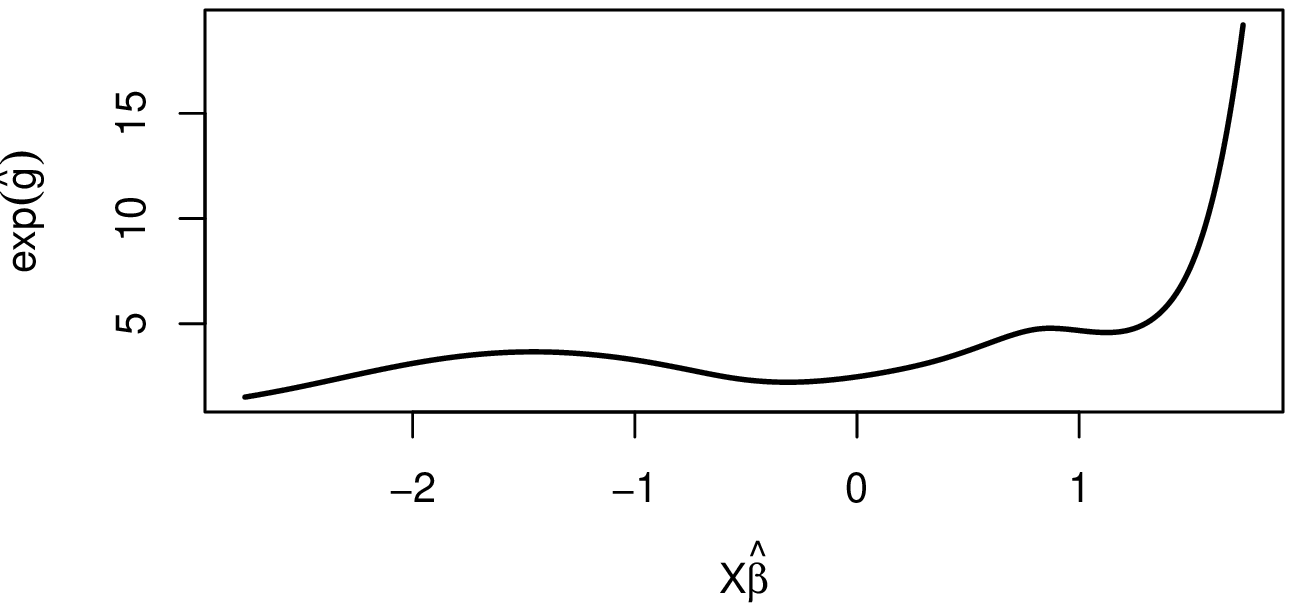}\\
\end{tabular}
\caption{Jensen Effect for reproduction by the grass \emph{Poa secunda} (POSE) in USSES data set, where we model the seedling count data using a Poisson single index model. Left: the Jensen Effect values $\delta$ plotted against the values of $\lambda$. Right: the curve of the exponential link function $\exp\left(\hat{g}\right)$, where $\lambda$ is selected by GCV.}
\label{fig:POSE}
 \end{figure}

\begin{figure}[H]
\centering
\begin{tabular}{cc}
\includegraphics[width =0.45\textwidth]{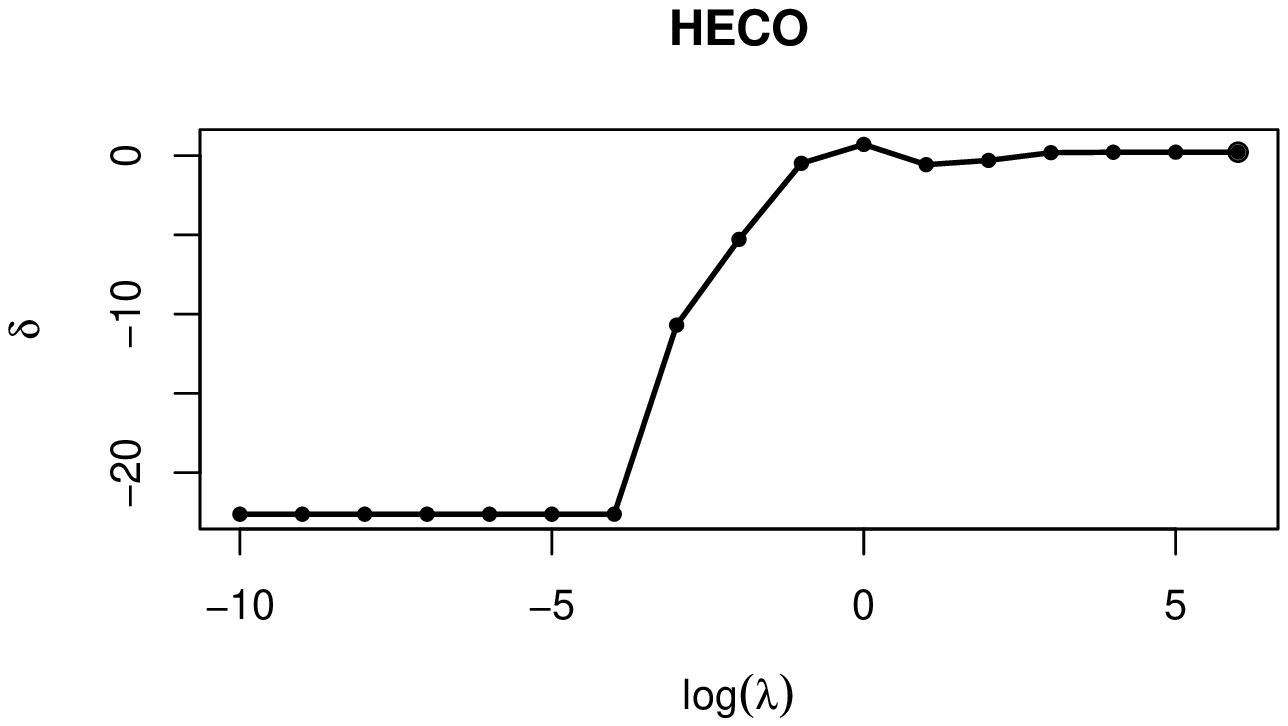} 
\includegraphics[width =0.45\textwidth]{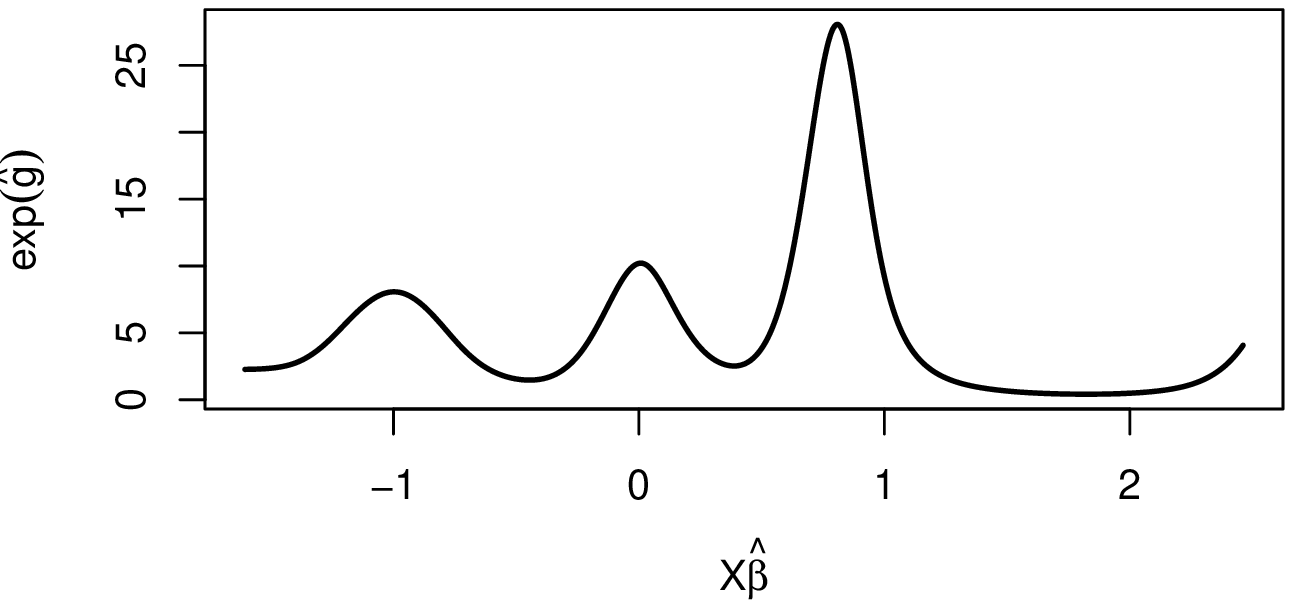}\\
\end{tabular}
\caption{Jensen Effect for reproduction by the grass \emph{Hesperostipa comata} (HECO) in USSES data set, where we model the seedling count data a Poisson single index model. Left: the Jensen Effect values $\delta$ plotted against the values of $\lambda$. Right: the curve of the exponential link function $\exp\left(\hat{g}\right)$, where $\lambda$ is corresponding to the minimum of $\delta$.}
\label{fig:HECO}
 \end{figure}

\section{Plots for USSES Plants data in Logistic Single Index Model} \label{appendix:real:logit}

\begin{figure}[H]
    \centering
    \begin{tabular}{cc}
    \includegraphics[height=8cm]{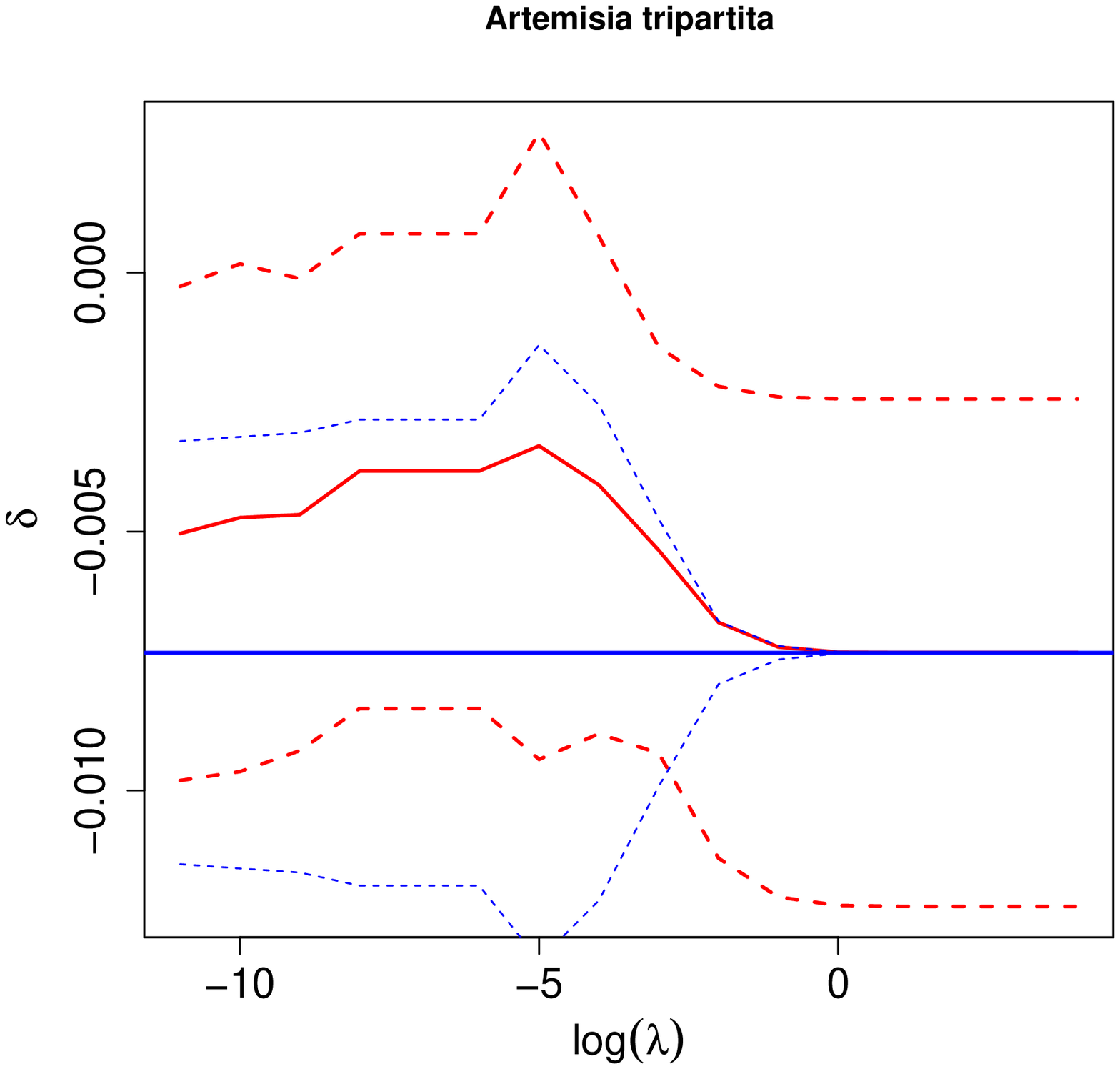} &
    \includegraphics[height=8cm]{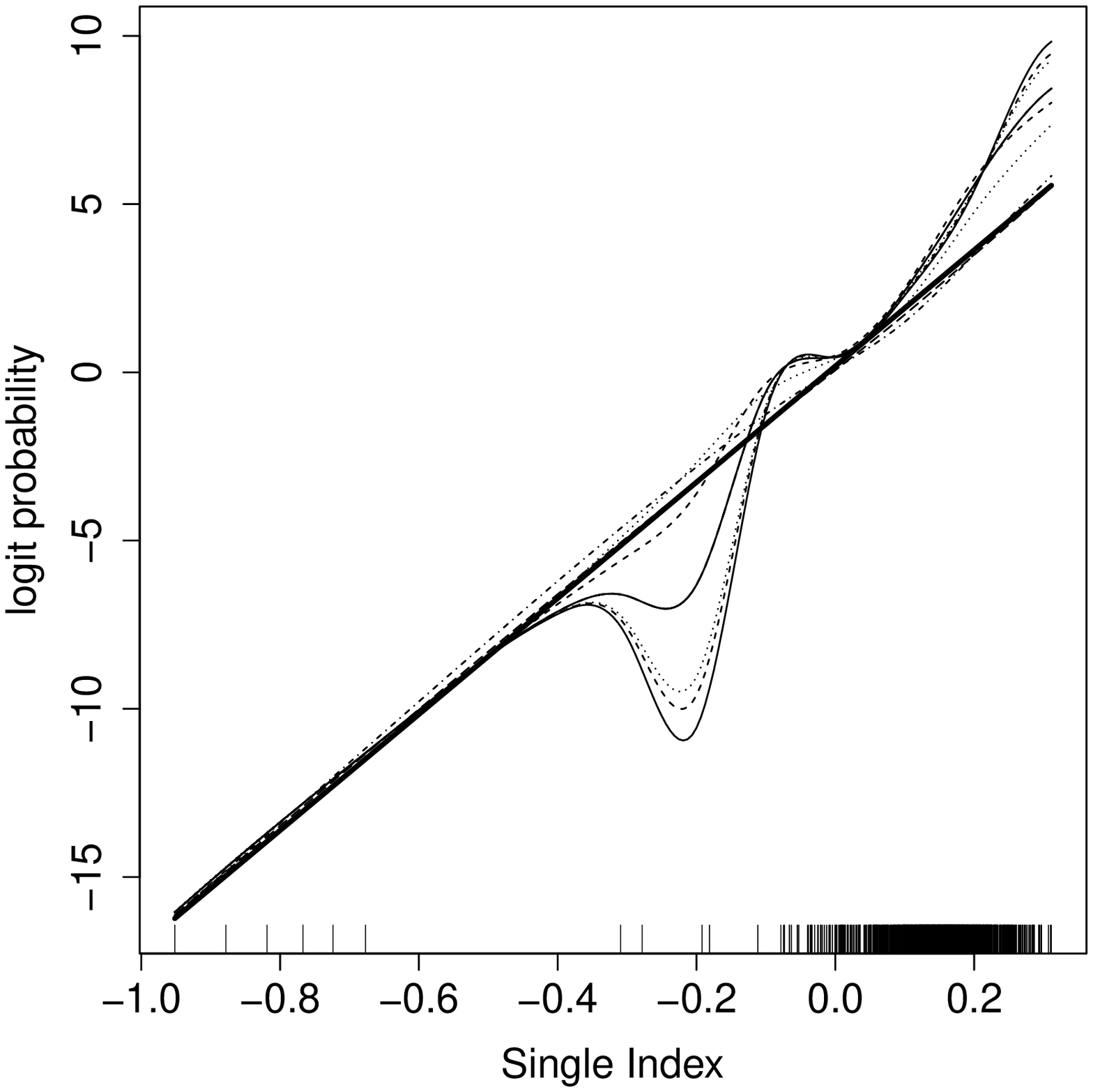}
    \end{tabular}
    \caption{Tests for the Jensen effect on survival for {\em Artemisia tripartita}. Left plots provide the value of $\delta$ estimated for each $\lambda$ (solid red lines) along with uniform confidence bands (dashed) as well as the value obtained by logistic regression (straight horizontal line) with uniform bands for the difference between that and the single index model for each $\lambda$ (blue dashed). Right plots estimate $g$ for the single index model with thick lines giving the estimate that minimizes GCV and the right plot giving the corresponding single index values. }
    \label{fig:artr_surv}
\end{figure}

\begin{figure}[H]
    \centering
    \begin{tabular}{cc}
    \includegraphics[height=8cm]{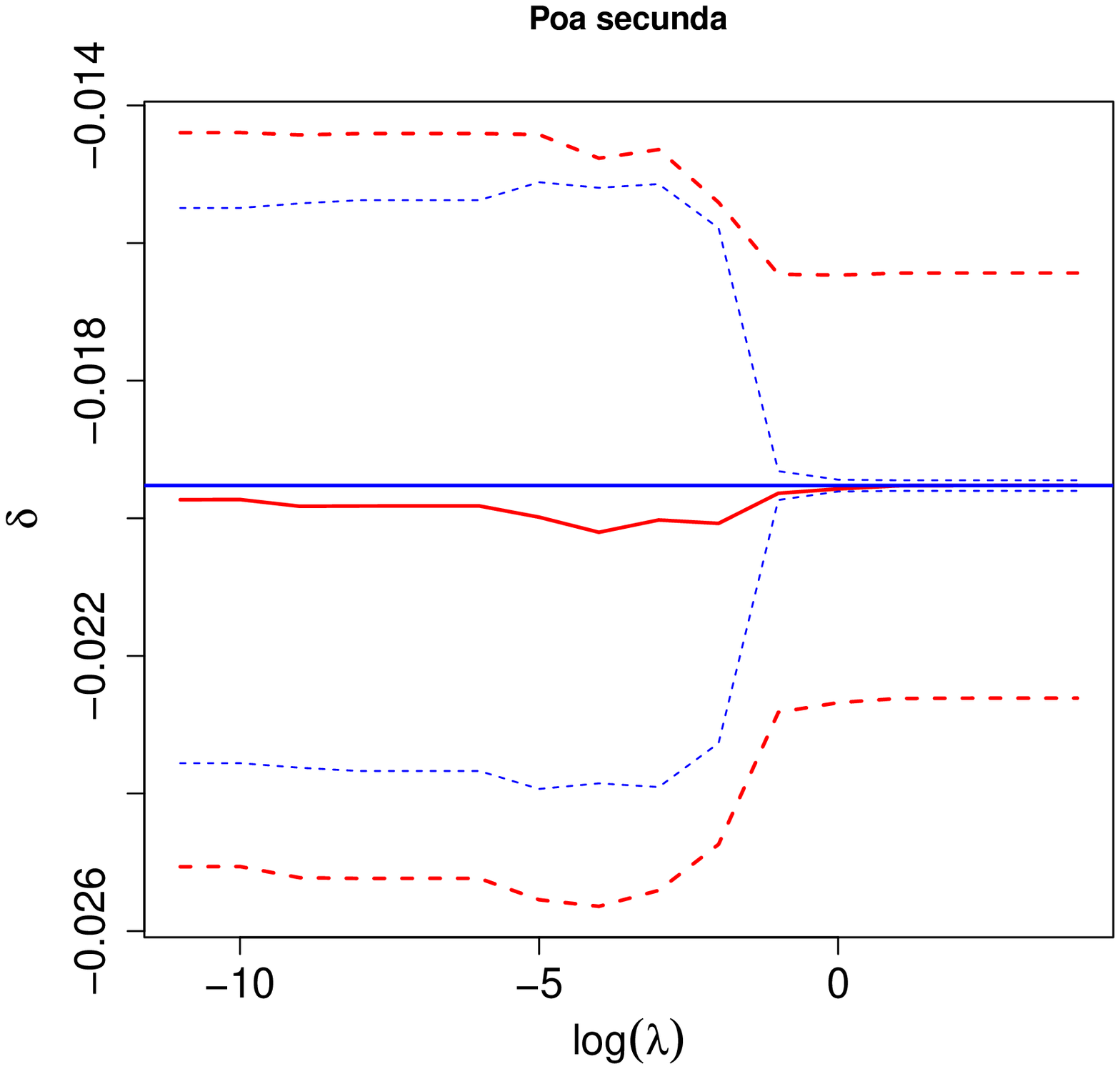} &
    \includegraphics[height=8cm]{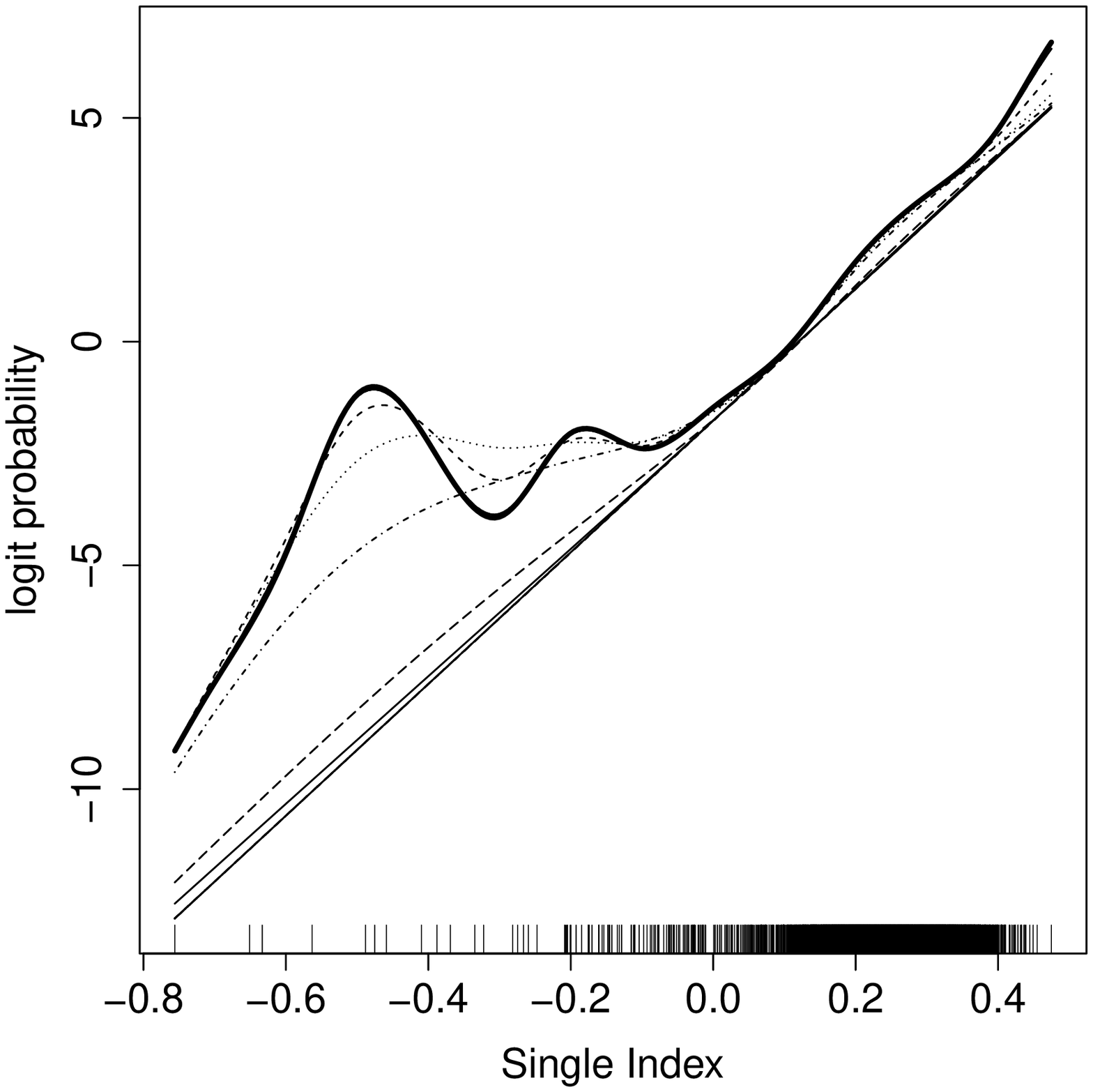}
    \end{tabular}
    \caption{Tests for the Jensen effect on survival for {\em Poa secunda}. Left plots provide the value of $\delta$ estimated for each $\lambda$ (solid red lines) along with uniform confidence bands (dashed) as well as the value obtained by logistic regression (straight horizontal line) with uniform bands for the difference between that and the single index model for each $\lambda$ (blue dashed). Right plots estimate $g$ for the single index model with thick lines giving the estimate that minimizes GCV and the right plot giving the corresponding single index values. }
    \label{fig:pose_surv}
\end{figure}

\begin{figure}[H]
    \centering
    \begin{tabular}{cc}
    \includegraphics[height=8cm]{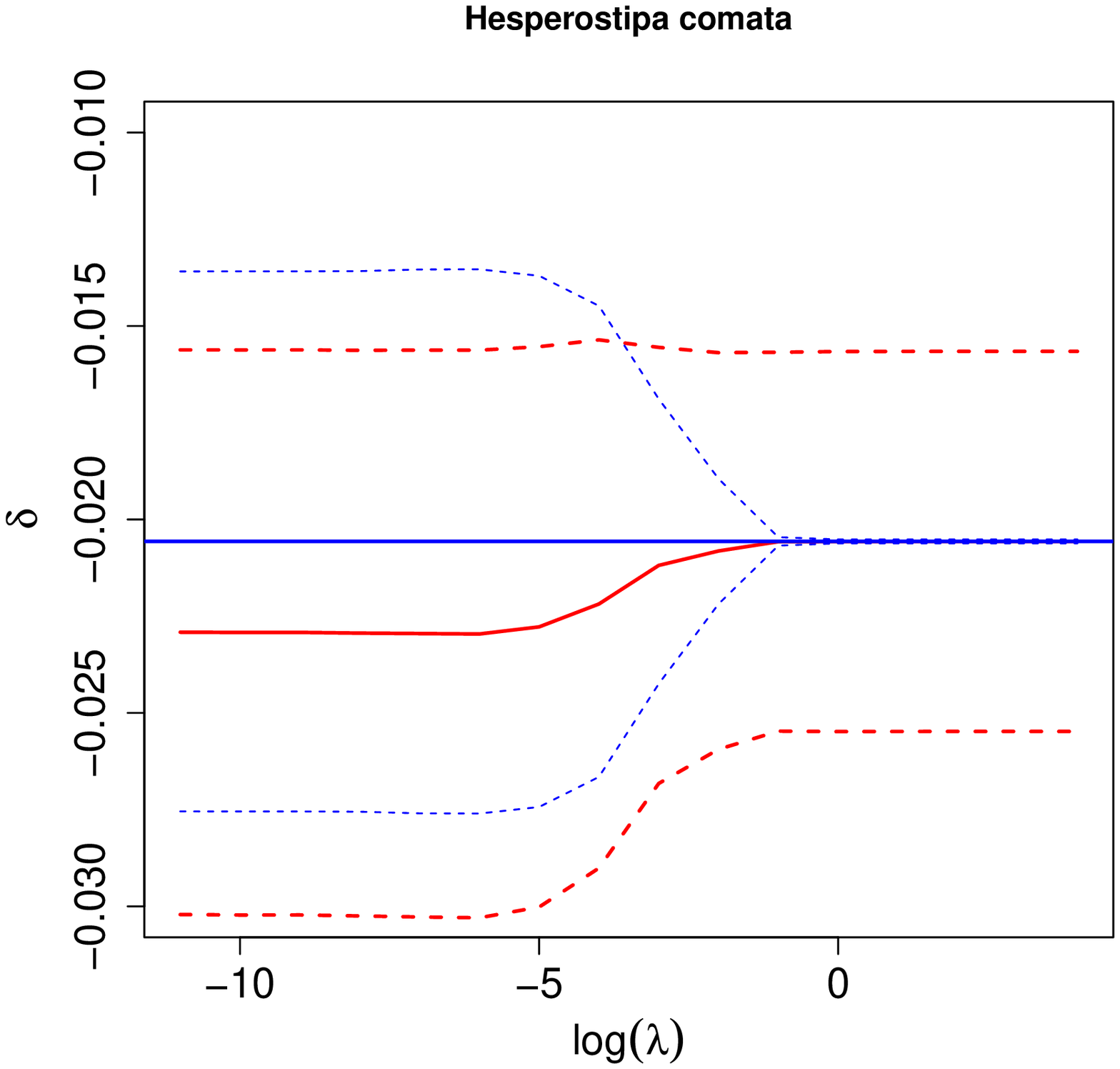} &
    \includegraphics[height=8cm]{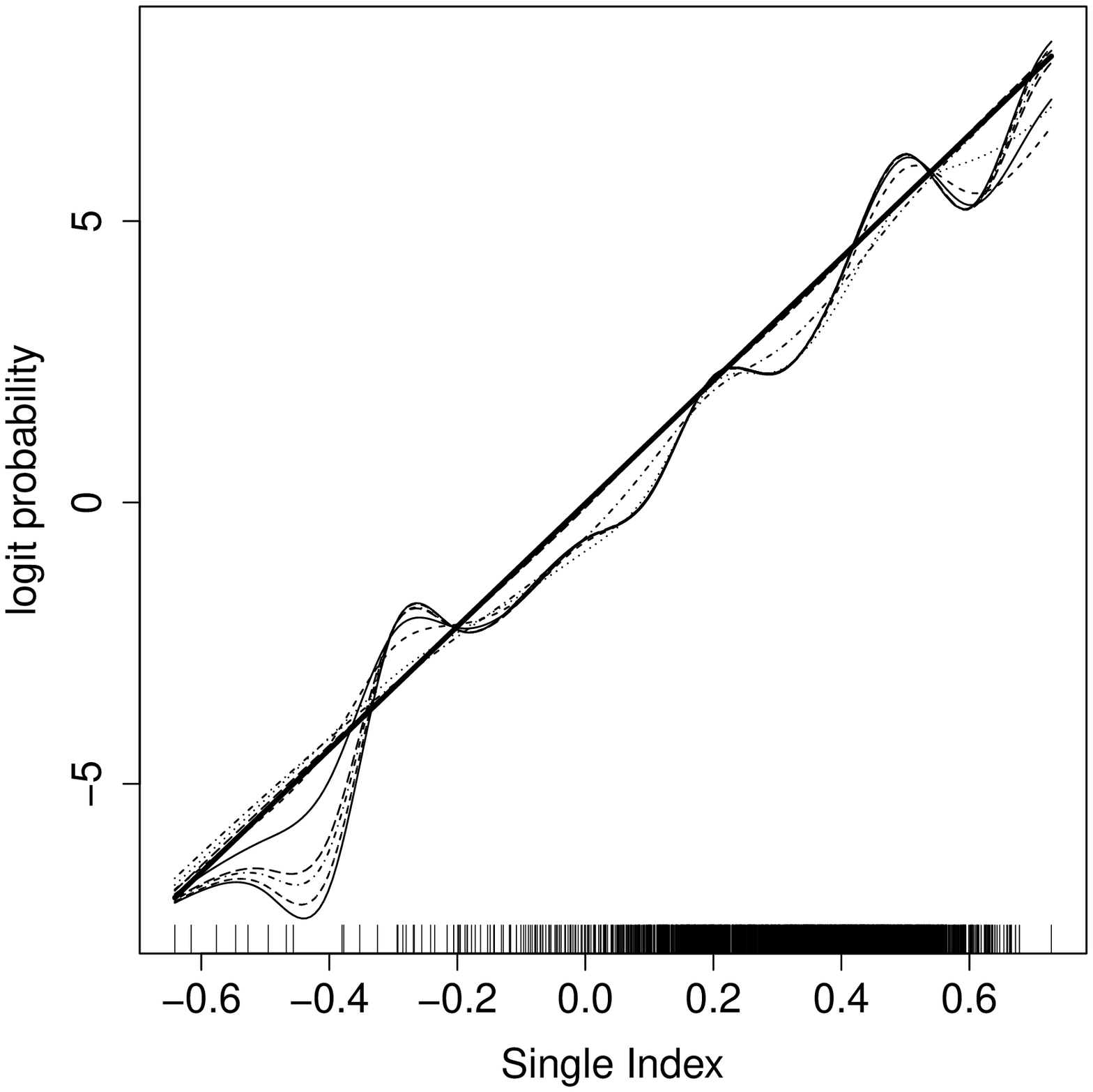}
    \end{tabular}
    \caption{Tests for the Jensen effect on survival for {\em Hesperostipa comata}. Left plots provide the value of $\delta$ estimated for each $\lambda$ (solid red lines) along with uniform confidence bands (dashed) as well as the value obtained by logistic regression (straight horizontal line) with uniform bands for the difference between that and the single index model for each $\lambda$ (blue dashed). Right plots estimate $g$ for the single index model with thick lines giving the estimate that minimizes GCV and the right plot giving the corresponding single index values. }
    \label{fig:heco_surv}
\end{figure}

\begin{figure}[H]
    \centering
    \begin{tabular}{cc}
    \includegraphics[height=8cm]{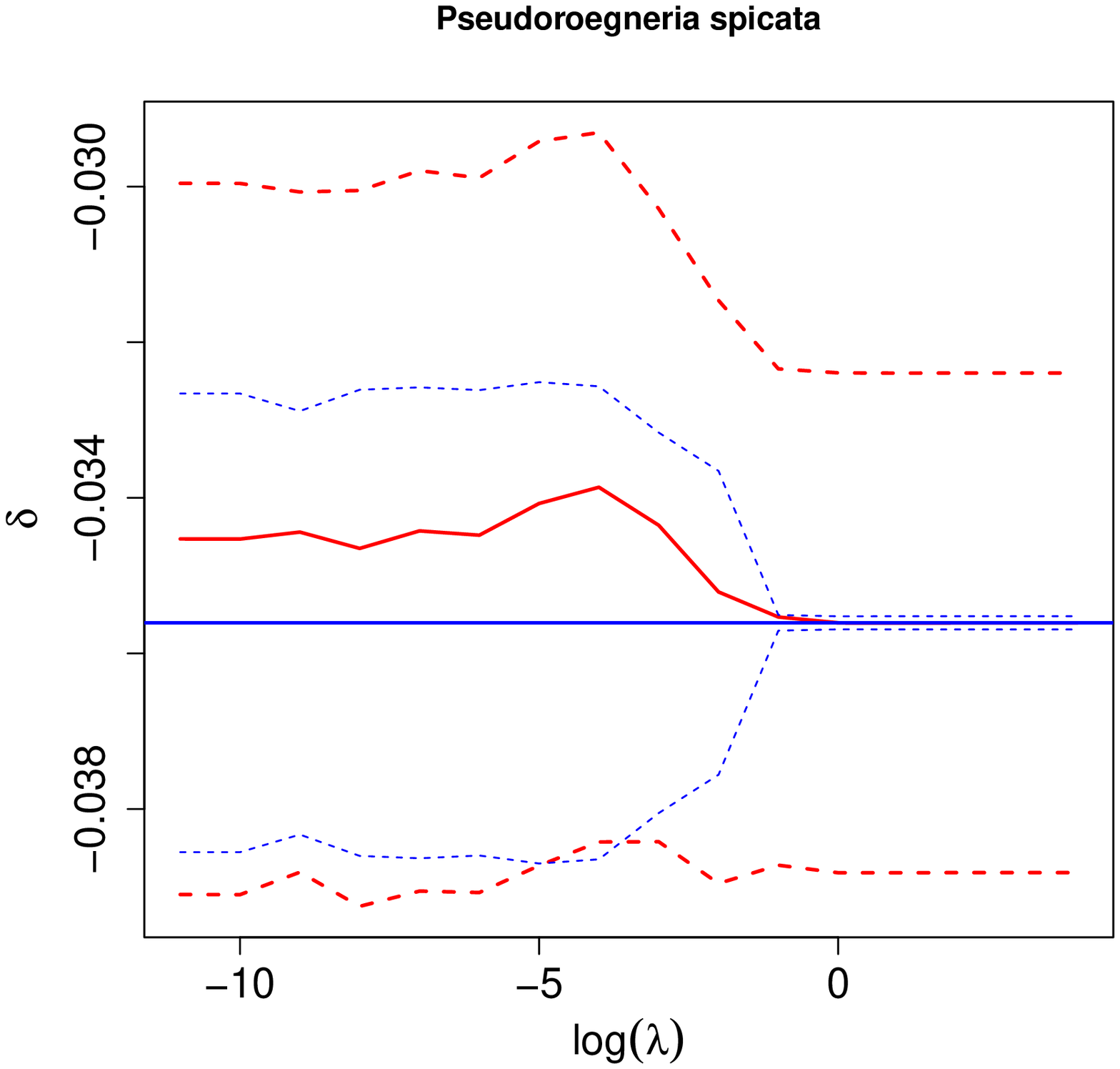} &
    \includegraphics[height=8cm]{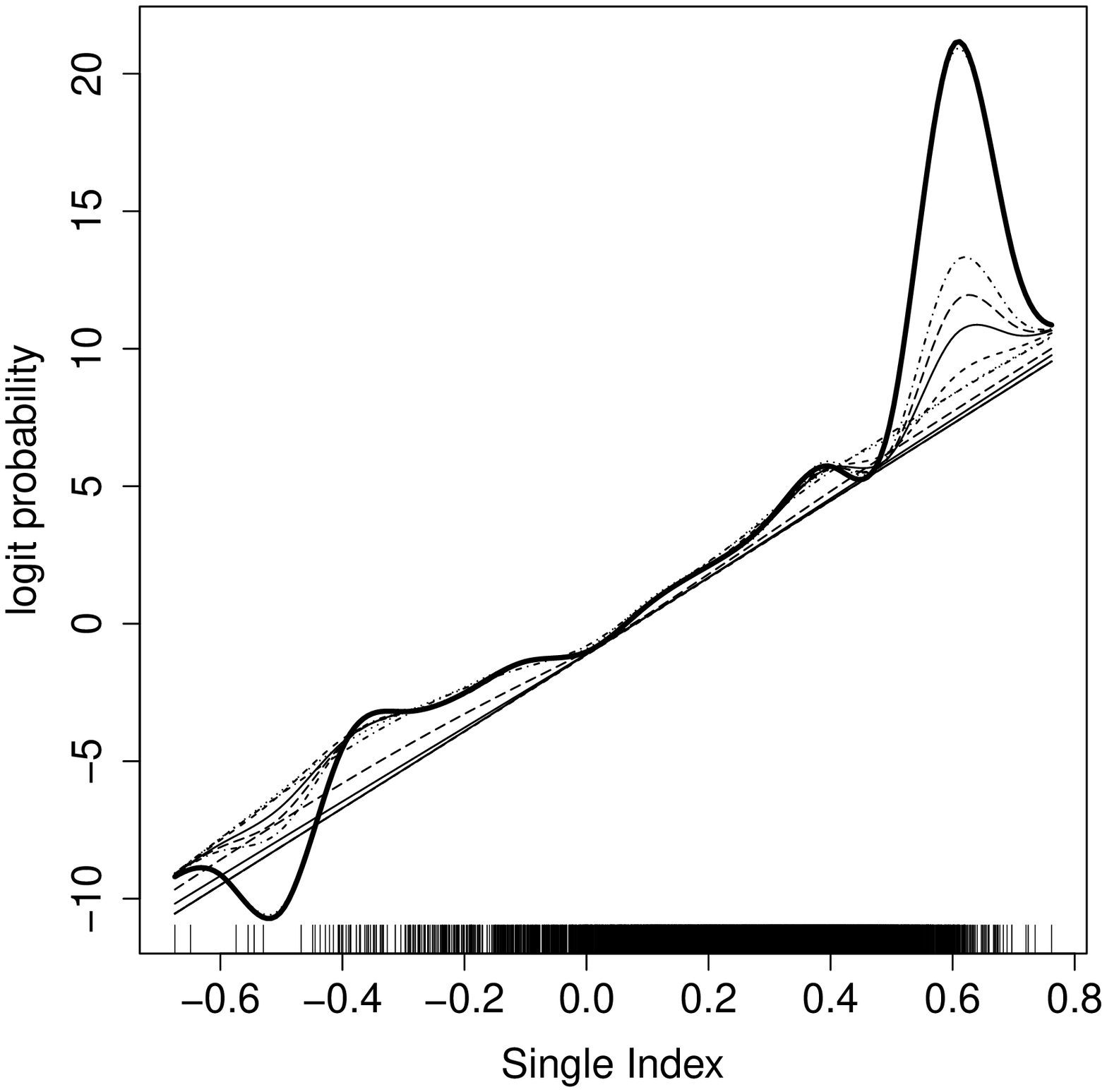}
    \end{tabular}
    \caption{Tests for the Jensen effect on survival for {\em Psuedoregneria spicata}. Left plots provide the value of $\delta$ estimated for each $\lambda$ (solid red lines) along with uniform confidence bands (dashed) as well as the value obtained by logistic regression (straight horizontal line) with uniform bands for the difference between that and the single index model for each $\lambda$ (blue dashed). Right plots estimate $g$ for the single index model with thick lines giving the estimate that minimizes GCV and the right plot giving the corresponding single index values. }
    \label{fig:pssp_surv}
\end{figure}

\end{document}